\documentclass[prb,reprint,groupedaddress,showpacs,floatfix,superscriptaddress]{revtex4-2}
\usepackage[pdftex]{graphicx}
\usepackage{dcolumn}
\usepackage{bm}
\usepackage{amssymb}
\usepackage{amsmath}
\usepackage{algorithm}
\usepackage{algpseudocode}
\usepackage{array}
\usepackage{dcolumn}
\usepackage{footmisc}
\usepackage[utf8]{inputenc}
\usepackage{tikz}
\usetikzlibrary{shapes,arrows}
\usepackage{color}
\usepackage[colorlinks=true,
            linkcolor=blue,
            urlcolor=blue,
            citecolor=blue]{hyperref}
\usepackage{mathtools}
\usepackage{soul}
\usepackage{csquotes}
\usepackage{graphicx}
\usepackage{xtab,afterpage}
\usepackage{slashbox}
\pdfoutput=1

\graphicspath{{mainFigs/}{./Figures/}}

\begin{document}
\title{Accurate Total Energies from the Adiabatic-Connection Fluctuation-Dissipation Theorem}
\author{N. D. Woods}
\email{nw361@cam.ac.uk}
\affiliation{Theory of Condensed Matter, Cavendish Laboratory, University of Cambridge, Cambridge, CB3 0HE, United Kingdom}
\author{M. T. Entwistle}
\affiliation{FU Berlin, Department of Mathematics and Computer Science, Arnimallee 12, 14195 Berlin, Germany}
\author{R. W. Godby}
\affiliation{Department of Physics, University of York, and European Theoretical Spectroscopy Facility, Heslington, York YO10 5DD, United Kingdom}

\date{\today}

\begin{abstract}
In the context of inhomogeneous one-dimensional finite systems, recent numerical advances [\href{https://journals.aps.org/prb/abstract/10.1103/PhysRevB.103.125155}{Phys$.$ Rev$.$ B \textbf{103}, 125155 (2021)}] allow us to compute the exact coupling-constant dependent exchange-correlation kernel $f^\lambda_\text{xc}(x,x',\omega)$ within linear response time-dependent density functional theory. This permits an improved understanding of ground-state total energies derived from the adiabatic-connection fluctuation-dissipation theorem (ACFDT). We consider both `one-shot' and `self-consistent' ACFDT calculations, and demonstrate that chemical accuracy is reliably preserved when the frequency dependence in the exact functional $f_\text{xc}[n](\omega=0)$ is neglected. This performance is understood on the grounds that the exact $f_\text{xc}[n]$ varies slowly over the most relevant $\omega$ range (but not in general), and hence the spatial structure in $f_\text{xc}[n](\omega=0)$ is able to largely remedy the principal issue in the present context: self-interaction (examined from the perspective of the exchange-correlation hole). Moreover, we find that the implicit orbitals contained within a self-consistent ACFDT calculation utilizing the adiabatic exact kernel $f_\text{xc}[n](\omega=0)$ are remarkably similar to the exact Kohn-Sham orbitals, thus further establishing that the majority of the physics required to capture the ground-state total energy resides in the spatial dependence of $f_\text{xc}[n]$ at $\omega = 0$.
\end{abstract}

\maketitle

\section{Introduction}

The \textit{adiabatic-connection fluctuation-dissipation theorem} (ACFDT) \citep{Ullrich2012,Olsen2019,Gorling2019,Marques2006,BurkeABC,Dobson1998} formalism is a distinctive, powerful  approach to calculating ground-state energies of molecular and solid-state systems. The underlying theory, which centers around the density-density linear response function $\chi$ across a range of electron-electron interaction strengths, is exact, but practical implementations utilize approximations within time-dependent density functional theory (DFT), resulting in imperfect ACFDT total energies.

Our strategy in this paper is to use recently developed techniques \citep{Woods2021} for obtaining the exact linear response time-dependent DFT kernel $f_\text{xc}(x,x',\omega)$ and ground-state xc potential $v_\text{xc}(x)$ to examine in depth the relationship between approximate kernels/potentials and their corresponding inexact ACFDT energies, indicating routes toward more accurate practical versions of the ACFDT scheme.

The ACFDT total energy method occupies the `fifth rung' on `Jacob's ladder of approximate density functionals' \citep{Constantin2019,Perdew2005} -- above local and semi-local approximations to the xc energy $E_\text{xc}[n]$,  since ACFDT-based calculations involve a full set of Kohn-Sham orbitals and energies $\{ |\phi_i\rangle, \varepsilon_i \}$  in order to construct the \textit{non-interacting} response function $\chi_0$ which is in turn used to solve the equations of linear response time-dependent DFT. Present-day practical implementations scale somewhere between $\mathcal{O}(N^3)$ and $\mathcal{O}(N^5)$ \cite{Wilhelm2016,Graf2018,Kaltak2014,2Kaltak2014,Furche2001,Fuchs2002,Bleiziffer2013,Nguyen2009} depending on the preferred approximate $f_\text{xc}[n]$ and whether one performs a so-called `one-shot' or `self-consistent' ACFDT calculation (see Section \ref{sec:Theory}), where $N$ can be taken as the number of constituent particles involved in the calculation.

The most striking successes of published ACFDT calculations have been in describing long-range correlations, e.g. van der Waals correlations between two disjoint subsystems within a larger system \citep{Marques2006,Dobson1998,Kohn1998}. Such correlations evade capture within Kohn-Sham DFT using local or semi-local xc approximations \citep{Burke1998,Yan2000,Engel2000}, whereas the ACFDT correlation functional includes inherent non-locality even at the lowest level of approximate $f_\text{xc}[n]$, the so-called random phase approximation (RPA) $f_\text{xc}^\text{RPA} = 0$. For example, ACFDT calculations utilizing the RPA are able to properly describe the dissociation limit of molecules such as N$_2$ \cite{Furche2001} -- a notorious challenge for conventional Kohn-Sham DFT \cite{Burke1998,Perdew1993}. Furthermore, the ACFDT framework in general is central to the  development of systematic van der Waals functionals \citep{Krogel2020,Berland2015}, which have been successful not just in determining dissociation curves, but also in computing van der Waals coefficients, bond lengths, bond energies, and so on. 

On the other hand, ACFDT approximations such as the RPA-ACFDT are known to be deficient in regard to \textit{absolute} total energies, owing to weaknesses in their treatment of short-range correlations \citep{Harl2010,Bleiziffer2013}, e.g. in the case of the homogeneous electron gas (HEG) \citep{Freeman1977,Perdew1992}. Indeed, individual RPA-ACFDT energies are often less accurate than those calculated using direct application of local/semi-local approximations to the correlation energy. The leading cause of this is thought to be the effect of spurious self-interaction in the Hartree kernel \citep{Eshuis2012,Ruan2021}.

A host of ACFDT approximations that venture beyond the RPA have been considered as possible remedies to this issue. Two such examples include the self-interaction-free exact-exchange kernel $f_\text{x}$ \citep{Fuchs2005,Hellgren2008,Bleiziffer2012,Dobson1998,Gorling1998,2Gorling1998,Kim2002,Hellgren2010,Hebelmann2010,Hebelmann2011,Colonna2014,Bleiziffer2015} and approaches that separate treatment of long-range and short-range contributions to the correlation energy, and use the RPA approximation for the former \citep{Toulouse2009,Janesko2009,Zhu2010,Toulouse2010,3Janesko2009,2Janesko2009,Toulouse2011,Angyan2011}. These improvements contribute toward alleviating some of the fundamental issues in the present context, e.g. in the calculation of atomization energies, and thus advance the ongoing effort to further establish the ACFDT approach as a total energy method, an effort with which this work is also concerned.

Although the RPA approximation within the ACFDT framework is amenable to a term-by-term analysis in the context of many-body perturbation theory \citep{Eshuis2012,Gonze1999,Hellgren2007}, this connection is lost when moving beyond the RPA approximation, and as such so is a certain degree of transparency. Since $f_\text{xc}$ implicitly contains all correlated many-body effects, including those required to describe excited-state phenomena such as the optical spectrum, it is imperative to better understand the connection between the various aspects of $f_\text{xc}$ and the ACFDT correlation energy. For example, a major consideration is the extent to which the frequency dependence in $f_\text{xc}$ is important here \citep{Entwistle2019,Lein2000} -- the exact $f_\text{xc}$ includes a drastic dependence on $\omega$ whereby singularities exist along the real $\omega$-axis that are critical for recovering the optical spectrum \citep{Woods2021}. 

In the context of finite one-dimensional systems we compute the exact $f^\lambda_\text{xc}(x,x',\omega)$ in order to elaborate and elucidate the connection between its spatial non-locality/frequency dependence and the ACFDT total energy. In particular, the so-called \textit{adiabatic exact} (AE) kernel \citep{Thiele2009} $f^\lambda_\text{xc}[n](x,x',\omega=0)$, i.e. the zero-frequency component of the exact $f^\lambda_\text{xc}[n]$ functional, is explored in relation to both one-shot and self-consistent ACFDT calculations.

\section{The Adiabatic-Connection Fluctuation-Dissipation Theorem}
\label{sec:Theory}

\subsection{Background}
\label{sec:Background}

The origin of the ACFDT in the context of inhomogenous systems dates back around five decades to the series of articles given in Refs$.$ \citep{Gunnarsson1976,Langreth1977,Langreth1975}. Since then, a number of resources have covered the derivation of the ACFDT \citep{Engel2011,Marques2006,BurkeABC, Ullrich2012,Eshuis2012,Ren2012,2Hebelmann2011,Harl2008}, a brief review is given here. The adiabatic connection establishes a link between the interacting many-body system and its corresponding non-interacting Kohn-Sham system, ultimately leading to an alternate expression for the xc energy. Toward this end, a one-parameter family of many-body Hamiltonians is defined,
\begin{align}
H(\lambda) = \hat{T} + \lambda \hat{v}_\text{ee} + \hat{v}_\text{ext} + \hat{v}_\text{dxm}(\lambda), \label{eq:CouplingDepHamiltonian}
\end{align}
such that the \textit{deus ex machina} potential \citep{Langreth1980} $\hat{v}_\text{dxm}(\lambda)$ is the unique \citep{Hohenberg1964} potential that ensures the ground-state density at all values of $\lambda \in [0,1]$ is equal to the ground-state electron density at $\lambda = 1$, labeled $n(x)$ -- this is the adiabatic connection. Hence, $\hat{v}_\text{dxm}(\lambda = 1) = 0$ and $\hat{v}_\text{dxm}(\lambda = 0) = \hat{v}_\text{H} + \hat{v}_\text{xc}$ where $\hat{v}_\text{H}$ is the Hartree potential. The  $\lambda$-interacting ground state of $H(\lambda)$ is denoted $| \Psi^\lambda \rangle$, allowing the total energy to be expressed as such,
\begin{align}
E &= \langle \Psi^{\lambda = 1} | H(1) |  \Psi^{\lambda = 1} \rangle \nonumber \\
&= \langle \Psi^{\lambda = 0} | H(0) |  \Psi^{\lambda = 0} \rangle \nonumber \\
&= T_0 + E_\text{H} + E_\text{ext} + E_\text{xc}, \label{eq:ConventionalExc}
\end{align}
where the latter two formulae constitute the conventional definition of the Kohn-Sham system \citep{Kohn1965}, i.e. $T_0$ is the non-interacting kinetic energy, $E_\text{H}$ is the Hartree energy, $E_\text{ext}$ is the external energy, and $E_\text{xc}$ is the xc energy. Moreover, $|\Psi^{\lambda = 0} \rangle$ represents the Kohn-Sham Slater determinant ground state. 

Rearrangement of the above expressions yields an alternate form for the Hxc energy $E_\text{Hxc} = E_\text{H} + E_\text{xc}$,
\begin{align}
E_\text{Hxc} &= \langle \Psi^{\lambda = 1} | H(1) |  \Psi^{\lambda = 1} \rangle - \langle \Psi^{\lambda = 0} | H(0) |  \Psi^{\lambda = 0} \rangle \nonumber \\
& -\langle \Psi^{\lambda = 1} | \hat{v}_\text{dxm}(1) |  \Psi^{\lambda = 1} \rangle + \langle \Psi^{\lambda = 0} | \hat{v}_\text{dxm}(0) |  \Psi^{\lambda = 0} \rangle, \nonumber
\end{align}
which becomes, upon use of the fundamental theorem of calculus and the Hellmann-Feynmann theorem,
\begin{align}
E_\text{Hxc} &= \int_0^1  \frac{d}{d\lambda} \left( \langle \Psi^{\lambda} | H(\lambda) |  \Psi^{\lambda} \rangle - \langle \Psi^{\lambda} | \hat{v}_\text{dxm}(\lambda) | \Psi^{\lambda} \rangle \right) \ d\lambda \nonumber \\
&= \int_0^1 \langle \Psi^\lambda | \hat{v}_\text{ee} |  \Psi^\lambda \rangle \ d \lambda. \label{eq:ACEnergy}
\end{align}
The expression in Eq$.$ (\ref{eq:ACEnergy}) contrasts with the conventional one Eq$.$ (\ref{eq:ConventionalExc}) as it does not involve the kinetic operator at the seemingly steep price of having to know the xc \textit{potential} energy \citep{BurkeABC},
\begin{align}
U_\text{xc}(\lambda) = \langle \Psi^\lambda | \lambda \hat{v}_\text{ee} | \Psi^\lambda \rangle - \lambda E_\text{H}, \label{eq:XCPotEnergy}
\end{align}
at each value of $\lambda$ along the adiabatic connection \footnote{The $\lambda$-dependent xc energy (and xc kernel) is zero when $\lambda=0$ because the particles are non-interacting.}. 

However, knowledge of the challenging expectation value in  Eq$.$ (\ref{eq:ACEnergy}) is tantamount to knowledge of the static (equal-time) two-point correlator $\langle \Psi^\lambda | \hat{n}(x) \hat{n}(x') | \Psi^\lambda \rangle$, which describes quantum statistical \textit{fluctuations} in the density inherent to the state $| \Psi^\lambda \rangle$ \citep{Martin2017}. The fluctuation-dissipation theorem \citep{Kubo1966} provides a relationship between the response of a system to these spontaneous \textit{internal} changes (fluctuations) in its density, and the response of that same system to \textit{external} perturbations in its density. The latter is described with the density-density linear response function,
\begin{align}
\chi^\lambda[n](x,x',\omega) = \left. \frac{\delta n}{\delta v_\text{ext}} \right|_{n_0},
\end{align}
i.e. the first-order change in the density due to a perturbation in the external potential within a system of $\lambda$-interacting particles described by $H(\lambda)$ in Eq$.$ (\ref{eq:CouplingDepHamiltonian}). In the present context, the fluctuation-dissipation theorem takes the form \citep{3Dobson1998}
\begin{align}
\langle \Psi^\lambda | \hat{n}(x) \hat{n}(x') | \Psi^\lambda \rangle = n(x)n(x') - \frac{2}{\pi} \int_0^\infty \chi^\lambda(x,x',i\omega) \ d\omega, \nonumber
\end{align}
thus connecting the ground-state xc energy with linear response theory. 

The above derivation outlines an in principle exact reformulation of conventional Kohn-Sham DFT \citep{BurkeABC}, meaning the total energy functional 
\begin{align}
E^\text{ACFD}[n] \coloneqq T_0[n] + E_\text{ext}[n] + E_\text{H}[n] + E_\text{xc}^\text{ACFD}[n] \label{eq:ACFDTTotalEnergy}
\end{align}
has the correct minimum, i.e. the exact ground-state energy, and this minimum is attained at the interacting ground-state density. Having performed the rearrangements given from the fluctuation-dissipation theorem, the ACFDT xc energy functional $E_\text{xc}^\text{ACFD}[n] =  E_\text{x}^\text{ACFD}[n] +  E_\text{c}^\text{ACFD}[n]$ becomes
\begin{align}
E^\text{ACFD}_\text{x}[n] =& E_\text{x}[\{ |\phi_i[n]\rangle \}], \\
E^\text{ACFD}_\text{c}[n] =& -\frac{1}{2\pi} \int_0^1 d\lambda \int_0^\infty d\omega \iint dx dx' \label{eq:EcACFDT}\\
& v_\text{ee}(x,x') [ \chi^\lambda[n](x,x',i\omega) - \chi_0[n](x,x',i\omega) ], \nonumber
\end{align}
where $E_\text{x}[\{ |\phi_i\rangle \}]$ is the exact-exchange functional evaluated at the Kohn-Sham orbitals $\{ |\phi_i\rangle \}$, and $\chi_0[n] = \delta n / \delta v_\text{KS}$ is the associated response function of the Kohn-Sham system. These are the central expressions around which the remainder of this paper is based. Namely, we evaluate the total energy functional Eq$.$ (\ref{eq:ACFDTTotalEnergy}) with its exact definition, and then introduce approximations into the functional through $f^\lambda_\text{xc}[n]$ (i.e. $\chi^\lambda[n]$).

Linear response time-dependent DFT establishes a unique map \citep{Runge1984,VanLeeuwen1999} from a tractable non-interacting (Kohn-Sham) response function $\chi_0[n]$ to an otherwise intractable $\lambda$-interacting response function $\chi^\lambda[n]$ through the xc kernel,
\begin{align}
f^\lambda_{\text{xc}}[n](x,x',\omega) = \left. \frac{\delta v^\lambda_{\text{xc}}}{\delta n} \right|_{n_0}, \label{eq:xcKernelDefinition}
\end{align}
i.e. the first-order change in the $\lambda$-dependent xc potential due to a perturbation in the density oscillating in time with frequency $\omega$. This map constitutes the Dyson equation of linear response time-dependent DFT,
\begin{align}
\chi^\lambda[n] = \chi_0[n] + \chi_0[n] * (\lambda f_\text{H} + f_\text{xc}^\lambda[n]) * \chi[n],  \label{eq:Dyson}
\end{align}
where $A * B = \int A(x,x')B(x',x'') dx'$, and $f_\text{H} = \delta v_\text{H} / \delta n$ is the Hartree kernel (the electron-electron interaction).

The xc kernel $f^\lambda_\text{xc}[n]$ is the central subject of approximation in linear response time-dependent DFT \citep{Ullrich2012}, and therefore the principal ingredients in an ACFDT total energy calculation are an approximation to the xc kernel functional $f^\lambda_\text{xc}[n]$ together with a prescription for determining the density $n$ at which to evaluate the ACFDT total energy functional $E^\text{ACFD}[n]$. In regard to the latter concern: there are two main approaches that are applied in practice to determine this density. The first is the most common, and is often referred to as a one-shot ACFDT calculation, wherein the density $n$ is obtained from a self-consistent solution of the ground-state Kohn-Sham equations with an approximate xc potential $v_\text{xc}[n]$ (see Ref$.$ \citep{Olsen2019} for a recent review). The second, often referred to as a self-consistent ACFDT calculation \citep{Niquet2003,Bleiziffer2013,Hellgren2012,Hellgren2010,Bleiziffer2015,Verma2012,Thierbach2020}, solves
\begin{align}
E_0 = \min_n E^\text{ACFD}[n], \label{eq:MinACFDT}
\end{align}
where the equations that yield a stationary (presumed to be minimizing) density are known \citep{Niquet2003,Gorling2005,Bleiziffer2013,Bleiziffer2012}. In the instance that both $v_\text{xc}[n]$ and $f^\lambda_\text{xc}[n]$ are exact, the output of a one-shot and self-consistent ACFDT calculation coincide, however this is not true when approximations are involved, as we shall explore in Section \ref{sec:Results}.

In practice, an approximate xc kernel functional $f^\lambda_\text{xc}[n]$ that is also parameterized with respect to $\lambda$ is \textit{not} required as it is possible to exploit a curious relationship between ground-state wavefunctions along the adiabatic connection and ground-state wavefunctions with  scaled spatial coordinates, see Refs$.$ \cite{Lein2000,BurkeABC}. This relationship permits us to specify a conventional functional $f_\text{xc}[n] \coloneqq f_\text{xc}^{\lambda=1}[n]$ and obtain its $\lambda$ dependence with little-to-no additional expense. Such an observation is central to practical ACFDT calculations, although we are unable to exploit it here due to using the softened Coulomb interaction.


\subsection{Implementation}

This work involves finite systems in one dimension interacting with a softened Coulomb electron-electron interaction
\begin{align}
v_\text{ee}(x,x') = \frac{1}{|x-x'| + \alpha}, \label{eq:softCoulomb}
\end{align}
where $\alpha$ is the softening parameter; $\alpha = 1$ a.u. is used in this work. A real-space grid of dimension $N$ discretizes the spatial domain $[-L,L]$ subject to Dirichlet boundary conditions. We consider four prototype systems, each of which include two like-spin electrons \footnote{Electrons of like spin obey the Pauli exclusion principle, and exhibit features that would need a larger number of spin-half electrons to become apparent, e.g. two like-spin electrons experience the exchange effect, which is not the case for two spin-half electrons in an $S=0$ state.} in the external potentials described below.

\begin{figure}[ht]
\begin{center}
\includegraphics[width=3.4in]{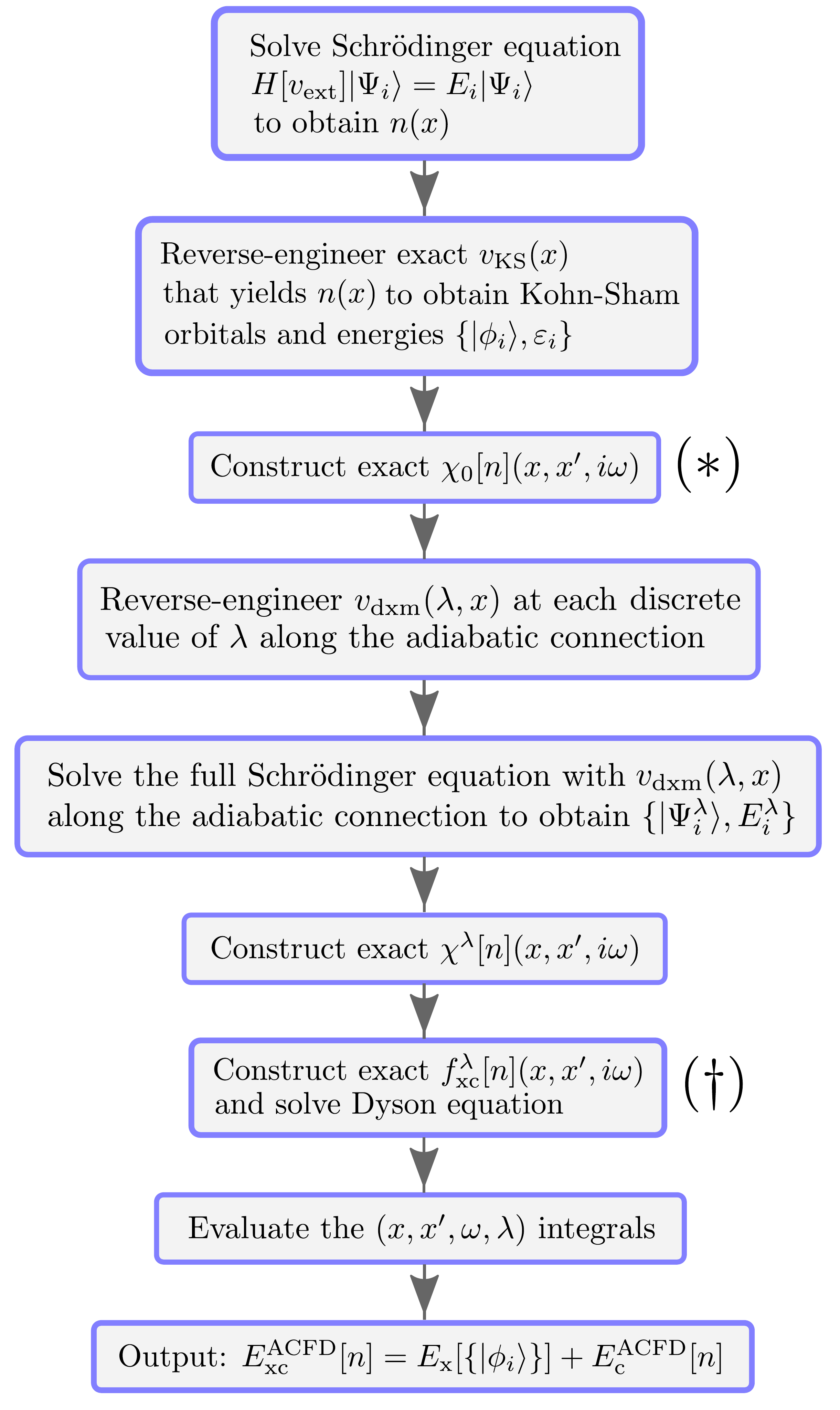}
\end{center}
\caption{A flow chart depicting the course of action taken to evaluate the exact ACFDT total energy $E^\text{ACFD}[n]$ at the exact interacting ground-state density $n$. This procedure is modified in order to investigate approximate ACFDT approaches as follows: $(*)$ a non-interacting Kohn-Sham calculation is performed with some approximate $v_\text{xc}[n]$, and then the algorithm proceeds with the corresponding approximate non-interacting response function, $(\dagger)$ rather than calculate and utilize the exact $f_\text{xc}^\lambda[n]$, an approximate functional $f_\text{xc}^\lambda[n]$ is chosen and used alongside $\chi_0[n]$ to solve the Dyson equation Eq$.$ (\ref{eq:Dyson}), ultimately yielding an approximate many-body response function $\chi^\lambda[n]$.}
\label{fig:ACFDTFlowCHart}
\end{figure}

The central complication when implementing the exact ACFDT total energy functional is evaluation of the ACFDT correlation functional Eq$.$ (\ref{eq:EcACFDT}), which we now proceed to elaborate, see also Fig$.$ \ref{fig:ACFDTFlowCHart}.

Having chosen some external potential $v_\text{ext}(x)$, the exact interacting density $n(x)$ is obtained through solution of the time-independent Schr\"odinger equation. The corresponding unique Kohn-Sham potential $v_\text{KS}(x)$ is then reverse-engineered by applying preconditioned root-finding techniques to an appropriate fixed-point map \citep{Ruggenthaler2015}. The Kohn-Sham orbitals and energies $\{  |\phi_i\rangle, \varepsilon_i \}$ are used to construct the non-interacting response function $\chi_0$ along $i\omega$ in the Lehmann representation \citep{Ullrich2012}.

In order to calculate the final ingredient $\chi^\lambda[n]$, we first obtain the $\lambda$-dependent wavefunctions $\{ |\Psi^\lambda_i \rangle \}$ along the adiabatic connection, as defined in Section \ref{sec:Background} -- this grossly impractical step is performed for investigative reasons, and constitutes the primary computational expense. For each value of the coupling constant on some discrete grid, the potential $v_\text{dxm}(\lambda, x)$ is obtained by yet again using root-finding techniques to target the $\lambda=1$ interacting density, $n(x)$ (see supplemental material for an example $v_\text{dxm}(\lambda, x)$ \footnote{URL to be inserted}). The full set of $\lambda$-dependent wavefunctions and energies $\{ |\Psi_i^\lambda \rangle, E_i^\lambda \}$ is then used to calculate the $\lambda$-interacting response functions in the Lehmann representation,
\begin{align}
\chi^\lambda(x,x',i\omega) = \sum_{n=1}^{\infty} -2 & \frac{\Omega^\lambda_n}{\omega^2 + (\Omega^\lambda_n)^2} \label{eq:ResponseLehman} \\ 
&\times \langle \Psi^\lambda_0 | \hat{n}(x) | \Psi^\lambda_n \rangle \langle \Psi^\lambda_n | \hat{n}(x') | \Psi^\lambda_0 \rangle, \nonumber
\end{align}
where $\Omega^\lambda_n = E^\lambda_n - E^\lambda_0$ is the $n-$th excitation energy of the $\lambda$-interacting Hamiltonian along the adiabatic connection.

At this stage it is possible to construct the exact $\lambda$-dependent xc kernel using the expression
\begin{align}
f^\lambda_\text{xc}(\omega) = \chi_0^{-1}(i\omega) - (\chi^\lambda)^{-1}(i\omega) - \lambda f_\text{H}, \label{eq:xcKernel}
\end{align}
which comes from inspection of the Dyson equation Eq$.$ (\ref{eq:Dyson}) (note that superscript ${-1}$ signifies the matrix inverse in a finite spatial basis). Construction of $f^\lambda_\text{xc}$ in this fashion is an intricate matter that has been dealt with in prior work \citep{Woods2021}. For our purposes, it suffices to observe that in all the cases presented below the exact $\lambda$-dependent response function is reconstructed to within machine precision when the exact $f^\lambda_\text{xc}$ and exact $\chi_0$ are used to solve the Dyson equation. This step is critical as ultimately we shall isolate certain features of the exact $f_\text{xc}^\lambda$ in order to examine their impact on the correlation energy, for example utilizing only the $\omega=0$ component of $f^\lambda_\text{xc}$ here defines the AE approximation. \\

The final step toward obtaining $E_c^\text{ACFD}$ involves evaluating the coupling constant and frequency integrals in Eq$.$ (\ref{eq:EcACFDT}). For the systems presented in Section \ref{sec:Results}, the $\lambda$-dependent integrand does not deviate much from a linear form, and thus Gauss-Legendre integration is able to reach machine precision with $N_\lambda=10$ grid points. However, the $\omega$-dependent integrand is not suited to the traditional Gauss-Legendre scheme, and so it is convention to utilize a change of coordinates in order to reduce the number of frequency grid points required to reach a desired accuracy \footnote{We note that the Gauss-Legendre scheme with $N$ grid points is able to exactly integrate a polynomial of degree less than or equal to $2N+1$. Therefore, an integral change of coordinates aims to transform the integrand into a low-degree polynomial.}. Upon careful comparison with a variety of methods from literature \citep{Harl2008}, we find the novel substitution
\begin{align}
\omega = a \tan \left( \frac{a\tilde{\omega}^2}{2} \right) \label{eq:ChangeOfCoords}
\end{align}
performs best, where $a$ is a numerical parameter and $\tilde{\omega}$ is the new coordinate. (The traditional Gauss-Legendre scheme is then applied to the transformed $\tilde{\omega}$-dependent integral.) This approach is able to reach more than sufficient accuracies with $N_\omega = 30$ grid points -- a comprehensive motivation and derivation can be found in the supplemental material. \\

The algorithm that has just been outlined captures the exact correlation energy to within $\mathcal{O}(10^{-10})$ a.u. across the systems studied in this work. This procedure can be suitably adapted, see Fig$.$ \ref{fig:ACFDTFlowCHart}, to include an approximate $f_\text{xc}[n]$ and/or an approximate $v_\text{xc}[n]$, where we recall the latter is used to determine the density at which $E^\text{ACFD}[n]$ is evaluated in a one-shot calculation (rather than evaluating $E^\text{ACFD}[n]$ at the exact interacting density as is done in the first and second panels of Fig$.$ \ref{fig:ACFDTFlowCHart}). 

On the other hand, a \textit{self-consistent} ACFDT total energy calculation comprises first specifying an initial guess Kohn-Sham potential $v_\text{KS}(x)$ in place of the first and second panels in Fig$.$ \ref{fig:ACFDTFlowCHart}. Note that since there is a one-to-one correspondence between the density and the Kohn-Sham potential, it is sufficient to minimize over variations in $v_\text{KS}$ in order to minimize the ACFDT total energy functional as in Eq$.$ (\ref{eq:MinACFDT}). We are then able to iterate the initial guess toward the minimizing Kohn-Sham potential by utilizing the BFGS optimization algorithm -- this involves looping over the flow chart in Fig$.$ \ref{fig:ACFDTFlowCHart}. In the event that the ACFDT total energy functional is specified with the exact $f_\text{xc}[n]$, the minimization procedure terminates at the exact Kohn-Sham potential $v_\text{KS}(x)$/the exact interacting density $n(x)$ \textit{without their explicit inclusion}. In general, iterations are terminated when the Jacobian norm is $\mathcal{O}(10^{-6})$, meaning the BFGS algorithm is making energy variations $\mathcal{O}(10^{-8})$ a.u. The calculations are parallelized over $\lambda$ grid points using \verb|dask| \citep{dask}.

Minimizing the ACFDT total energy functional is able to circumvent the troublesome `starting-point dependence' inherent to a one-shot calculation. In practice, the minimization is accomplished by solving a set of \textit{optimized effective potential} equations in order to generate the minimizing density \citep{Niquet2003,Bleiziffer2013,Hellgren2012,Hellgren2010,Bleiziffer2015,Verma2012,Thierbach2020}, rather than direct minimization of the functional as is considered in this work. Despite the latter being much more expensive, we are required to take these measures as the AE kernel $f_\text{xc}[n](\omega=0)$ has no analytic representation in terms of the density/orbitals. In either case, self-consistent calculations are more computationally demanding than one-shot calculations. However, the accuracy of self-consistent ACFDT total energies is entirely determined by the approximate $f_\text{xc}[n]$, which in certain circumstances can be advantageous, as we shall examine in Section \ref{sec:Results}.

\section{Results}
\label{sec:Results}

Let us begin by first examining one-shot and self-consistent ACFDT total energies in order to explore the significance of the spatial/frequency dependence in $f_\text{xc}(x,x',i\omega)$ \textit{and} the influence of the density $n$ at which $E^{\text{ACFD}}[n]$ is evaluated. An atomic system is used to illustrate the results concerning short-ranged correlations, whereas a double well system is used in the subsequent section on long-ranged correlations. A system with a flat slab-like density profile assists in determining various sources of error in Section \ref{sec:SourceOfError}. Finally, all four systems, including the infinite potential well, enable us to ensure sufficient generality in the conclusions drawn.

This work focuses on three approximations to the xc kernel: the RPA $f_\text{xc}^\text{RPA}[n] = 0$, an adiabatic local density approximation (LDA) $f^\text{ALDA}_\text{xc}[n](x,x',\omega=0) \propto \delta(x-x')$, and the AE xc kernel $f_\text{xc}[n](x,x',\omega=0)$. Since this work utilizes the softened Coulomb interaction, we are prohibited from using the scaling relationship to obtain $f_\text{xc}^\lambda[n]$ from $f_\text{xc}[n]$, and therefore our LDA energy functional from which $f^{\text{ALDA},\lambda}_\text{xc}[n]$ is constructed is parameterized \textit{at each value of} $\lambda$ with reference to both the HEG and one-dimensional slab systems \citep{Entwistle2018} (see supplemental material for details regarding this parameterization).

\subsection{Short-Range Correlations}

\begin{figure}[ht]
\begin{center}
\includegraphics[width=3.4in]{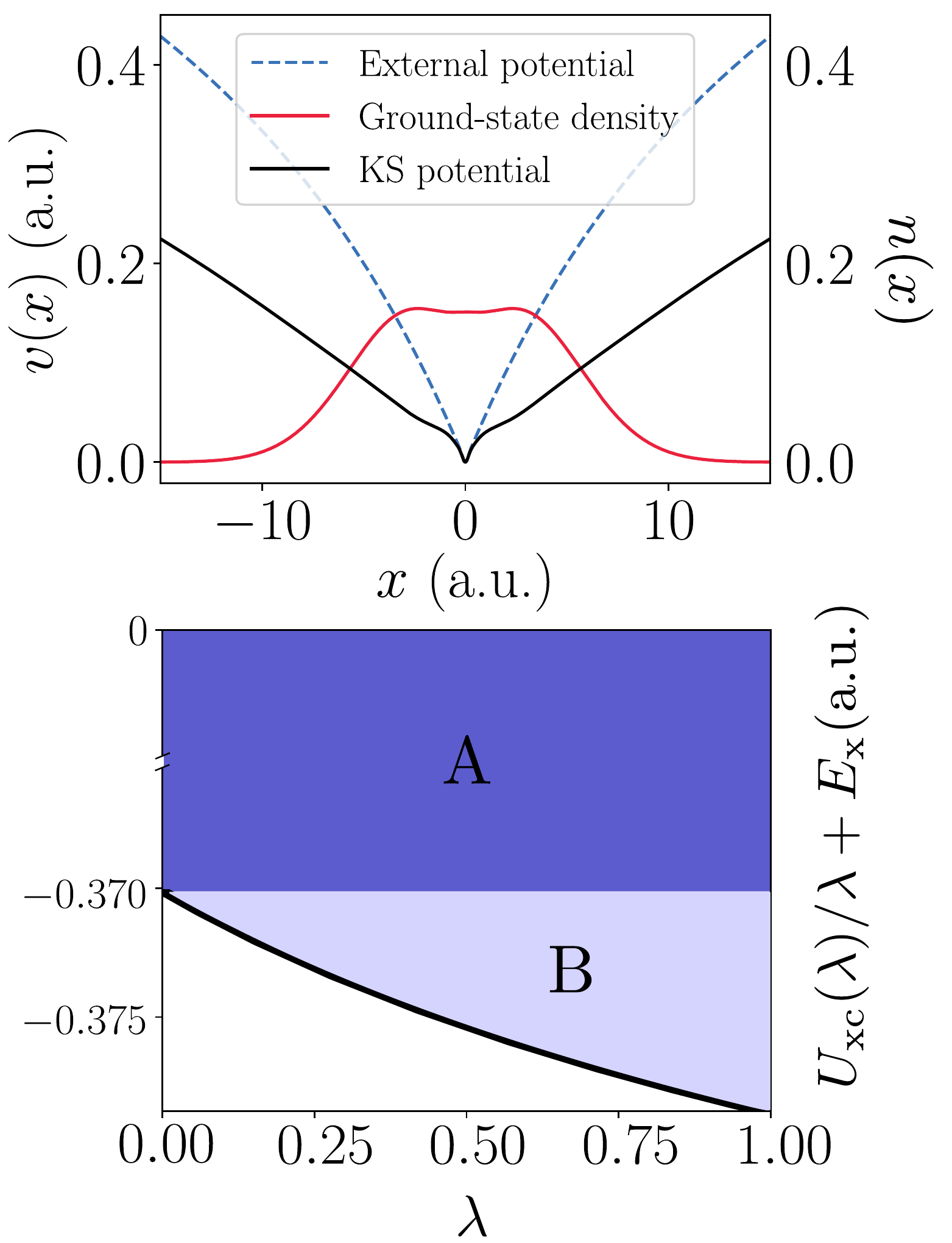}
\end{center}
\caption{(Upper) The ground-state density, external potential, and reverse-engineered Kohn-Sham potential for the atomic system. The external and Kohn-Sham potentials have been shifted for illustrative purposes. (Lower) The exact adiabatic connection curve, i.e. the $\lambda$-dependent integrand of the ACFDT formula Eq$.$ (\ref{eq:ACFDTLambdaIntegrand}), where the exchange energy (area of shaded region $A$) and correlation energy (area of shaded region $B$) are given geometric context. The white region that compliments the shaded regions has area equal to minus the kinetic correlation energy, $T_\text{c}$.}
\label{fig:AC_Curve}
\end{figure}

We shall now consider two like-spin electrons confined in an atom-like potential $v_\text{ext} = -1 / (|0.05 x| + 1)$ within the domain $[-15,15]$ a.u. which is discretized over $N_x = 121$ grid points -- see upper panel of Fig$.$ \ref{fig:AC_Curve}. The so-called exact \textit{adiabatic connection curve} \citep{BurkeABC,Teale2009} is given in the lower panel of Fig$.$ \ref{fig:AC_Curve} which provides a geometric interpretation of the $\lambda$-dependent ACFDT integrand, 
\begin{align}
E_\text{xc}^\text{ACFD}[n] = E_\text{x}[\{ |\phi_i\rangle \}] + \int_0^1 \frac{U_\text{xc}(\lambda)}{\lambda} \ d\lambda, \label{eq:ACFDTLambdaIntegrand}
\end{align}
i.e. the $\lambda$-dependent xc potential energy, see Eq$.$ (\ref{eq:ACEnergy}) and Eq$.$ (\ref{eq:XCPotEnergy}). The slight convex bend in the adiabatic connection curve that is observed here implies a modest static correlation \citep{BurkeABC}. In this instance, the correlation energy is $1.3 \%$ of the xc energy, and the xc energy is $24 \%$ of the total energy, $E_\text{tot} = -1.510$ a.u.

\begin{figure}[ht]
\begin{center}
\includegraphics[width=3.4in]{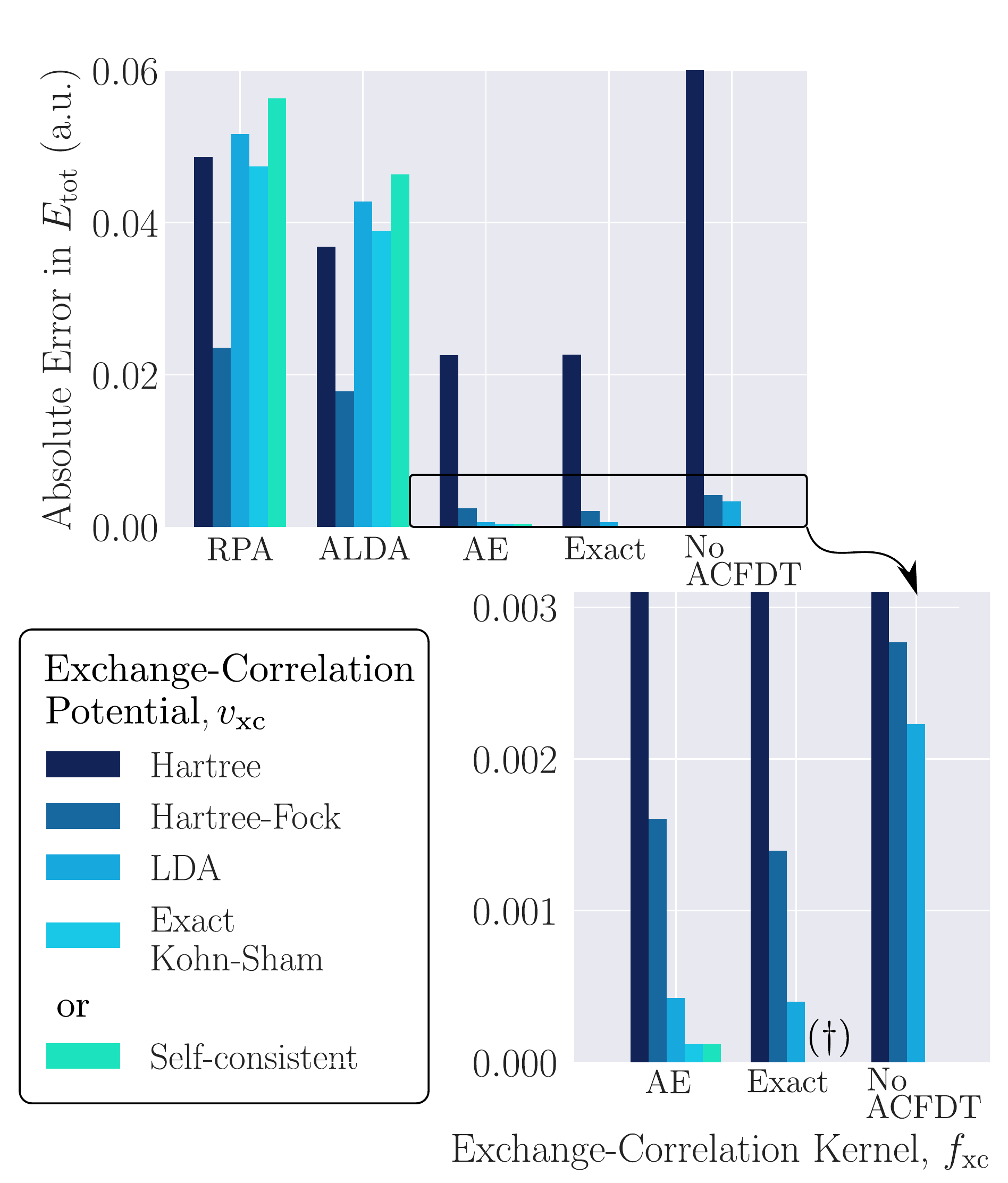}
\end{center}
\caption{Absolute error in the atomic total energy across a variety of approximate ACFDT total energies. Each of the first four groups of colored bars is labeled with the approximate $f_\text{xc}[n]$ used to specify $E^\text{ACFD}[n]$, while the fifth group (`No ACFDT') denotes a conventional ground-state Kohn-Sham calculation, for comparison. The first four bars of each group indicate the approximate $v_\text{xc}[n]$ used to generate the density (and orbitals) at which $E^\text{ACFD}[n]$ is evaluated, while the fifth bar (cyan) indicates that the input density was determined self-consistently (i.e. it minimizes the applicable ACFDT total energy functional). The lower bar chart enlarges the outlined region in the upper bar chart. $(\dagger)$ At this position, there are two bars of zero height, i.e. the exact interacting ground-state density coincides with the density that minimizes the energy functional, both of which return the exact ground-state energy.}
\label{fig:AtomEtotRelativeError}
\end{figure}

The relative error in the atomic total energy is illustrated in Fig$.$ \ref{fig:AtomEtotRelativeError} across the whole range of approximate ground-state xc potentials and xc kernels considered herein. 

With the exception of the energies calculated utilizing the notoriously poor Hartree orbitals, the predominant clustering in error appears to be according to the approximate $f_\text{xc}[n]$, rather than the approximate $v_\text{xc}[n]$ used to generate the input density. This suggests that there is a fairly general insensitivity to the density at which $E^\text{ACFD}[n]$ is evaluated after having been specified with some $f_\text{xc}[n]$. Arguments have been made that this must be the case in the context of the RPA \citep{Dobson1998}, and we are now able to demonstrate that it is also the case when using the adiabatic LDA and the AE approximations. This conclusion translates to all other systems studied here as can be seen in the supplemental material, wherein similar figures can be found for each of the three remaining systems: an infinite potential well, a slab, and a double well.

Secondly, the AE kernel, i.e. ignoring the frequency dependence in the otherwise exact functional $f_\text{xc}[n](x,x',\omega=0)$, reduces the error in the total energy by orders of magnitude when compared to the RPA and adiabatic LDA kernels. For example, the \textit{absolute} error in the AE-ACFDT total energy is $0.0006$ a.u. when ground-state LDA density/orbitals are used in a one-shot calculation -- significantly better than the usual measure of chemical accuracy, $0.0016$ a.u. (1 kcal/mol). Moreover, we find that this level of accuracy is reliably achieved across differing ground-state orbitals (with the exception of the Hartree orbitals) and differing systems.

There is a sense in which this performance can be said to be inherent to the AE approximation $f_\text{xc}[n](x,x',\omega=0)$. Namely, in using the exact Kohn-Sham potential $v_\text{KS}(x)$ to generate the one-shot ground-state orbitals, we are able to isolate error coming solely from the fact that the approximate xc kernel is not exact. Furthermore, minimizing the ACFDT total energy functional serves a similar purpose, i.e. error is entirely due to the approximate $f_\text{xc}[n]$. It can be observed in Fig$.$ \ref{fig:AtomEtotRelativeError} and in its corresponding tabulated data (see supplemental material) that these two methods produce relative errors in the total energy to within $\mathcal{O}(10^{-6})$ of each other. In other words, the implicit (minimizing) potential/orbitals that are contained within the self-consistent AE-ACFDT calculation are close to the exact Kohn-Sham potential/orbitals, both of which yield ACFDT total energies well below chemical accuracy. \\

\begin{figure}[ht]
\begin{center}
\includegraphics[width=2.5in]{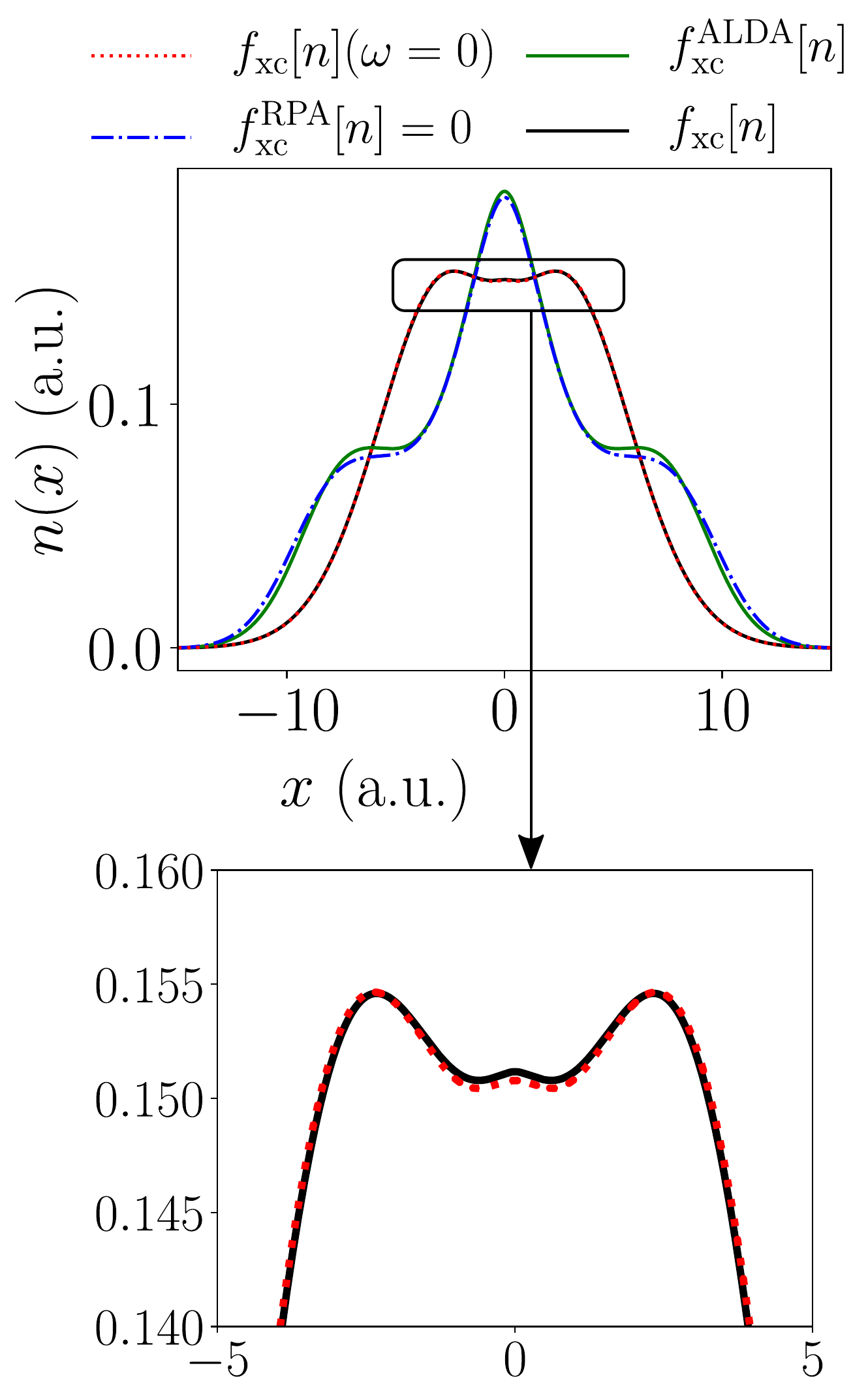}
\end{center}
\caption{(Upper panel) The density $n$ that minimizes the atomic ACFDT energy functional $E^\text{ACFD}[n]$ specified with some approximate $f_\text{xc}[n]$. The minimizing density when the \textit{exact} $f_\text{xc}[n]$ is used is the interacting ground-state density (black solid). (Lower panel) Zooming in such that the subtle difference between the minimizing densities that comes from ignoring the frequency dependence in the exact $f_\text{xc}[n]$ can be seen.}
\label{fig:AtomMinimisingDensities}
\end{figure}

\begin{figure}[ht]
\begin{center}
\includegraphics[width=2.8in]{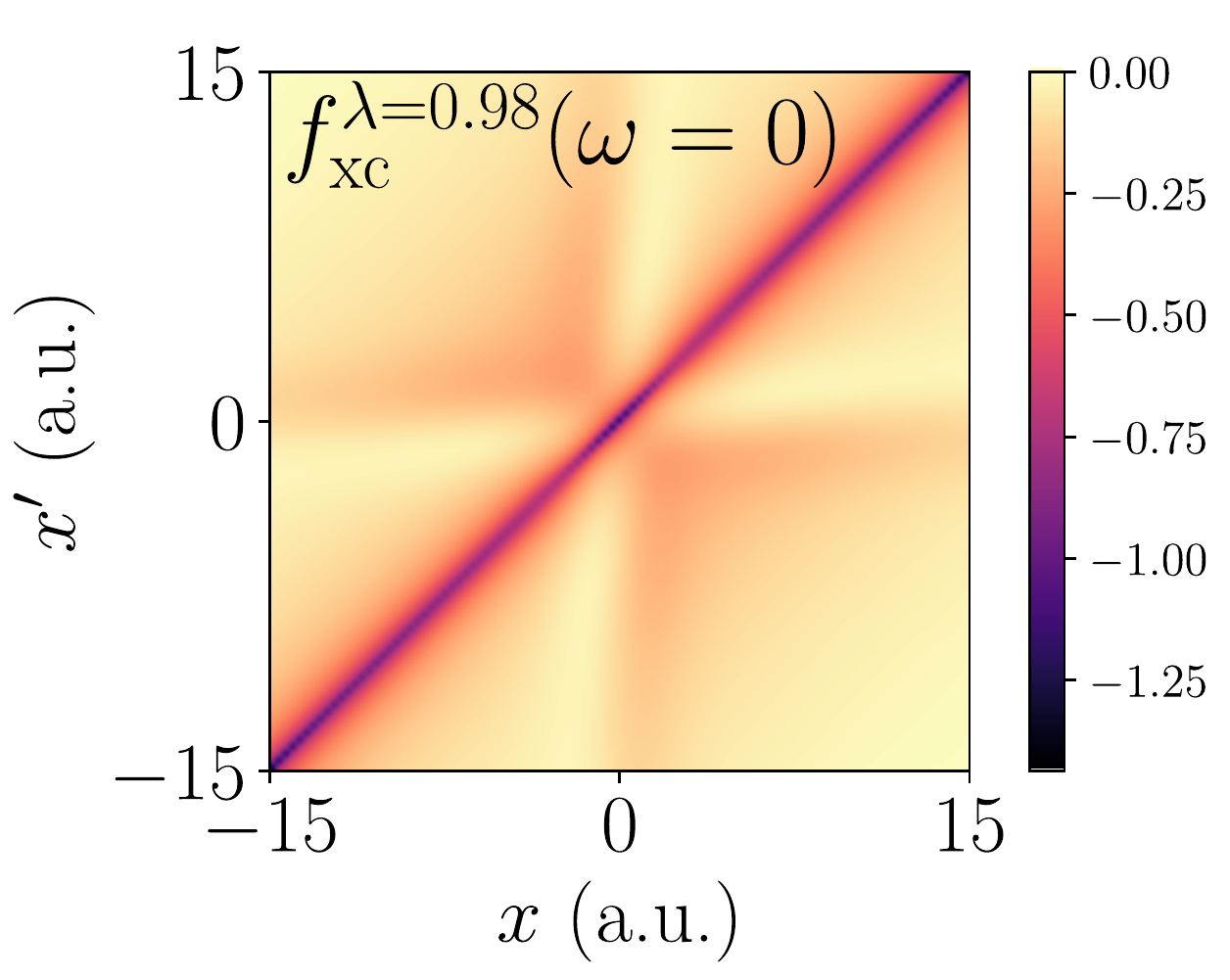}
\end{center}
\caption{The AE $f^\lambda_\text{xc}(x,x',\omega=0)$ for the atomic system at $\lambda = 0.98$, i.e. at a discrete value of $\lambda$ that is sampled along the adiabatic connection.}
\label{fig:AtomAdiabaticXCKernel}
\end{figure}

This similarity between the self-consistent AE-ACFDT orbitals and the exact Kohn-Sham orbitals is manifest when comparing the associated minimizing density with the exact interacting density, Fig$.$ \ref{fig:AtomMinimisingDensities}. These observations point toward a central conclusion of this work: the spatial structure in the exact xc kernel at $\omega = 0$, see Fig$.$ \ref{fig:AtomAdiabaticXCKernel}, is sufficient to almost entirely determine the exact total/correlation energy, and in turn the exact interacting density and exact Kohn-Sham potential. Hence, the intricate difference between the non-local spatial structure in Fig$.$ \ref{fig:AtomAdiabaticXCKernel} versus the local spatial structure in, for example, an adiabatic LDA kernel gives rise to a significant difference in the respective total energies, the reasons for which we shall now explore.

Turning now toward the RPA and adiabatic LDA kernels, inspection of Fig$.$ \ref{fig:AtomEtotRelativeError} and Fig$.$ \ref{fig:AtomMinimisingDensities} leads one to conjecture that these approximate kernels share the same fundamental issues, with the adiabatic LDA suffering slightly less. Both approximations are in serious error when it comes to the correlation energy: an order of magnitude too negative in the case of the atom, and more so in the case of the infinite potential well and double well. Therefore, \textit{minimizing} the corresponding ACFDT total energy functionals necessarily makes matters worse -- the RPA and adiabatic LDA minimizing densities deviate significantly from the interacting ground-state density, and even deviate from most approximations to it, Fig$.$ \ref{fig:AtomMinimisingDensities}. 

The source of this error is understood here in the context of the so-called $\lambda$-averaged xc hole. In conventional wavefunction-based theories, the statistical hole $n^{\text{hole}}(x,x')$ describes how the probability distribution of particle positions $n(x)$ changes upon measurement of a particle at $x'$ \footnote{The probability that a particle resides at position $x$ given that a particle has been measured at $x'$ is thus given by the conditional probability $n(x|x') = n(x) - n^\text{hole}(x,x')$.}. In density-based theories, however, the xc hole is redefined and is the object with which the density undergoes Coulomb interaction to produce the xc energy \citep{BurkeABC,Burke1998,Yan2000,Gunnarsson1976,2Dobson1998}, 
\begin{align}
E_\text{xc} &= \iint \frac{n(x)n^{\text{hole}}_\text{xc}(x,x') }{|x-x'|} \ dx dx' \\
&=  \iiint \ \frac{n(x)n^{\text{hole},\lambda}(x,x')}{|x-x'|} \ d\lambda dx dx', \label{eq:XCHole}
\end{align}
where $n^{\text{hole},\lambda}$ is the traditional statistical hole of the $\lambda$-interacting systems along the adiabatic connection. Comparing Eq$.$ (\ref{eq:XCHole}) with the ACFDT xc energy expression provides a unique definition of the xc hole in the present context, and moreover it defines an \textit{approximate} xc hole in terms of $f_\text{xc}[n]$ \citep{Gunnarsson1976}. Nevertheless, the two definitions are closely related, and it is instructive to view the $\lambda$-averaged xc hole in both contexts, not least because certain important sum rules are shared, e.g. $\int n^\text{hole}_\text{xc}(x,x') \ dx = -1$.

\begin{figure}[ht]
\begin{center}
\includegraphics[width=2.7in]{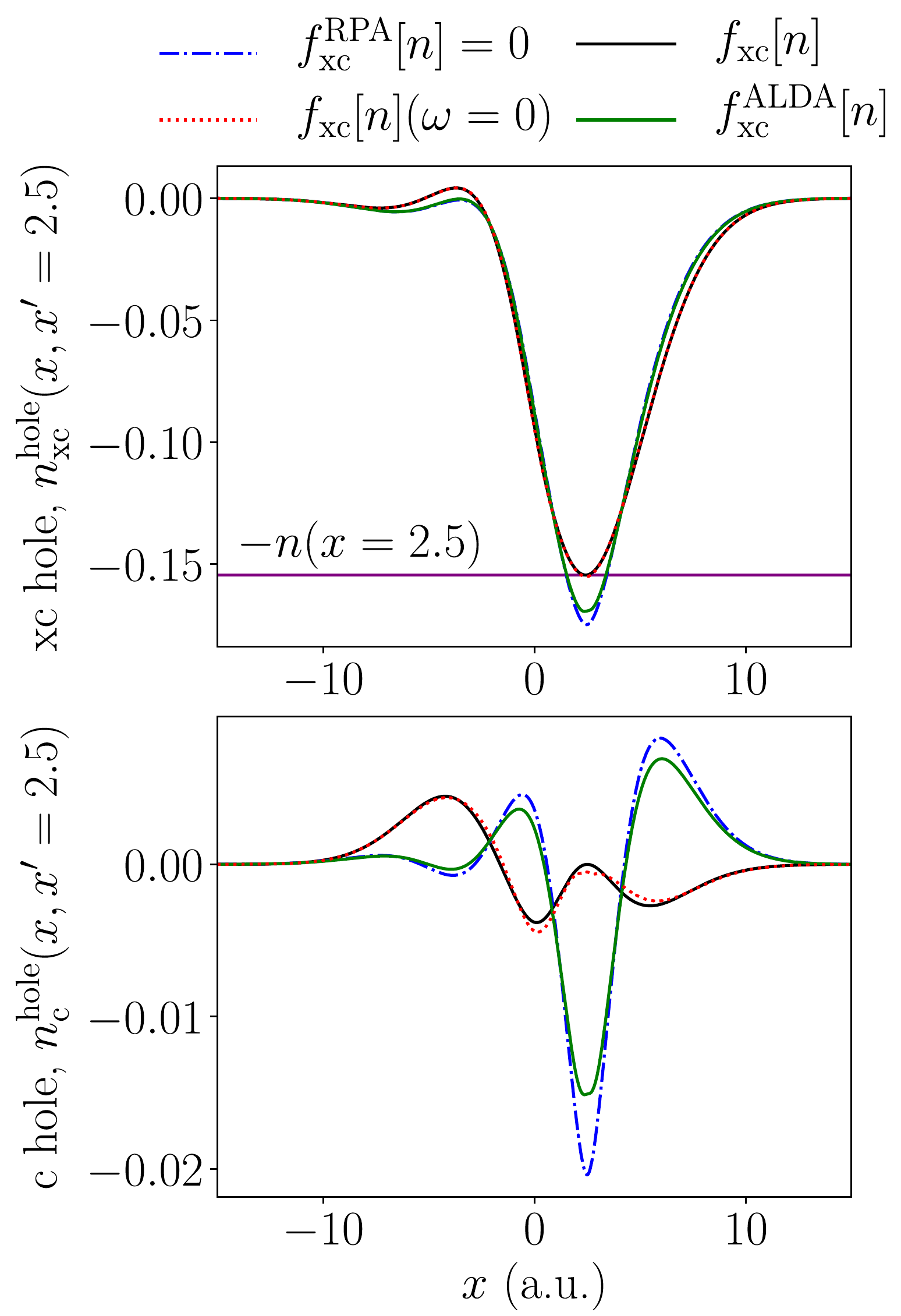}
\end{center}
\caption{(Upper panel) The atomic xc hole at $x'=2.5$ a.u. derived using the approximate xc kernels considered in this work. (Note that in all cases the exchange hole is exact because the exact ground-state Kohn-Sham orbitals were used to evaluate the ACFDT functional.) The density at $x'=2.5$ a.u. (purple) and the xc hole should satisfy the following relation: $n^\text{hole}_\text{xc}(x=2.5, x'=2.5) = -n(x=2.5)$, i.e. there is zero chance the second particle is at the same position as the first. (Lower panel) The correlation hole at $x'=2.5$ a.u., where it can be seen that the adiabatic LDA and RPA on-top correlation holes are much too deep, giving rise to the excessively negative energies seen in Fig$.$ \ref{fig:AtomEtotRelativeError}.}
\label{fig:AtomXCHoles}
\end{figure}

As discussed above, in order to isolate deficiencies in the approximate $f_\text{xc}[n]$, we shall hereafter consider xc holes that utilize the exact Kohn-Sham orbitals and density in the corresponding one-shot ACFDT calculations \footnote{This means that the exchange hole will be exact.}. The upper panel in Fig$.$ \ref{fig:AtomXCHoles} demonstrates that the so-called \textit{on-top} xc hole is far too deep in the case of the RPA and the adiabatic LDA. This can be interpreted as the second particle having \textit{negative} probability to be measured at the position of the first -- an artifact due to the original particle interacting with itself at $x' = 2.5$ a.u. This \textit{self-interaction} at the level of the xc kernel plagues both the RPA and adiabatic LDA similarly, which can be seen more clearly in the correlation hole \citep{BurkeABC} (lower panel of Fig$.$ \ref{fig:AtomXCHoles}). In fact, the RPA and adiabatic LDA correlation holes reach a minimum at $x=2.5$ a.u. where they should be identically zero, that is to say, exchange is entirely responsible for the fact that two fermions cannot be measured at the same position. Minimizing the ACFDT energy functional defined with $f_\text{xc}^{\text{ALDA}}$ or $f_\text{xc}^{\text{RPA}}=0$ accentuates the self-interaction thereby making the on-top correlation hole even deeper.

Note that the traditional on-top LDA xc hole, that is, the on-top hole defined with the exact LDA pair-correlation function \citep{Gunnarsson1976}, is known to be accurate and therefore central to the success of conventional ground-state LDA calculations \citep{Burke1998}. However, the ALDA-ACFDT approximate xc hole, which utilizes an adiabatic xc kernel derived from an LDA functional, is distinct from the traditional LDA xc hole, leading to errors of the kind discussed in the previous paragraph.

A simple quantitative measure of self-interaction is given using one-particle calculations, wherein any correlation present is necessarily due to self-interaction. For the atomic system, the one-particle adiabatic LDA correlation energy is $-0.016$ a.u., whereas the corresponding adiabatic LDA two-particle correlation energy is $-0.05$ a.u. Doubling the spurious one-particle energy reveals that the majority of the two-particle correlation energy constitutes self-interaction. The AE approximation has \textit{no} spurious one-particle correlation energy, because the exact one-particle $f_\text{xc}[n]$ is itself adiabatic. Whilst this line of reasoning suggests that the AE approximation is self-interaction free, this is not quite the case: the exact exchange kernel $f_\text{x}[n](x,x',\omega)$ provides a more rigorous definition of a self-interaction-free kernel, and this includes an $\omega$ dependence \citep{Kim2002,Hellgren2008}. Nonetheless, it can be seen in Fig$.$ \ref{fig:AtomXCHoles} that the non-local spatial structure in the exact $f_\text{xc}[n]$ at $\omega=0$ is able to largely remedy the deficiencies due to self-interaction -- Section \ref{sec:SourceOfError} provides more insight on this matter.

\subsection{Long-Range Correlations}

\begin{figure}[ht]
\begin{center}
\includegraphics[width=3.2in]{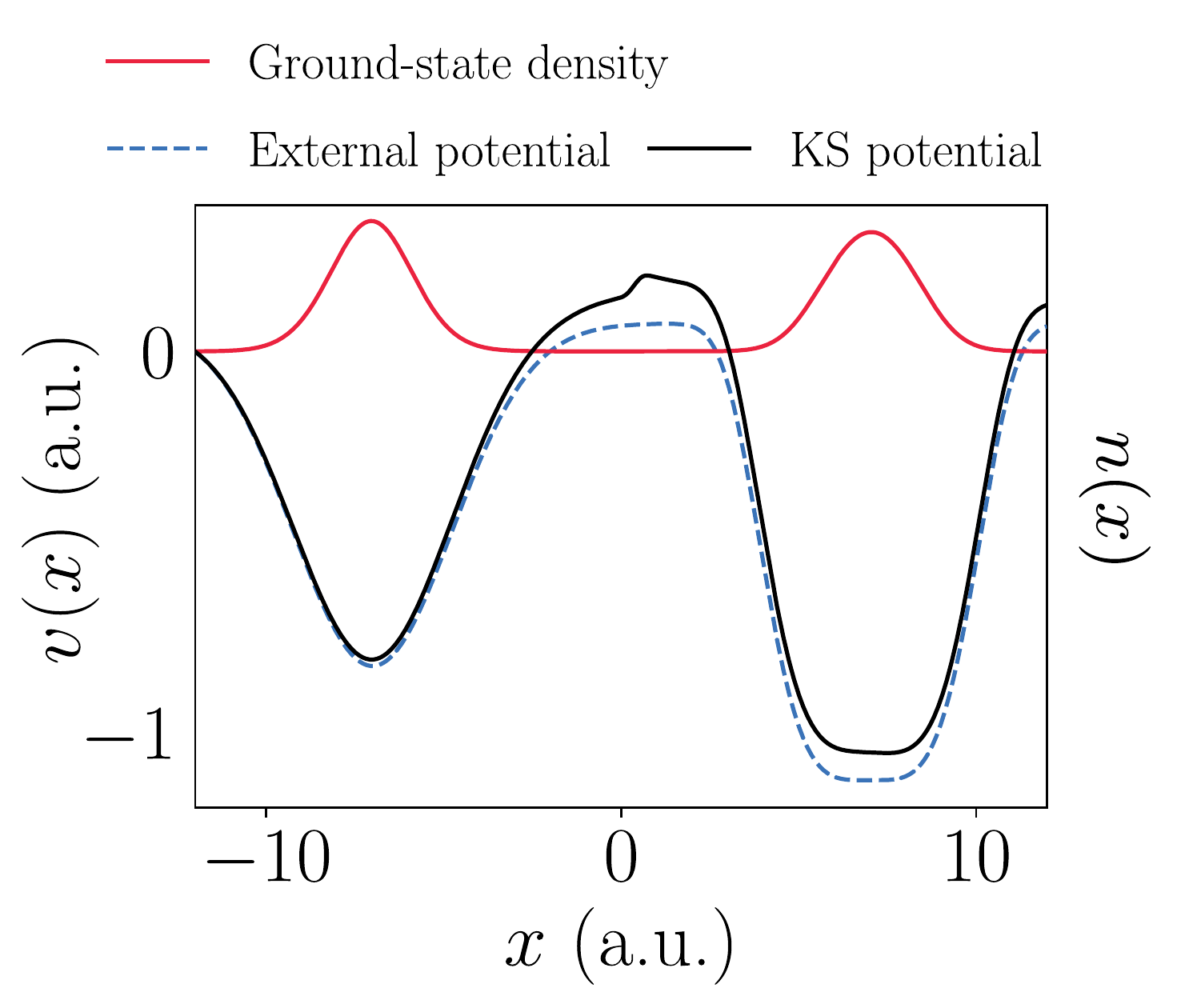}
\end{center}
\caption{The ground-state density, external potential, and reverse-engineered Kohn-Sham potential for the double well system. The external and Kohn-Sham potentials have been shifted for illustrative purposes.}
\label{fig:DoubleWell}
\end{figure}

We now examine a system containing long-range correlations, i.e. van der Waals correlations \citep{3Dobson1998}: a double well, see Fig$.$ \ref{fig:DoubleWell}. This system is also known to exhibit a step-like feature in the Kohn-Sham potential \citep{Hodgson2016,Hodgson2021,Almbladh1985} in order to localize one electron in each well -- without such a step, the non-interacting particles would spuriously collapse into the same well, as is the case in Hartree theory. 

The wells are separated at distance $R = 14$ a.u., meaning the correlation energy is low $\mathcal{O}(10^{-6})$ a.u., and as such the RPA and adiabatic LDA suffer even more as a result of self-interaction: the relative error in $E_\text{c}$ is now $\mathcal{O}(10^{5})$. However, one might expect such error is alleviated to a degree when computing quantities that involve energy differences, e.g. dissociation curves. In the context of dissociation curves, it is imperative that the total energy method in question provides at least some account of long-range correlations, and it is this phenomenon toward which we now focus our attention.

\begin{figure}[ht]
\begin{center}
\includegraphics[width=2.7in]{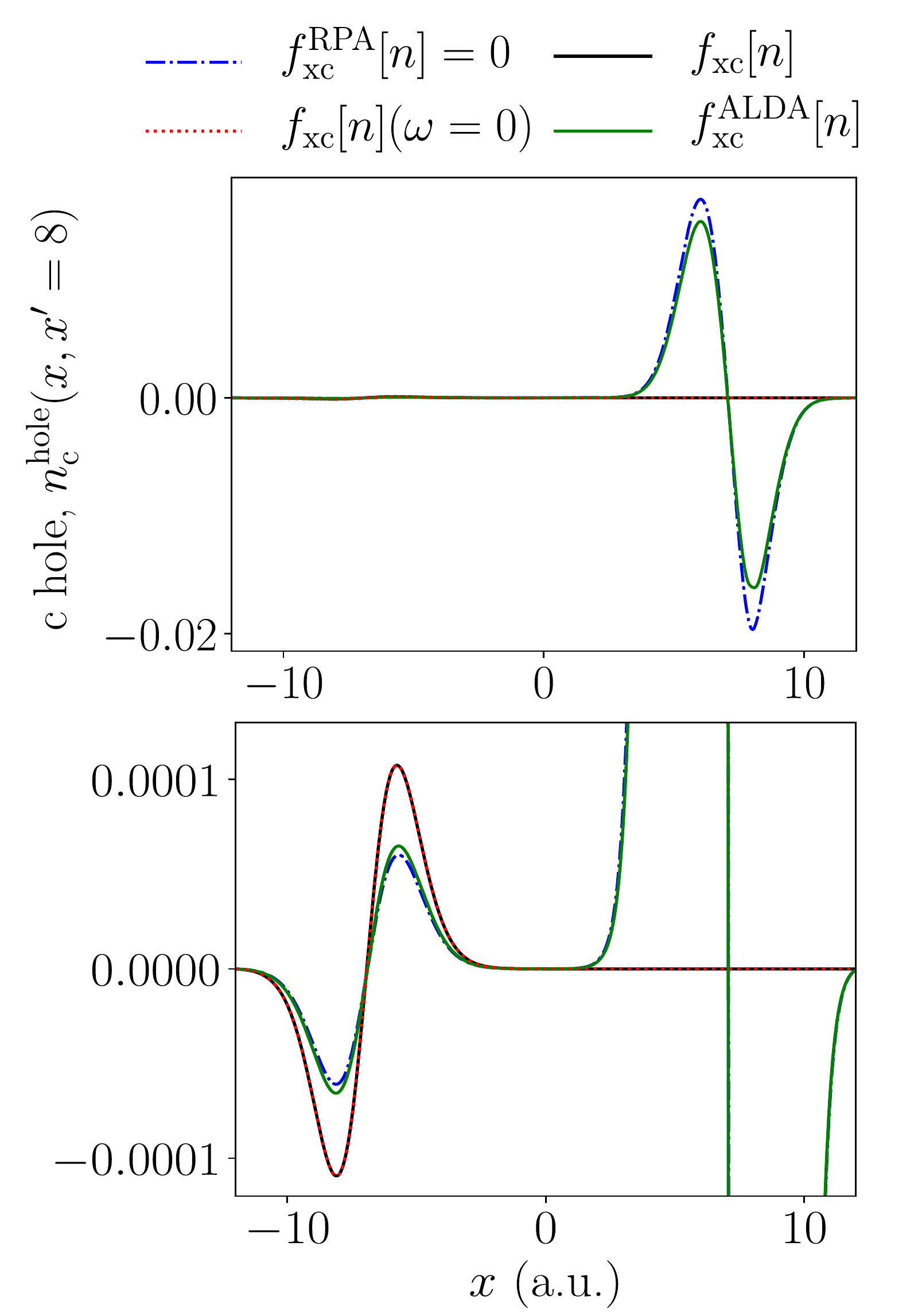}
\end{center}
\caption{The double well correlation hole $n^\text{hole}_\text{c}(x,x'=8)$ derived using the approximate xc kernels considered in this work. A particle is measured in the right-hand well, see Fig$.$ \ref{fig:DoubleWell}, at $x' = 8$ a.u., in which a significant spurious correlation hole emerges due to self-interaction when using the adiabatic LDA and RPA approximations. The physical correlation is entirely long range here, and the lower panel zooms in on the long-range contribution from the upper panel in order to demonstrate that all three approximations are indeed able to capture this long-range correlation hole.}
\label{fig:DoubleWellCHoles}
\end{figure}

The correlation hole for the double well is given in Fig$.$ \ref{fig:DoubleWellCHoles}. As expected, upon measuring an electron in the right-hand well at $x' = 8$ a.u., there is a strong erroneous contribution to RPA and adiabatic LDA correlation holes \textit{in the same well where the electron was measured} due to the electron Coulomb interacting with itself -- the correlation hole should largely reside in the \textit{opposite} well. Inclusion of the non-local spatial structure in the exact xc kernel at $\omega = 0$ is able to almost entirely correct this issue, as discussed in the previous section.

On the other hand, it can also be observed in the lower panel of Fig$.$ \ref{fig:DoubleWellCHoles} that the correlation hole in the opposing well, a necessarily long-range feature, is modeled successfully in all three of the approximations considered. This observation represents the systematic non-locality introduced via the Dyson equation, and constitutes the principal advantage of the ACFDT formalism over conventional Kohn-Sham calculations. We expect that these considerations account for the fact that, even in the case of the RPA, the limit of large separation in molecules can be described \citep{Furche2001}. We further note that the AE approximation $f_\text{xc}[n](\omega=0)$ is able to \textit{quantitatively} capture the long-range correlation hole, thus further establishing the notion that the majority of both short-range and long-range correlations reside in the spatial dependence of $f_\text{xc}[n](x,x',\omega=0)$ when using the ACFDT framework.

Despite self-consistent RPA-ACFDT and adiabatic LDA-ACFDT calculations accentuating self-interaction error by making the on-top correlation hole even deeper, we find that the long-range correlation hole is \textit{improved} by minimizing the ACFDT functional. Therefore, even though these approximate self-consistent ACFDT calculations yield poorer absolute energies than their one-shot counterparts, it appears that the long-range properties are improved.

The exact double well $f_\text{xc}$ computed here contains the step-like features discussed in \citep{2Hellgren2012,Hellgren2013,Hellgren2018} that relate to the derivative discontinuity (see supplemental material). In fact, all three approximations to $f_\text{xc}[n]$ recover the step in the Kohn-Sham potential when minimizing the corresponding ACFDT energy functional. This is due to the fact that the exact-exchange optimized effective potential contains the step, and the ACFDT total energy functionals comprise in large part the Hartree-Fock functional.

\subsection{Sources of Error Explained}
\label{sec:SourceOfError}

\begin{figure}[ht]
\begin{center}
\includegraphics[width=3.5in]{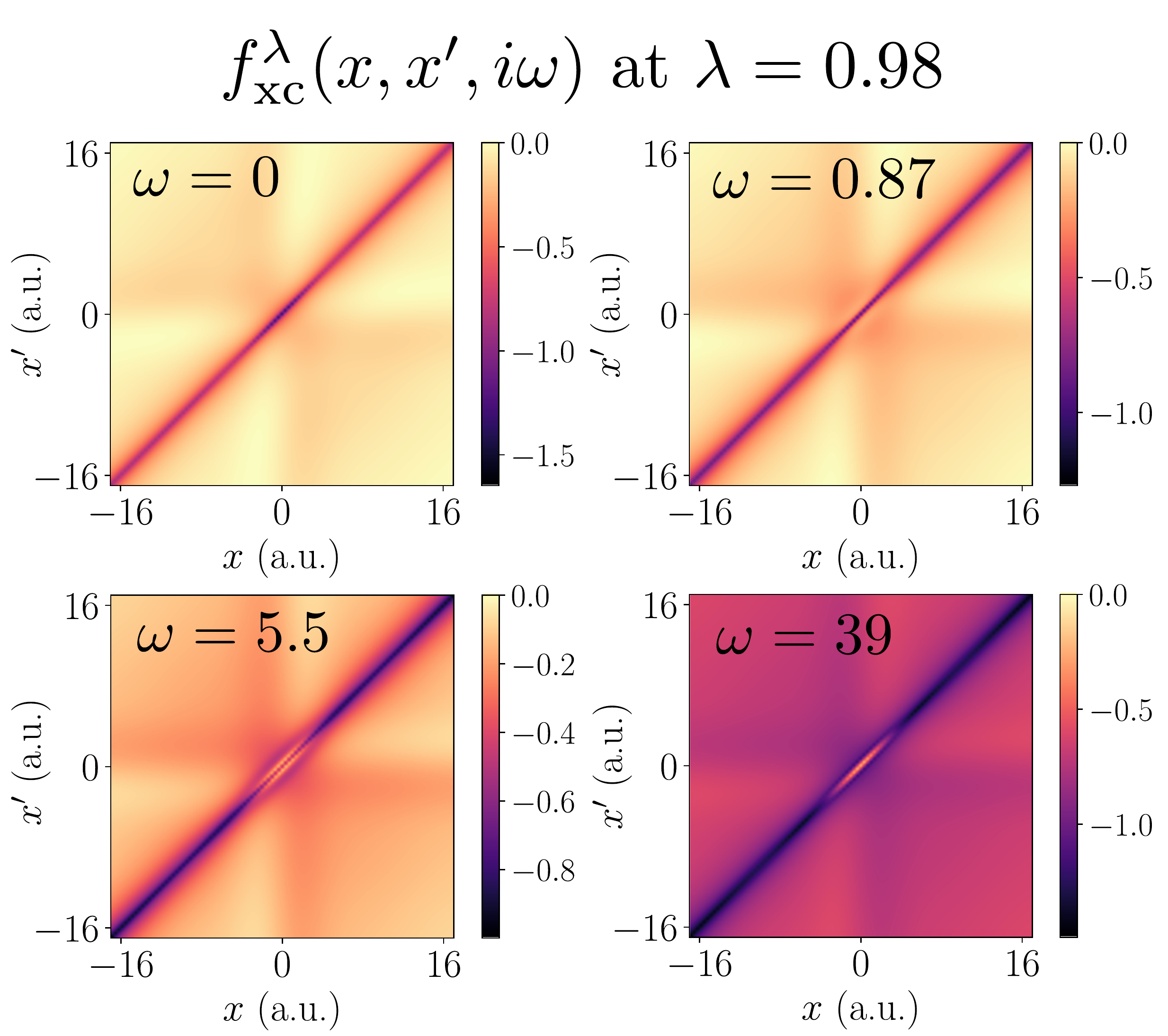}
\end{center}
\caption{The exact numerical xc kernel $f^\lambda_\text{xc}(x,x',i\omega)$ for the slab system at $\lambda = 0.98$ along the adiabatic connection. The xc kernel is depicted at different stages along the $i\omega$ axis, demonstrating that there is little frequency dependence below $\omega = 1$ a.u., whereupon the xc kernel then deviates from its adiabatic form.}
\label{fig:SlabXCKernel}
\end{figure}

The central aim of this section is to carefully examine the reasons behind the exceptional performance of the AE approximation $f_\text{xc}[n](\omega=0)$ as demonstrated in the previous sections. The slab system, see supplemental material, is used to illustrate the forthcoming conclusions. First, it is possible to observe that the exact $f_\text{xc}[n](x,x',i\omega)$ along $i\omega$ undergoes considerable change away from its adiabatic $\omega = 0$ limit, see Fig$.$ \ref{fig:SlabXCKernel}. (Note, however, that this change is much more regular than along the real $\omega$-axis due to circumventing the singularities described in \citep{Woods2021}). Therefore, the performance of the AE approximation \textit{cannot} be attributed to xc kernels generally possessing modest frequency dependence along $i\omega$.

\begin{figure}[ht]
\begin{center}
\includegraphics[width=3.2in]{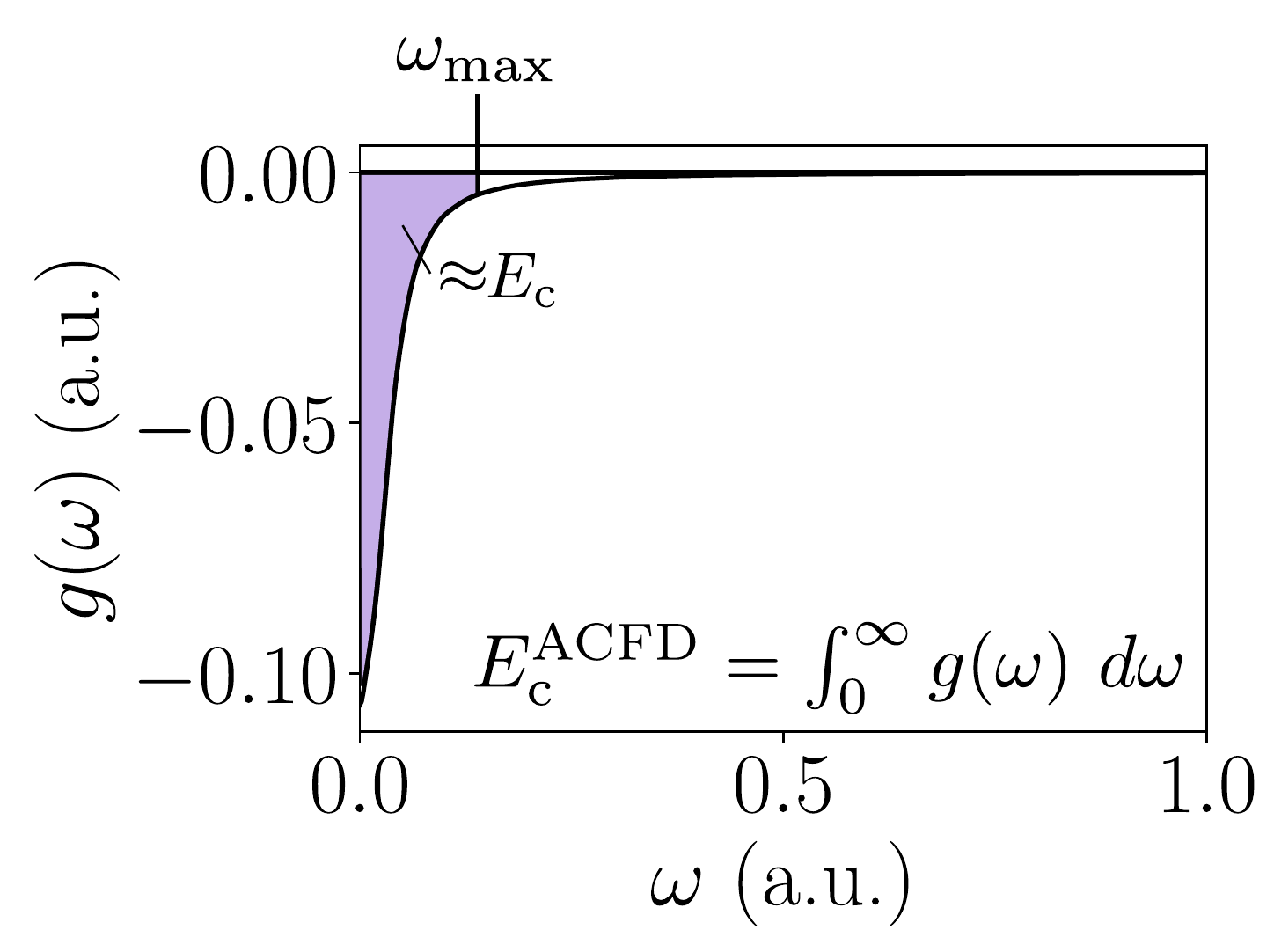}
\end{center}
\caption{The exact $\omega$-dependent integrand $g(\omega)$ of the ACFDT correlation energy expression is depicted for the slab system. Analytic integration allows us to terminate the integration at some finite $\omega_\text{max}$ in order to determine the amount of correlation energy contained in the curve at frequencies $\omega \leq \omega_\text{max}$.}
\label{fig:SlabFreqIntegration}
\end{figure}

Since we have access to the exact $\lambda$-interacting adiabatic connection wavefunctions/energies $\{ |\Psi_i^\lambda\rangle, E_i^\lambda \}$ and the exact Kohn-Sham single-particle wavefunctions/energies $\{ |\Phi_i\rangle, \varepsilon_i \}$, the $\omega$-dependent integral in $E_\text{c}^\text{ACFD}[n]$ can be evaluated analytically up to some arbitrary $\omega_\text{max}$, see supplemental material. It is thus possible to isolate error whose exclusive source is a finite $\omega_\text{max}$, as depicted in Fig$.$ \ref{fig:SlabFreqIntegration} -- another perspective on this is that the exact $f_\text{xc}[n]$ is used for $\omega \leq \omega_\text{max}$, and $f_\text{xc}[n] = -\lambda f_\text{H}$, i.e. $\chi = \chi_0$, is used for $\omega > \omega_\text{max}$.  

The absolute error in the correlation energy is defined, 
\begin{align}
\Delta E_c(\omega_\text{max}) = |E^\text{ACFD}_c(\infty) - E^\text{ACFD}_c(\omega_\text{max})|, \label{eq:AbsErrorEc}
\end{align}
where $E_\text{c}^\text{ACFD}(\omega_\text{max})$ is the ACFDT correlation energy whose analytic $\omega$ integration has been terminated at $\omega_\text{max}$ ($E^\text{ACFD}_c(\infty)$ is therefore the exact correlation energy). We find that chemical accuracy is surpassed prior to $\omega = 1$ a.u., i.e. the capacity for an approximate $f_\text{xc}[n]$ to yield accurate correlation energies predominantly resides in its structure \textit{below} some characteristic $\omega_\text{max}$ ($\omega_\text{max} \approx 1$ in this case) -- see supplemental material for a plot. In fact, beyond this $\omega_\text{max}$, the AE approximation $f_\text{xc}[n](\omega = 0)$  produces approximate interacting response functions upon solution of the Dyson equation that are \textit{as poor as} the RPA, adiabatic LDA, or simply setting $\chi^\lambda = \chi_0$. 

In all systems studied here, whilst the exact $f_\text{xc}[n]$ is not adiabatic in general, it \textit{is} adiabatic over the most relevant $\omega$ range, and therefore the AE approximation performs accordingly. Such an observation is commensurate with prior findings \citep{Woods2021}: in the context of the optical spectrum, $f_\text{xc}[n](\omega)$ is required \textit{at the $\omega$ corresponding to a transition energy}, and thus at frequencies beyond the lowest lying excitations, the AE approximation ceases to perform.  

In light of these observations, it is imperative that practical approximate integration schemes target the relevant $\omega$ region, whereas the long-range tail of the $\omega$-dependent integrand is less important. Whilst the integration scheme proposed in this work Eq$.$ (\ref{eq:ChangeOfCoords}) compresses the domain $[0,\infty]$ to $[0,\sqrt{\pi / a}]$, thus allowing us to capture the tail, its central advantage instead comes from an explicit treatment of the terms responsible for the low-$\omega$ structure in $g(\omega)$, see Fig$.$ \ref{fig:SlabFreqIntegration}. Namely, the scheme defines an integral change of coordinates such that a single term in the Lehmann representation of $\chi$ (or $\chi_0$) Eq$.$ (\ref{eq:ResponseLehman}) is \textit{linear}, meaning Gauss-Legendre quadrature is exact with just one $\omega$ grid point, see supplemental material. The characteristic range $\omega_\text{max}$ within which most of the correlation energy resides will depend on a number of factors in practice, such as the size of the interacting and Kohn-Sham gap, and number of excitations clustered around this gap \citep{Harl2008}. 

\section{Conclusion}

The exact coupling constant-dependent xc kernel $f^\lambda_\text{xc}[n](x,x',i\omega)$ for four prototype one-dimensional finite systems has been calculated \citep{Woods2021} and utilized to better understand the sources of error in practical ACFDT total energy calculations.

Whilst the frequency-dependence in the exact $f_\text{xc}[n]$ along the real $\omega$-axis is critical for recovering the optical spectrum \citep{Woods2021,Maitra2004,Maitra2021}, this is not the case for ACFDT total energies, where it is demonstrated that chemical accuracy can be consistently attained using the AE kernel $f_\text{xc}[n](\omega=0)$, i.e. neglecting the frequency dependence in the exact xc kernel, but otherwise retaining its non-local spatial structure when solving the Dyson equation of linear response time-dependent DFT (this was also the case for the HEG in Refs$.$ \citep{Lein2000,Ruzsinszky2020}). The exact kernel $f_\text{xc}[n]$ is shown to exhibit little change within the $\omega$ interval that contains the majority of the ground-state correlation energy. Therefore, it is crucial that approximate kernels are accurate within this interval, as is the case for the AE kernel, thereby explaining its success. In light of these findings, a novel change of coordinates is proposed for the ACFDT $\omega$-dependent integral that directly targets the relevant term(s) in the Lehmann representation of the response function in order to reduce the number of Gauss-Legendre grid points required. An interesting course for future work would involve testing this integration scheme in practical settings.


The coupling-constant averaged correlation hole is used, alongside one-particle calculations, to illustrate that strong self-interaction is present in the RPA and adiabatic LDA kernels -- in both cases, the on-top correlation hole is far too deep, and the two-particle correlation energy differs little from the twice the spurious one-particle correlation energy. Due to the observations outlined in the previous paragraph, the spatial non-locality in the AE kernel almost entirely remedies the problem of self-interaction, despite the self-interaction-free exact-exchange kernel $f_\text{x}$ possessing a frequency dependence. 

In the case of a double-well system, we show that all three of the approximate kernels considered in this work -- the RPA, adiabatic LDA, and AE kernels -- are able to capture the long-range correlation hole. This observation further evidences the central advantage of ACFDT calculations, i.e. in describing long-range van der Waals correlations. Moreover, we find that minimizing the ACFDT total energy functional, while yielding less accurate absolute energies in the case of the RPA and adiabatic LDA, \textit{is} able to improve the description of long-range correlations. 

The distinction between one-shot and self-consistent ACFDT total energies is considered in depth, where we recall that the latter minimizes the ACFDT total energy functional, whereas the former evaluates the ACFDT total energy functional at some density $n$ obtained from an approximate ground-state Kohn-Sham calculation. The ACFDT total energy functional is found to be somewhat insensitive to the density at which the functional is evaluated (within reason), meaning the dominant factor that dictates the effectiveness of an ACFDT calculation is the approximate $f_\text{xc}[n]$ with which $E^\text{ACFD}[n]$ is defined. Where self-interaction is present, minimizing the ACFDT functional accentuates issues, thus making the erroneous on-top correlation hole even deeper. However, in the context of self-consistent AE-ACFDT calculations, the exact Kohn-Sham potential is faithfully recovered as the implicit optimized effective potential contained within the calculation, and therefore so are chemically accurate total energies. 

The scope for obtaining both improved total energies and improved Kohn-Sham potentials using the ACFDT approach appears to be significant, and depends critically on capturing the spatial non-locality present in the AE kernel. For example, the energies and potentials that would result from combining modern adiabatic non-local kernels \citep{Gorling2019, Olsen2019} with self-consistent ACFDT calculations \citep{Bleiziffer2013,Bleiziffer2015,Thierbach2020,Niquet2003,Hellgren2012,Hellgren2010,Verma2012,Nguyen2009} offer an interesting prospect.

\begin{acknowledgments}
The authors thank Micheal Hutcheon for helpful discussions. N.D.W is supported by the EPSRC Centre for Doctoral Training in Computational Methods for Materials Science for funding under grant number EP/L015552/1. We are grateful for computational support from the UK national high performance computing service, ARCHER, through the UKCP consortium under EPSRC grant number EP/P022596/1. Data created during this research is available through the Cambridge Apollo research repository \footnote{\url{https://doi.org/10.17863/CAM.75013}}.
\end{acknowledgments}

\bibliographystyle{apsrev4-2}
\bibliography{mainReferences}

\begin{thebibliography}{94}%
\makeatletter
\providecommand \@ifxundefined [1]{%
 \@ifx{#1\undefined}
}%
\providecommand \@ifnum [1]{%
 \ifnum #1\expandafter \@firstoftwo
 \else \expandafter \@secondoftwo
 \fi
}%
\providecommand \@ifx [1]{%
 \ifx #1\expandafter \@firstoftwo
 \else \expandafter \@secondoftwo
 \fi
}%
\providecommand \natexlab [1]{#1}%
\providecommand \enquote  [1]{``#1''}%
\providecommand \bibnamefont  [1]{#1}%
\providecommand \bibfnamefont [1]{#1}%
\providecommand \citenamefont [1]{#1}%
\providecommand \href@noop [0]{\@secondoftwo}%
\providecommand \href [0]{\begingroup \@sanitize@url \@href}%
\providecommand \@href[1]{\@@startlink{#1}\@@href}%
\providecommand \@@href[1]{\endgroup#1\@@endlink}%
\providecommand \@sanitize@url [0]{\catcode `\\12\catcode `\$12\catcode
  `\&12\catcode `\#12\catcode `\^12\catcode `\_12\catcode `\%12\relax}%
\providecommand \@@startlink[1]{}%
\providecommand \@@endlink[0]{}%
\providecommand \url  [0]{\begingroup\@sanitize@url \@url }%
\providecommand \@url [1]{\endgroup\@href {#1}{\urlprefix }}%
\providecommand \urlprefix  [0]{URL }%
\providecommand \Eprint [0]{\href }%
\providecommand \doibase [0]{https://doi.org/}%
\providecommand \selectlanguage [0]{\@gobble}%
\providecommand \bibinfo  [0]{\@secondoftwo}%
\providecommand \bibfield  [0]{\@secondoftwo}%
\providecommand \translation [1]{[#1]}%
\providecommand \BibitemOpen [0]{}%
\providecommand \bibitemStop [0]{}%
\providecommand \bibitemNoStop [0]{.\EOS\space}%
\providecommand \EOS [0]{\spacefactor3000\relax}%
\providecommand \BibitemShut  [1]{\csname bibitem#1\endcsname}%
\let\auto@bib@innerbib\@empty
\bibitem [{\citenamefont {Ullrich}(2012)}]{Ullrich2012}%
  \BibitemOpen
  \bibfield  {author} {\bibinfo {author} {\bibfnamefont {C.~A.}\ \bibnamefont
  {Ullrich}},\ }\href@noop {} {\emph {\bibinfo {title} {{Time-dependent
  density-functional theory: concepts and applications}}}}\ (\bibinfo
  {publisher} {Oxford University Press},\ \bibinfo {year} {2012})\ p.\ \bibinfo
  {pages} {334}\BibitemShut {NoStop}%
\bibitem [{\citenamefont {Olsen}\ \emph {et~al.}(2019)\citenamefont {Olsen},
  \citenamefont {Patrick}, \citenamefont {Bates}, \citenamefont {Ruzsinszky},\
  and\ \citenamefont {Thygesen}}]{Olsen2019}%
  \BibitemOpen
  \bibfield  {author} {\bibinfo {author} {\bibfnamefont {T.}~\bibnamefont
  {Olsen}}, \bibinfo {author} {\bibfnamefont {C.~E.}\ \bibnamefont {Patrick}},
  \bibinfo {author} {\bibfnamefont {J.~E.}\ \bibnamefont {Bates}}, \bibinfo
  {author} {\bibfnamefont {A.}~\bibnamefont {Ruzsinszky}},\ and\ \bibinfo
  {author} {\bibfnamefont {K.~S.}\ \bibnamefont {Thygesen}},\ }\href
  {https://doi.org/10.1038/s41524-019-0242-8} {\bibinfo {title} {{Beyond the
  RPA and GW methods with adiabatic xc-kernels for accurate ground state and
  quasiparticle energies}}} (\bibinfo {year} {2019})\BibitemShut {NoStop}%
\bibitem [{\citenamefont {G{\"{o}}rling}(2019)}]{Gorling2019}%
  \BibitemOpen
  \bibfield  {author} {\bibinfo {author} {\bibfnamefont {A.}~\bibnamefont
  {G{\"{o}}rling}},\ }\href {https://doi.org/10.1103/PhysRevB.99.235120}
  {\bibfield  {journal} {\bibinfo  {journal} {Physical Review B}\ }\textbf
  {\bibinfo {volume} {99}},\ \bibinfo {pages} {235120} (\bibinfo {year}
  {2019})}\BibitemShut {NoStop}%
\bibitem [{\citenamefont {Marques}\ \emph {et~al.}(2006)\citenamefont
  {Marques}, \citenamefont {Ullrich}, \citenamefont {Nogueira}, \citenamefont
  {Rubio}, \citenamefont {Burke},\ and\ \citenamefont {Gross}}]{Marques2006}%
  \BibitemOpen
  \bibinfo {editor} {\bibfnamefont {M.~A.}\ \bibnamefont {Marques}}, \bibinfo
  {editor} {\bibfnamefont {C.~A.}\ \bibnamefont {Ullrich}}, \bibinfo {editor}
  {\bibfnamefont {F.}~\bibnamefont {Nogueira}}, \bibinfo {editor}
  {\bibfnamefont {A.}~\bibnamefont {Rubio}}, \bibinfo {editor} {\bibfnamefont
  {K.}~\bibnamefont {Burke}},\ and\ \bibinfo {editor} {\bibfnamefont
  {E.~K.~U.}\ \bibnamefont {Gross}},\ eds.,\ \href
  {https://doi.org/10.1007/b11767107} {\emph {\bibinfo {title} {{Time-Dependent
  Density Functional Theory}}}},\ \bibinfo {series} {Lecture Notes in Physics},
  Vol.\ \bibinfo {volume} {706}\ (\bibinfo  {publisher} {Springer Berlin
  Heidelberg},\ \bibinfo {address} {Berlin, Heidelberg},\ \bibinfo {year}
  {2006})\ pp.\ \bibinfo {pages} {443--462}\BibitemShut {NoStop}%
\bibitem [{\citenamefont {Burke}(2007)}]{BurkeABC}%
  \BibitemOpen
  \bibinfo {editor} {\bibfnamefont {K.}~\bibnamefont {Burke}},\ ed.,\ \href
  {https://doi.org/https://dft.uci.edu/doc/g1.pdf} {\emph {\bibinfo {title}
  {{ABC of DFT}}}}\ (\bibinfo  {publisher} {Unpublished},\ \bibinfo {year}
  {2007})\ pp.\ \bibinfo {pages} {93--111}\BibitemShut {NoStop}%
\bibitem [{\citenamefont {Dobson}\ \emph
  {et~al.}(1998{\natexlab{a}})\citenamefont {Dobson}, \citenamefont {Vignale},\
  and\ \citenamefont {Mukunda}}]{Dobson1998}%
  \BibitemOpen
  \bibinfo {editor} {\bibfnamefont {J.}~\bibnamefont {Dobson}}, \bibinfo
  {editor} {\bibfnamefont {G.}~\bibnamefont {Vignale}},\ and\ \bibinfo {editor}
  {\bibfnamefont {D.}~\bibnamefont {Mukunda}},\ eds.,\ \href
  {https://link.springer.com/book/10.1007/978-1-4899-0316-7} {\emph {\bibinfo
  {title} {{Electronic Density Functional Theory: Recent Progress and New
  Directions}}}}\ (\bibinfo  {publisher} {Springer Science},\ \bibinfo {year}
  {1998})\BibitemShut {NoStop}%
\bibitem [{\citenamefont {Woods}\ \emph {et~al.}(2021)\citenamefont {Woods},
  \citenamefont {Entwistle},\ and\ \citenamefont {Godby}}]{Woods2021}%
  \BibitemOpen
  \bibfield  {author} {\bibinfo {author} {\bibfnamefont {N.~D.}\ \bibnamefont
  {Woods}}, \bibinfo {author} {\bibfnamefont {M.~T.}\ \bibnamefont
  {Entwistle}},\ and\ \bibinfo {author} {\bibfnamefont {R.~W.}\ \bibnamefont
  {Godby}},\ }\href {https://doi.org/10.1103/PhysRevB.103.125155} {\bibfield
  {journal} {\bibinfo  {journal} {Physical Review B}\ }\textbf {\bibinfo
  {volume} {103}},\ \bibinfo {pages} {125155} (\bibinfo {year} {2021})},\
  \Eprint {https://arxiv.org/abs/2101.12207} {arXiv:2101.12207} \BibitemShut
  {NoStop}%
\bibitem [{\citenamefont {Constantin}(2019)}]{Constantin2019}%
  \BibitemOpen
  \bibfield  {author} {\bibinfo {author} {\bibfnamefont {L.~A.}\ \bibnamefont
  {Constantin}},\ }\href {https://doi.org/10.1103/PhysRevB.99.085117}
  {\bibfield  {journal} {\bibinfo  {journal} {Physical Review B}\ }\textbf
  {\bibinfo {volume} {99}},\ \bibinfo {pages} {085117} (\bibinfo {year}
  {2019})}\BibitemShut {NoStop}%
\bibitem [{\citenamefont {Perdew}\ \emph {et~al.}(2005)\citenamefont {Perdew},
  \citenamefont {Ruzsinszky}, \citenamefont {Tao}, \citenamefont {Staroverov},
  \citenamefont {Scuseria},\ and\ \citenamefont {Csonka}}]{Perdew2005}%
  \BibitemOpen
  \bibfield  {author} {\bibinfo {author} {\bibfnamefont {J.~P.}\ \bibnamefont
  {Perdew}}, \bibinfo {author} {\bibfnamefont {A.}~\bibnamefont {Ruzsinszky}},
  \bibinfo {author} {\bibfnamefont {J.}~\bibnamefont {Tao}}, \bibinfo {author}
  {\bibfnamefont {V.~N.}\ \bibnamefont {Staroverov}}, \bibinfo {author}
  {\bibfnamefont {G.~E.}\ \bibnamefont {Scuseria}},\ and\ \bibinfo {author}
  {\bibfnamefont {G.~I.}\ \bibnamefont {Csonka}},\ }\href
  {https://doi.org/10.1063/1.1904565} {\bibfield  {journal} {\bibinfo
  {journal} {Journal of Chemical Physics}\ }\textbf {\bibinfo {volume} {123}},\
  \bibinfo {pages} {6158} (\bibinfo {year} {2005})}\BibitemShut {NoStop}%
\bibitem [{\citenamefont {Wilhelm}\ \emph {et~al.}(2016)\citenamefont
  {Wilhelm}, \citenamefont {Seewald}, \citenamefont {Ben},\ and\ \citenamefont
  {Hutter}}]{Wilhelm2016}%
  \BibitemOpen
  \bibfield  {author} {\bibinfo {author} {\bibfnamefont {J.}~\bibnamefont
  {Wilhelm}}, \bibinfo {author} {\bibfnamefont {P.}~\bibnamefont {Seewald}},
  \bibinfo {author} {\bibfnamefont {M.~D.}\ \bibnamefont {Ben}},\ and\ \bibinfo
  {author} {\bibfnamefont {J.}~\bibnamefont {Hutter}},\ }\href
  {https://doi.org/10.1021/ACS.JCTC.6B00840} {\bibfield  {journal} {\bibinfo
  {journal} {Journal of Chemical Theory and Computation}\ }\textbf {\bibinfo
  {volume} {12}},\ \bibinfo {pages} {5851} (\bibinfo {year}
  {2016})}\BibitemShut {NoStop}%
\bibitem [{\citenamefont {Graf}\ \emph {et~al.}(2018)\citenamefont {Graf},
  \citenamefont {Beuerle}, \citenamefont {Schurkus}, \citenamefont {Luenser},
  \citenamefont {Savasci},\ and\ \citenamefont {Ochsenfeld}}]{Graf2018}%
  \BibitemOpen
  \bibfield  {author} {\bibinfo {author} {\bibfnamefont {D.}~\bibnamefont
  {Graf}}, \bibinfo {author} {\bibfnamefont {M.}~\bibnamefont {Beuerle}},
  \bibinfo {author} {\bibfnamefont {H.~F.}\ \bibnamefont {Schurkus}}, \bibinfo
  {author} {\bibfnamefont {A.}~\bibnamefont {Luenser}}, \bibinfo {author}
  {\bibfnamefont {G.}~\bibnamefont {Savasci}},\ and\ \bibinfo {author}
  {\bibfnamefont {C.}~\bibnamefont {Ochsenfeld}},\ }\href
  {https://doi.org/10.1021/ACS.JCTC.8B00177} {\bibfield  {journal} {\bibinfo
  {journal} {Journal of Chemical Theory and Computation}\ }\textbf {\bibinfo
  {volume} {14}},\ \bibinfo {pages} {2505} (\bibinfo {year}
  {2018})}\BibitemShut {NoStop}%
\bibitem [{\citenamefont {Kaltak}\ \emph
  {et~al.}(2014{\natexlab{a}})\citenamefont {Kaltak}, \citenamefont
  {Klime{\v{s}}},\ and\ \citenamefont {Kresse}}]{Kaltak2014}%
  \BibitemOpen
  \bibfield  {author} {\bibinfo {author} {\bibfnamefont {M.}~\bibnamefont
  {Kaltak}}, \bibinfo {author} {\bibfnamefont {J.}~\bibnamefont
  {Klime{\v{s}}}},\ and\ \bibinfo {author} {\bibfnamefont {G.}~\bibnamefont
  {Kresse}},\ }\href {https://doi.org/10.1021/CT5001268} {\bibfield  {journal}
  {\bibinfo  {journal} {Journal of Chemical Theory and Computation}\ }\textbf
  {\bibinfo {volume} {10}},\ \bibinfo {pages} {2498} (\bibinfo {year}
  {2014}{\natexlab{a}})}\BibitemShut {NoStop}%
\bibitem [{\citenamefont {Kaltak}\ \emph
  {et~al.}(2014{\natexlab{b}})\citenamefont {Kaltak}, \citenamefont
  {Klime{\v{s}}},\ and\ \citenamefont {Kresse}}]{2Kaltak2014}%
  \BibitemOpen
  \bibfield  {author} {\bibinfo {author} {\bibfnamefont {M.}~\bibnamefont
  {Kaltak}}, \bibinfo {author} {\bibfnamefont {J.}~\bibnamefont
  {Klime{\v{s}}}},\ and\ \bibinfo {author} {\bibfnamefont {G.}~\bibnamefont
  {Kresse}},\ }\href {https://doi.org/10.1103/PhysRevB.90.054115} {\bibfield
  {journal} {\bibinfo  {journal} {Physical Review B}\ }\textbf {\bibinfo
  {volume} {90}},\ \bibinfo {pages} {054115} (\bibinfo {year}
  {2014}{\natexlab{b}})}\BibitemShut {NoStop}%
\bibitem [{\citenamefont {Furche}(2001)}]{Furche2001}%
  \BibitemOpen
  \bibfield  {author} {\bibinfo {author} {\bibfnamefont {F.}~\bibnamefont
  {Furche}},\ }\href {https://doi.org/10.1103/PhysRevB.64.195120} {\bibfield
  {journal} {\bibinfo  {journal} {Physical Review B - Condensed Matter and
  Materials Physics}\ }\textbf {\bibinfo {volume} {64}},\ \bibinfo {pages}
  {195120} (\bibinfo {year} {2001})}\BibitemShut {NoStop}%
\bibitem [{\citenamefont {Fuchs}\ and\ \citenamefont
  {Gonze}(2002)}]{Fuchs2002}%
  \BibitemOpen
  \bibfield  {author} {\bibinfo {author} {\bibfnamefont {M.}~\bibnamefont
  {Fuchs}}\ and\ \bibinfo {author} {\bibfnamefont {X.}~\bibnamefont {Gonze}},\
  }\href {https://doi.org/10.1103/PhysRevB.65.235109} {\bibfield  {journal}
  {\bibinfo  {journal} {Physical Review B - Condensed Matter and Materials
  Physics}\ }\textbf {\bibinfo {volume} {65}},\ \bibinfo {pages} {1} (\bibinfo
  {year} {2002})}\BibitemShut {NoStop}%
\bibitem [{\citenamefont {Bleiziffer}\ \emph {et~al.}(2013)\citenamefont
  {Bleiziffer}, \citenamefont {He{\ss}elmann},\ and\ \citenamefont
  {G{\"{o}}rling}}]{Bleiziffer2013}%
  \BibitemOpen
  \bibfield  {author} {\bibinfo {author} {\bibfnamefont {P.}~\bibnamefont
  {Bleiziffer}}, \bibinfo {author} {\bibfnamefont {A.}~\bibnamefont
  {He{\ss}elmann}},\ and\ \bibinfo {author} {\bibfnamefont {A.}~\bibnamefont
  {G{\"{o}}rling}},\ }\href {https://doi.org/10.1063/1.4818984} {\bibfield
  {journal} {\bibinfo  {journal} {Journal of Chemical Physics}\ }\textbf
  {\bibinfo {volume} {139}},\ \bibinfo {pages} {84113} (\bibinfo {year}
  {2013})}\BibitemShut {NoStop}%
\bibitem [{\citenamefont {Nguyen}\ and\ \citenamefont {{De
  Gironcoli}}(2009)}]{Nguyen2009}%
  \BibitemOpen
  \bibfield  {author} {\bibinfo {author} {\bibfnamefont {H.~V.}\ \bibnamefont
  {Nguyen}}\ and\ \bibinfo {author} {\bibfnamefont {S.}~\bibnamefont {{De
  Gironcoli}}},\ }\href {https://doi.org/10.1103/PhysRevB.79.205114} {\bibfield
   {journal} {\bibinfo  {journal} {Physical Review B - Condensed Matter and
  Materials Physics}\ }\textbf {\bibinfo {volume} {79}},\ \bibinfo {pages}
  {205114} (\bibinfo {year} {2009})}\BibitemShut {NoStop}%
\bibitem [{\citenamefont {Kohn}\ \emph {et~al.}(1998)\citenamefont {Kohn},
  \citenamefont {Meir},\ and\ \citenamefont {Makarov}}]{Kohn1998}%
  \BibitemOpen
  \bibfield  {author} {\bibinfo {author} {\bibfnamefont {W.}~\bibnamefont
  {Kohn}}, \bibinfo {author} {\bibfnamefont {Y.}~\bibnamefont {Meir}},\ and\
  \bibinfo {author} {\bibfnamefont {D.~E.}\ \bibnamefont {Makarov}},\ }\href
  {https://doi.org/10.1103/PhysRevLett.80.4153} {\bibfield  {journal} {\bibinfo
   {journal} {Physical Review Letters}\ }\textbf {\bibinfo {volume} {80}},\
  \bibinfo {pages} {4153} (\bibinfo {year} {1998})},\ \Eprint
  {https://arxiv.org/abs/9707328} {arXiv:9707328 [cond-mat]} \BibitemShut
  {NoStop}%
\bibitem [{\citenamefont {Burke}\ \emph {et~al.}(1998)\citenamefont {Burke},
  \citenamefont {Perdew},\ and\ \citenamefont {Ernzerhof}}]{Burke1998}%
  \BibitemOpen
  \bibfield  {author} {\bibinfo {author} {\bibfnamefont {K.}~\bibnamefont
  {Burke}}, \bibinfo {author} {\bibfnamefont {J.~P.}\ \bibnamefont {Perdew}},\
  and\ \bibinfo {author} {\bibfnamefont {M.}~\bibnamefont {Ernzerhof}},\ }\href
  {https://doi.org/10.1063/1.476976} {\bibfield  {journal} {\bibinfo  {journal}
  {Journal of Chemical Physics}\ }\textbf {\bibinfo {volume} {109}},\ \bibinfo
  {pages} {3760} (\bibinfo {year} {1998})}\BibitemShut {NoStop}%
\bibitem [{\citenamefont {Yan}\ \emph {et~al.}(2000)\citenamefont {Yan},
  \citenamefont {Perdew},\ and\ \citenamefont {Kurth}}]{Yan2000}%
  \BibitemOpen
  \bibfield  {author} {\bibinfo {author} {\bibfnamefont {Z.}~\bibnamefont
  {Yan}}, \bibinfo {author} {\bibfnamefont {J.~P.}\ \bibnamefont {Perdew}},\
  and\ \bibinfo {author} {\bibfnamefont {S.}~\bibnamefont {Kurth}},\ }\href
  {https://doi.org/10.1103/PhysRevB.61.16430} {\bibfield  {journal} {\bibinfo
  {journal} {Physical Review B - Condensed Matter and Materials Physics}\
  }\textbf {\bibinfo {volume} {61}},\ \bibinfo {pages} {16430} (\bibinfo {year}
  {2000})}\BibitemShut {NoStop}%
\bibitem [{\citenamefont {Engel}\ \emph {et~al.}(2000)\citenamefont {Engel},
  \citenamefont {H{\"{o}}ck},\ and\ \citenamefont {Dreizler}}]{Engel2000}%
  \BibitemOpen
  \bibfield  {author} {\bibinfo {author} {\bibfnamefont {E.}~\bibnamefont
  {Engel}}, \bibinfo {author} {\bibfnamefont {A.}~\bibnamefont {H{\"{o}}ck}},\
  and\ \bibinfo {author} {\bibfnamefont {R.~M.}\ \bibnamefont {Dreizler}},\
  }\href {https://doi.org/10.1103/PhysRevA.61.032502} {\bibfield  {journal}
  {\bibinfo  {journal} {Physical Review A - Atomic, Molecular, and Optical
  Physics}\ }\textbf {\bibinfo {volume} {61}},\ \bibinfo {pages} {5} (\bibinfo
  {year} {2000})}\BibitemShut {NoStop}%
\bibitem [{\citenamefont {Perdew}(1993)}]{Perdew1993}%
  \BibitemOpen
  \bibfield  {author} {\bibinfo {author} {\bibfnamefont {J.~P.}\ \bibnamefont
  {Perdew}},\ }\href {https://doi.org/10.1002/qua.560480813} {\bibfield
  {journal} {\bibinfo  {journal} {International Journal of Quantum Chemistry}\
  }\textbf {\bibinfo {volume} {48}},\ \bibinfo {pages} {93} (\bibinfo {year}
  {1993})}\BibitemShut {NoStop}%
\bibitem [{\citenamefont {Krogel}\ \emph {et~al.}(2020)\citenamefont {Krogel},
  \citenamefont {Yuk}, \citenamefont {Kent},\ and\ \citenamefont
  {Cooper}}]{Krogel2020}%
  \BibitemOpen
  \bibfield  {author} {\bibinfo {author} {\bibfnamefont {J.~T.}\ \bibnamefont
  {Krogel}}, \bibinfo {author} {\bibfnamefont {S.~F.}\ \bibnamefont {Yuk}},
  \bibinfo {author} {\bibfnamefont {P.~R.~C.}\ \bibnamefont {Kent}},\ and\
  \bibinfo {author} {\bibfnamefont {V.~R.}\ \bibnamefont {Cooper}},\ }\href
  {https://doi.org/10.1021/ACS.JPCA.0C05973} {\bibfield  {journal} {\bibinfo
  {journal} {The Journal of Physical Chemistry A}\ }\textbf {\bibinfo {volume}
  {124}},\ \bibinfo {pages} {9867} (\bibinfo {year} {2020})}\BibitemShut
  {NoStop}%
\bibitem [{\citenamefont {Berland}\ \emph {et~al.}(2015)\citenamefont
  {Berland}, \citenamefont {Cooper}, \citenamefont {Lee}, \citenamefont
  {Schr{\"{o}}der}, \citenamefont {Thonhauser}, \citenamefont {Hyldgaard},\
  and\ \citenamefont {Lundqvist}}]{Berland2015}%
  \BibitemOpen
  \bibfield  {author} {\bibinfo {author} {\bibfnamefont {K.}~\bibnamefont
  {Berland}}, \bibinfo {author} {\bibfnamefont {V.~R.}\ \bibnamefont {Cooper}},
  \bibinfo {author} {\bibfnamefont {K.}~\bibnamefont {Lee}}, \bibinfo {author}
  {\bibfnamefont {E.}~\bibnamefont {Schr{\"{o}}der}}, \bibinfo {author}
  {\bibfnamefont {T.}~\bibnamefont {Thonhauser}}, \bibinfo {author}
  {\bibfnamefont {P.}~\bibnamefont {Hyldgaard}},\ and\ \bibinfo {author}
  {\bibfnamefont {B.~I.}\ \bibnamefont {Lundqvist}},\ }\href
  {https://doi.org/10.1088/0034-4885/78/6/066501} {\bibfield  {journal}
  {\bibinfo  {journal} {Reports on Progress in Physics}\ }\textbf {\bibinfo
  {volume} {78}},\ \bibinfo {pages} {066501} (\bibinfo {year}
  {2015})}\BibitemShut {NoStop}%
\bibitem [{\citenamefont {Harl}\ \emph {et~al.}(2010)\citenamefont {Harl},
  \citenamefont {Schimka},\ and\ \citenamefont {Kresse}}]{Harl2010}%
  \BibitemOpen
  \bibfield  {author} {\bibinfo {author} {\bibfnamefont {J.}~\bibnamefont
  {Harl}}, \bibinfo {author} {\bibfnamefont {L.}~\bibnamefont {Schimka}},\ and\
  \bibinfo {author} {\bibfnamefont {G.}~\bibnamefont {Kresse}},\ }\href
  {https://doi.org/10.1103/PhysRevB.81.115126} {\bibfield  {journal} {\bibinfo
  {journal} {Physical Review B - Condensed Matter and Materials Physics}\
  }\textbf {\bibinfo {volume} {81}},\ \bibinfo {pages} {115126} (\bibinfo
  {year} {2010})}\BibitemShut {NoStop}%
\bibitem [{\citenamefont {Freeman}(1977)}]{Freeman1977}%
  \BibitemOpen
  \bibfield  {author} {\bibinfo {author} {\bibfnamefont {D.~L.}\ \bibnamefont
  {Freeman}},\ }\href {https://doi.org/10.1103/PhysRevB.15.5512} {\bibfield
  {journal} {\bibinfo  {journal} {Physical Review B}\ }\textbf {\bibinfo
  {volume} {15}},\ \bibinfo {pages} {5512} (\bibinfo {year}
  {1977})}\BibitemShut {NoStop}%
\bibitem [{\citenamefont {Perdew}\ and\ \citenamefont
  {Wang}(1992)}]{Perdew1992}%
  \BibitemOpen
  \bibfield  {author} {\bibinfo {author} {\bibfnamefont {J.~P.}\ \bibnamefont
  {Perdew}}\ and\ \bibinfo {author} {\bibfnamefont {Y.}~\bibnamefont {Wang}},\
  }\href {https://doi.org/10.1103/PhysRevB.45.13244} {\bibfield  {journal}
  {\bibinfo  {journal} {Physical Review B}\ }\textbf {\bibinfo {volume} {45}},\
  \bibinfo {pages} {13244} (\bibinfo {year} {1992})}\BibitemShut {NoStop}%
\bibitem [{\citenamefont {Eshuis}\ \emph {et~al.}(2012)\citenamefont {Eshuis},
  \citenamefont {Bates},\ and\ \citenamefont {Furche}}]{Eshuis2012}%
  \BibitemOpen
  \bibfield  {author} {\bibinfo {author} {\bibfnamefont {H.}~\bibnamefont
  {Eshuis}}, \bibinfo {author} {\bibfnamefont {J.~E.}\ \bibnamefont {Bates}},\
  and\ \bibinfo {author} {\bibfnamefont {F.}~\bibnamefont {Furche}},\ }\href
  {https://doi.org/10.1007/s00214-011-1084-8} {\bibfield  {journal} {\bibinfo
  {journal} {Theoretical Chemistry Accounts}\ }\textbf {\bibinfo {volume}
  {131}},\ \bibinfo {pages} {1} (\bibinfo {year} {2012})}\BibitemShut {NoStop}%
\bibitem [{\citenamefont {Ruan}\ \emph {et~al.}(2021)\citenamefont {Ruan},
  \citenamefont {Ren}, \citenamefont {Gould},\ and\ \citenamefont
  {Ruzsinszky}}]{Ruan2021}%
  \BibitemOpen
  \bibfield  {author} {\bibinfo {author} {\bibfnamefont {S.}~\bibnamefont
  {Ruan}}, \bibinfo {author} {\bibfnamefont {X.}~\bibnamefont {Ren}}, \bibinfo
  {author} {\bibfnamefont {T.}~\bibnamefont {Gould}},\ and\ \bibinfo {author}
  {\bibfnamefont {A.}~\bibnamefont {Ruzsinszky}},\ }\href
  {https://doi.org/10.1021/ACS.JCTC.0C01079} {\bibfield  {journal} {\bibinfo
  {journal} {Journal of Chemical Theory and Computation}\ }\textbf {\bibinfo
  {volume} {17}},\ \bibinfo {pages} {2107} (\bibinfo {year}
  {2021})}\BibitemShut {NoStop}%
\bibitem [{\citenamefont {Fuchs}\ \emph {et~al.}(2005)\citenamefont {Fuchs},
  \citenamefont {Niquet}, \citenamefont {Gonze},\ and\ \citenamefont
  {Burke}}]{Fuchs2005}%
  \BibitemOpen
  \bibfield  {author} {\bibinfo {author} {\bibfnamefont {M.}~\bibnamefont
  {Fuchs}}, \bibinfo {author} {\bibfnamefont {Y.~M.}\ \bibnamefont {Niquet}},
  \bibinfo {author} {\bibfnamefont {X.}~\bibnamefont {Gonze}},\ and\ \bibinfo
  {author} {\bibfnamefont {K.}~\bibnamefont {Burke}},\ }\href
  {https://doi.org/10.1063/1.1858371} {\bibfield  {journal} {\bibinfo
  {journal} {Journal of Chemical Physics}\ }\textbf {\bibinfo {volume} {122}},\
  \bibinfo {pages} {121104} (\bibinfo {year} {2005})}\BibitemShut {NoStop}%
\bibitem [{\citenamefont {Hellgren}\ and\ \citenamefont {{Von
  Barth}}(2008)}]{Hellgren2008}%
  \BibitemOpen
  \bibfield  {author} {\bibinfo {author} {\bibfnamefont {M.}~\bibnamefont
  {Hellgren}}\ and\ \bibinfo {author} {\bibfnamefont {U.}~\bibnamefont {{Von
  Barth}}},\ }\href {https://doi.org/10.1103/PhysRevB.78.115107} {\bibfield
  {journal} {\bibinfo  {journal} {Physical Review B - Condensed Matter and
  Materials Physics}\ }\textbf {\bibinfo {volume} {78}},\ \bibinfo {pages}
  {115107} (\bibinfo {year} {2008})}\BibitemShut {NoStop}%
\bibitem [{\citenamefont {Bleiziffer}\ \emph {et~al.}(2012)\citenamefont
  {Bleiziffer}, \citenamefont {Heelmann},\ and\ \citenamefont
  {G{\"{o}}rling}}]{Bleiziffer2012}%
  \BibitemOpen
  \bibfield  {author} {\bibinfo {author} {\bibfnamefont {P.}~\bibnamefont
  {Bleiziffer}}, \bibinfo {author} {\bibfnamefont {A.}~\bibnamefont
  {Heelmann}},\ and\ \bibinfo {author} {\bibfnamefont {A.}~\bibnamefont
  {G{\"{o}}rling}},\ }\href {https://doi.org/10.1063/1.3697845} {\bibfield
  {journal} {\bibinfo  {journal} {Journal of Chemical Physics}\ }\textbf
  {\bibinfo {volume} {136}},\ \bibinfo {pages} {134102} (\bibinfo {year}
  {2012})}\BibitemShut {NoStop}%
\bibitem [{\citenamefont {G\"orling}(1998)}]{Gorling1998}%
  \BibitemOpen
  \bibfield  {author} {\bibinfo {author} {\bibfnamefont {A.}~\bibnamefont
  {G\"orling}},\ }\href {https://doi.org/https://doi.org/10.1063/1.3179756}
  {\bibfield  {journal} {\bibinfo  {journal} {Int J Quant Chem}\ }\textbf
  {\bibinfo {volume} {69}},\ \bibinfo {pages} {265277} (\bibinfo {year}
  {1998})}\BibitemShut {NoStop}%
\bibitem [{\citenamefont {G{\"{o}}rling}(1998)}]{2Gorling1998}%
  \BibitemOpen
  \bibfield  {author} {\bibinfo {author} {\bibfnamefont {A.}~\bibnamefont
  {G{\"{o}}rling}},\ }\href {https://doi.org/10.1103/PhysRevA.57.3433}
  {\bibfield  {journal} {\bibinfo  {journal} {Physical Review A - Atomic,
  Molecular, and Optical Physics}\ }\textbf {\bibinfo {volume} {57}},\ \bibinfo
  {pages} {3433} (\bibinfo {year} {1998})}\BibitemShut {NoStop}%
\bibitem [{\citenamefont {Kim}\ and\ \citenamefont
  {G{\"{o}}rling}(2002)}]{Kim2002}%
  \BibitemOpen
  \bibfield  {author} {\bibinfo {author} {\bibfnamefont {Y.~H.}\ \bibnamefont
  {Kim}}\ and\ \bibinfo {author} {\bibfnamefont {A.}~\bibnamefont
  {G{\"{o}}rling}},\ }\href {https://doi.org/10.1103/PhysRevB.66.035114}
  {\bibfield  {journal} {\bibinfo  {journal} {Physical Review B - Condensed
  Matter and Materials Physics}\ }\textbf {\bibinfo {volume} {66}},\ \bibinfo
  {pages} {1} (\bibinfo {year} {2002})}\BibitemShut {NoStop}%
\bibitem [{\citenamefont {Hellgren}\ and\ \citenamefont {{Von
  Barth}}(2010)}]{Hellgren2010}%
  \BibitemOpen
  \bibfield  {author} {\bibinfo {author} {\bibfnamefont {M.}~\bibnamefont
  {Hellgren}}\ and\ \bibinfo {author} {\bibfnamefont {U.}~\bibnamefont {{Von
  Barth}}},\ }\href {https://doi.org/10.1063/1.3290947} {\bibfield  {journal}
  {\bibinfo  {journal} {Journal of Chemical Physics}\ }\textbf {\bibinfo
  {volume} {132}},\ \bibinfo {pages} {34106} (\bibinfo {year}
  {2010})}\BibitemShut {NoStop}%
\bibitem [{\citenamefont {He{\ss}elmann}\ and\ \citenamefont
  {G{\"{o}}rling}(2010)}]{Hebelmann2010}%
  \BibitemOpen
  \bibfield  {author} {\bibinfo {author} {\bibfnamefont {A.}~\bibnamefont
  {He{\ss}elmann}}\ and\ \bibinfo {author} {\bibfnamefont {A.}~\bibnamefont
  {G{\"{o}}rling}},\ }\href {https://doi.org/10.1080/00268970903476662}
  {\bibfield  {journal} {\bibinfo  {journal} {Molecular Physics}\ }\textbf
  {\bibinfo {volume} {108}},\ \bibinfo {pages} {359} (\bibinfo {year}
  {2010})}\BibitemShut {NoStop}%
\bibitem [{\citenamefont {He{\ss}elmann}\ and\ \citenamefont
  {G{\"{o}}rling}(2011{\natexlab{a}})}]{Hebelmann2011}%
  \BibitemOpen
  \bibfield  {author} {\bibinfo {author} {\bibfnamefont {A.}~\bibnamefont
  {He{\ss}elmann}}\ and\ \bibinfo {author} {\bibfnamefont {A.}~\bibnamefont
  {G{\"{o}}rling}},\ }\href {https://doi.org/10.1103/PhysRevLett.106.093001}
  {\bibfield  {journal} {\bibinfo  {journal} {Physical Review Letters}\
  }\textbf {\bibinfo {volume} {106}},\ \bibinfo {pages} {093001} (\bibinfo
  {year} {2011}{\natexlab{a}})}\BibitemShut {NoStop}%
\bibitem [{\citenamefont {Colonna}\ \emph {et~al.}(2014)\citenamefont
  {Colonna}, \citenamefont {Hellgren},\ and\ \citenamefont {{De
  Gironcoli}}}]{Colonna2014}%
  \BibitemOpen
  \bibfield  {author} {\bibinfo {author} {\bibfnamefont {N.}~\bibnamefont
  {Colonna}}, \bibinfo {author} {\bibfnamefont {M.}~\bibnamefont {Hellgren}},\
  and\ \bibinfo {author} {\bibfnamefont {S.}~\bibnamefont {{De Gironcoli}}},\
  }\href {https://doi.org/10.1103/PhysRevB.90.125150} {\bibfield  {journal}
  {\bibinfo  {journal} {Physical Review B - Condensed Matter and Materials
  Physics}\ }\textbf {\bibinfo {volume} {90}},\ \bibinfo {pages} {125150}
  (\bibinfo {year} {2014})}\BibitemShut {NoStop}%
\bibitem [{\citenamefont {Bleiziffer}\ \emph {et~al.}(2015)\citenamefont
  {Bleiziffer}, \citenamefont {Krug},\ and\ \citenamefont
  {G{\"{o}}rling}}]{Bleiziffer2015}%
  \BibitemOpen
  \bibfield  {author} {\bibinfo {author} {\bibfnamefont {P.}~\bibnamefont
  {Bleiziffer}}, \bibinfo {author} {\bibfnamefont {M.}~\bibnamefont {Krug}},\
  and\ \bibinfo {author} {\bibfnamefont {A.}~\bibnamefont {G{\"{o}}rling}},\
  }\href {https://doi.org/10.1063/1.4922517} {\bibfield  {journal} {\bibinfo
  {journal} {Journal of Chemical Physics}\ }\textbf {\bibinfo {volume} {142}},\
  \bibinfo {pages} {244108} (\bibinfo {year} {2015})}\BibitemShut {NoStop}%
\bibitem [{\citenamefont {Toulouse}\ \emph {et~al.}(2009)\citenamefont
  {Toulouse}, \citenamefont {Gerber}, \citenamefont {Jansen}, \citenamefont
  {Savin},\ and\ \citenamefont {{\'{A}}ngy{\'{a}}n}}]{Toulouse2009}%
  \BibitemOpen
  \bibfield  {author} {\bibinfo {author} {\bibfnamefont {J.}~\bibnamefont
  {Toulouse}}, \bibinfo {author} {\bibfnamefont {I.~C.}\ \bibnamefont
  {Gerber}}, \bibinfo {author} {\bibfnamefont {G.}~\bibnamefont {Jansen}},
  \bibinfo {author} {\bibfnamefont {A.}~\bibnamefont {Savin}},\ and\ \bibinfo
  {author} {\bibfnamefont {J.~G.}\ \bibnamefont {{\'{A}}ngy{\'{a}}n}},\ }\href
  {https://doi.org/10.1103/PhysRevLett.102.096404} {\bibfield  {journal}
  {\bibinfo  {journal} {Physical Review Letters}\ }\textbf {\bibinfo {volume}
  {102}},\ \bibinfo {pages} {096404} (\bibinfo {year} {2009})},\ \Eprint
  {https://arxiv.org/abs/0812.3302} {arXiv:0812.3302} \BibitemShut {NoStop}%
\bibitem [{\citenamefont {Janesko}\ \emph
  {et~al.}(2009{\natexlab{a}})\citenamefont {Janesko}, \citenamefont
  {Henderson},\ and\ \citenamefont {Scuseria}}]{Janesko2009}%
  \BibitemOpen
  \bibfield  {author} {\bibinfo {author} {\bibfnamefont {B.~G.}\ \bibnamefont
  {Janesko}}, \bibinfo {author} {\bibfnamefont {T.~M.}\ \bibnamefont
  {Henderson}},\ and\ \bibinfo {author} {\bibfnamefont {G.~E.}\ \bibnamefont
  {Scuseria}},\ }\href {https://doi.org/10.1063/1.3090814} {\bibfield
  {journal} {\bibinfo  {journal} {Journal of Chemical Physics}\ }\textbf
  {\bibinfo {volume} {130}},\ \bibinfo {pages} {94103} (\bibinfo {year}
  {2009}{\natexlab{a}})}\BibitemShut {NoStop}%
\bibitem [{\citenamefont {Zhu}\ \emph {et~al.}(2010)\citenamefont {Zhu},
  \citenamefont {Toulouse}, \citenamefont {Savin},\ and\ \citenamefont
  {{\'{A}}ngy{\'{a}}n}}]{Zhu2010}%
  \BibitemOpen
  \bibfield  {author} {\bibinfo {author} {\bibfnamefont {W.}~\bibnamefont
  {Zhu}}, \bibinfo {author} {\bibfnamefont {J.}~\bibnamefont {Toulouse}},
  \bibinfo {author} {\bibfnamefont {A.}~\bibnamefont {Savin}},\ and\ \bibinfo
  {author} {\bibfnamefont {J.~G.}\ \bibnamefont {{\'{A}}ngy{\'{a}}n}},\ }\href
  {https://doi.org/10.1063/1.3431616} {\bibfield  {journal} {\bibinfo
  {journal} {Journal of Chemical Physics}\ }\textbf {\bibinfo {volume} {132}},\
  \bibinfo {pages} {84119} (\bibinfo {year} {2010})}\BibitemShut {NoStop}%
\bibitem [{\citenamefont {Toulouse}\ \emph {et~al.}(2010)\citenamefont
  {Toulouse}, \citenamefont {Zhu}, \citenamefont {{\'{A}}ngy{\'{a}}n},\ and\
  \citenamefont {Savin}}]{Toulouse2010}%
  \BibitemOpen
  \bibfield  {author} {\bibinfo {author} {\bibfnamefont {J.}~\bibnamefont
  {Toulouse}}, \bibinfo {author} {\bibfnamefont {W.}~\bibnamefont {Zhu}},
  \bibinfo {author} {\bibfnamefont {J.~G.}\ \bibnamefont
  {{\'{A}}ngy{\'{a}}n}},\ and\ \bibinfo {author} {\bibfnamefont
  {A.}~\bibnamefont {Savin}},\ }\href
  {https://doi.org/10.1103/PhysRevA.82.032502} {\bibfield  {journal} {\bibinfo
  {journal} {Physical Review A - Atomic, Molecular, and Optical Physics}\
  }\textbf {\bibinfo {volume} {82}},\ \bibinfo {pages} {032502} (\bibinfo
  {year} {2010})},\ \Eprint {https://arxiv.org/abs/1006.2061} {arXiv:1006.2061}
  \BibitemShut {NoStop}%
\bibitem [{\citenamefont {Janesko}\ \emph
  {et~al.}(2009{\natexlab{b}})\citenamefont {Janesko}, \citenamefont
  {Henderson},\ and\ \citenamefont {Scuseria}}]{3Janesko2009}%
  \BibitemOpen
  \bibfield  {author} {\bibinfo {author} {\bibfnamefont {B.~G.}\ \bibnamefont
  {Janesko}}, \bibinfo {author} {\bibfnamefont {T.~M.}\ \bibnamefont
  {Henderson}},\ and\ \bibinfo {author} {\bibfnamefont {G.~E.}\ \bibnamefont
  {Scuseria}},\ }\href {https://doi.org/10.1063/1.3176514} {\bibfield
  {journal} {\bibinfo  {journal} {Journal of Chemical Physics}\ }\textbf
  {\bibinfo {volume} {131}},\ \bibinfo {pages} {114105} (\bibinfo {year}
  {2009}{\natexlab{b}})}\BibitemShut {NoStop}%
\bibitem [{\citenamefont {Janesko}\ and\ \citenamefont
  {Scuseria}(2009)}]{2Janesko2009}%
  \BibitemOpen
  \bibfield  {author} {\bibinfo {author} {\bibfnamefont {B.~G.}\ \bibnamefont
  {Janesko}}\ and\ \bibinfo {author} {\bibfnamefont {G.~E.}\ \bibnamefont
  {Scuseria}},\ }\href {https://doi.org/10.1063/1.3250834} {\bibfield
  {journal} {\bibinfo  {journal} {Journal of Chemical Physics}\ }\textbf
  {\bibinfo {volume} {131}},\ \bibinfo {pages} {154106} (\bibinfo {year}
  {2009})}\BibitemShut {NoStop}%
\bibitem [{\citenamefont {Toulouse}\ \emph {et~al.}(2011)\citenamefont
  {Toulouse}, \citenamefont {Zhu}, \citenamefont {Savin}, \citenamefont
  {Jansen},\ and\ \citenamefont {{\'{A}}ngy{\'{a}}n}}]{Toulouse2011}%
  \BibitemOpen
  \bibfield  {author} {\bibinfo {author} {\bibfnamefont {J.}~\bibnamefont
  {Toulouse}}, \bibinfo {author} {\bibfnamefont {W.}~\bibnamefont {Zhu}},
  \bibinfo {author} {\bibfnamefont {A.}~\bibnamefont {Savin}}, \bibinfo
  {author} {\bibfnamefont {G.}~\bibnamefont {Jansen}},\ and\ \bibinfo {author}
  {\bibfnamefont {J.~G.}\ \bibnamefont {{\'{A}}ngy{\'{a}}n}},\ }\href
  {https://doi.org/10.1063/1.3626551} {\bibfield  {journal} {\bibinfo
  {journal} {Journal of Chemical Physics}\ }\textbf {\bibinfo {volume} {135}},\
  \bibinfo {pages} {84119} (\bibinfo {year} {2011})}\BibitemShut {NoStop}%
\bibitem [{\citenamefont {{\'{A}}ngy{\'{a}}n}\ \emph
  {et~al.}(2011)\citenamefont {{\'{A}}ngy{\'{a}}n}, \citenamefont {Liu},
  \citenamefont {Toulouse},\ and\ \citenamefont {Jansen}}]{Angyan2011}%
  \BibitemOpen
  \bibfield  {author} {\bibinfo {author} {\bibfnamefont {J.~G.}\ \bibnamefont
  {{\'{A}}ngy{\'{a}}n}}, \bibinfo {author} {\bibfnamefont {R.~F.}\ \bibnamefont
  {Liu}}, \bibinfo {author} {\bibfnamefont {J.}~\bibnamefont {Toulouse}},\ and\
  \bibinfo {author} {\bibfnamefont {G.}~\bibnamefont {Jansen}},\ }\href
  {https://doi.org/10.1021/ct200501r} {\bibfield  {journal} {\bibinfo
  {journal} {Journal of Chemical Theory and Computation}\ }\textbf {\bibinfo
  {volume} {7}},\ \bibinfo {pages} {3116} (\bibinfo {year} {2011})},\ \Eprint
  {https://arxiv.org/abs/1404.1663} {arXiv:1404.1663} \BibitemShut {NoStop}%
\bibitem [{\citenamefont {Gonze}\ and\ \citenamefont
  {Scheffler}(1999)}]{Gonze1999}%
  \BibitemOpen
  \bibfield  {author} {\bibinfo {author} {\bibfnamefont {X.}~\bibnamefont
  {Gonze}}\ and\ \bibinfo {author} {\bibfnamefont {M.}~\bibnamefont
  {Scheffler}},\ }\href {https://doi.org/10.1103/PhysRevLett.82.4416}
  {\bibfield  {journal} {\bibinfo  {journal} {Physical Review Letters}\
  }\textbf {\bibinfo {volume} {82}},\ \bibinfo {pages} {4416} (\bibinfo {year}
  {1999})}\BibitemShut {NoStop}%
\bibitem [{\citenamefont {Hellgren}\ and\ \citenamefont {{Von
  Barth}}(2007)}]{Hellgren2007}%
  \BibitemOpen
  \bibfield  {author} {\bibinfo {author} {\bibfnamefont {M.}~\bibnamefont
  {Hellgren}}\ and\ \bibinfo {author} {\bibfnamefont {U.}~\bibnamefont {{Von
  Barth}}},\ }\href {https://doi.org/10.1103/PhysRevB.76.075107} {\bibfield
  {journal} {\bibinfo  {journal} {Physical Review B - Condensed Matter and
  Materials Physics}\ }\textbf {\bibinfo {volume} {76}},\ \bibinfo {pages}
  {075107} (\bibinfo {year} {2007})},\ \Eprint {https://arxiv.org/abs/0703819}
  {arXiv:0703819 [cond-mat]} \BibitemShut {NoStop}%
\bibitem [{\citenamefont {Entwistle}\ and\ \citenamefont
  {Godby}(2019)}]{Entwistle2019}%
  \BibitemOpen
  \bibfield  {author} {\bibinfo {author} {\bibfnamefont {M.~T.}\ \bibnamefont
  {Entwistle}}\ and\ \bibinfo {author} {\bibfnamefont {R.~W.}\ \bibnamefont
  {Godby}},\ }\href {https://doi.org/10.1103/PhysRevB.99.161102} {\bibfield
  {journal} {\bibinfo  {journal} {Physical Review B}\ }\textbf {\bibinfo
  {volume} {99}},\ \bibinfo {pages} {161102} (\bibinfo {year}
  {2019})}\BibitemShut {NoStop}%
\bibitem [{\citenamefont {Lein}\ \emph {et~al.}(2000)\citenamefont {Lein},
  \citenamefont {Gross},\ and\ \citenamefont {Perdew}}]{Lein2000}%
  \BibitemOpen
  \bibfield  {author} {\bibinfo {author} {\bibfnamefont {M.}~\bibnamefont
  {Lein}}, \bibinfo {author} {\bibfnamefont {E.}~\bibnamefont {Gross}},\ and\
  \bibinfo {author} {\bibfnamefont {J.~P.}\ \bibnamefont {Perdew}},\ }\href
  {https://doi.org/10.1103/PhysRevB.61.13431} {\bibfield  {journal} {\bibinfo
  {journal} {Physical Review B - Condensed Matter and Materials Physics}\
  }\textbf {\bibinfo {volume} {61}},\ \bibinfo {pages} {13431} (\bibinfo {year}
  {2000})}\BibitemShut {NoStop}%
\bibitem [{\citenamefont {Thiele}\ and\ \citenamefont
  {K{\"{u}}mmel}(2009)}]{Thiele2009}%
  \BibitemOpen
  \bibfield  {author} {\bibinfo {author} {\bibfnamefont {M.}~\bibnamefont
  {Thiele}}\ and\ \bibinfo {author} {\bibfnamefont {S.}~\bibnamefont
  {K{\"{u}}mmel}},\ }\href {https://doi.org/10.1103/PhysRevA.80.012514}
  {\bibfield  {journal} {\bibinfo  {journal} {Physical Review A}\ }\textbf
  {\bibinfo {volume} {80}},\ \bibinfo {pages} {012514} (\bibinfo {year}
  {2009})}\BibitemShut {NoStop}%
\bibitem [{\citenamefont {Gunnarsson}\ and\ \citenamefont
  {Lundqvist}(1976)}]{Gunnarsson1976}%
  \BibitemOpen
  \bibfield  {author} {\bibinfo {author} {\bibfnamefont {O.}~\bibnamefont
  {Gunnarsson}}\ and\ \bibinfo {author} {\bibfnamefont {B.~I.}\ \bibnamefont
  {Lundqvist}},\ }\href {https://doi.org/10.1103/PhysRevB.13.4274} {\bibfield
  {journal} {\bibinfo  {journal} {Physical Review B}\ }\textbf {\bibinfo
  {volume} {13}},\ \bibinfo {pages} {4274} (\bibinfo {year}
  {1976})}\BibitemShut {NoStop}%
\bibitem [{\citenamefont {Langreth}\ and\ \citenamefont
  {Perdew}(1977)}]{Langreth1977}%
  \BibitemOpen
  \bibfield  {author} {\bibinfo {author} {\bibfnamefont {D.~C.}\ \bibnamefont
  {Langreth}}\ and\ \bibinfo {author} {\bibfnamefont {J.~P.}\ \bibnamefont
  {Perdew}},\ }\href {https://doi.org/10.1103/PhysRevB.15.2884} {\bibfield
  {journal} {\bibinfo  {journal} {Physical Review B}\ }\textbf {\bibinfo
  {volume} {15}},\ \bibinfo {pages} {2884} (\bibinfo {year}
  {1977})}\BibitemShut {NoStop}%
\bibitem [{\citenamefont {Langreth}\ and\ \citenamefont
  {Perdew}(1975)}]{Langreth1975}%
  \BibitemOpen
  \bibfield  {author} {\bibinfo {author} {\bibfnamefont {D.~C.}\ \bibnamefont
  {Langreth}}\ and\ \bibinfo {author} {\bibfnamefont {J.~P.}\ \bibnamefont
  {Perdew}},\ }\href {https://doi.org/10.1016/0038-1098(75)90618-3} {\bibfield
  {journal} {\bibinfo  {journal} {Solid State Communications}\ }\textbf
  {\bibinfo {volume} {17}},\ \bibinfo {pages} {1425} (\bibinfo {year}
  {1975})}\BibitemShut {NoStop}%
\bibitem [{\citenamefont {Eberhard}\ and\ \citenamefont
  {Reiner}(2011)}]{Engel2011}%
  \BibitemOpen
  \bibfield  {author} {\bibinfo {author} {\bibfnamefont {E.}~\bibnamefont
  {Eberhard}}\ and\ \bibinfo {author} {\bibfnamefont {D.~M.}\ \bibnamefont
  {Reiner}},\ }\href@noop {} {\emph {\bibinfo {title} {{Density Functional
  Theory: An Advanced Course}}}}\ (\bibinfo  {publisher} {Springer-Verlag
  Berlin Heidelberg},\ \bibinfo {year} {2011})\BibitemShut {NoStop}%
\bibitem [{\citenamefont {Ren}\ \emph {et~al.}(2012)\citenamefont {Ren},
  \citenamefont {Rinke}, \citenamefont {Joas},\ and\ \citenamefont
  {Scheffler}}]{Ren2012}%
  \BibitemOpen
  \bibfield  {author} {\bibinfo {author} {\bibfnamefont {X.}~\bibnamefont
  {Ren}}, \bibinfo {author} {\bibfnamefont {P.}~\bibnamefont {Rinke}}, \bibinfo
  {author} {\bibfnamefont {C.}~\bibnamefont {Joas}},\ and\ \bibinfo {author}
  {\bibfnamefont {M.}~\bibnamefont {Scheffler}},\ }\href
  {https://doi.org/10.1007/s10853-012-6570-4} {\bibfield  {journal} {\bibinfo
  {journal} {Journal of Materials Science}\ }\textbf {\bibinfo {volume} {47}},\
  \bibinfo {pages} {7447} (\bibinfo {year} {2012})},\ \Eprint
  {https://arxiv.org/abs/1203.5536} {arXiv:1203.5536} \BibitemShut {NoStop}%
\bibitem [{\citenamefont {He{\ss}elmann}\ and\ \citenamefont
  {G{\"{o}}rling}(2011{\natexlab{b}})}]{2Hebelmann2011}%
  \BibitemOpen
  \bibfield  {author} {\bibinfo {author} {\bibfnamefont {A.}~\bibnamefont
  {He{\ss}elmann}}\ and\ \bibinfo {author} {\bibfnamefont {A.}~\bibnamefont
  {G{\"{o}}rling}},\ }\href {https://doi.org/10.1080/00268976.2011.614282}
  {\bibfield  {journal} {\bibinfo  {journal} {Molecular Physics}\ }\textbf
  {\bibinfo {volume} {109}},\ \bibinfo {pages} {2473} (\bibinfo {year}
  {2011}{\natexlab{b}})}\BibitemShut {NoStop}%
\bibitem [{\citenamefont {Harl}(2008)}]{Harl2008}%
  \BibitemOpen
  \bibfield  {author} {\bibinfo {author} {\bibfnamefont {J.}~\bibnamefont
  {Harl}},\ }\emph {\bibinfo {title} {The linear response function in density
  functional theory}},\ \href@noop {} {Ph.D. thesis},\ \bibinfo  {school}
  {University of Vienna} (\bibinfo {year} {2008})\BibitemShut {NoStop}%
\bibitem [{\citenamefont {Langreth}\ and\ \citenamefont
  {Perdew}(1980)}]{Langreth1980}%
  \BibitemOpen
  \bibfield  {author} {\bibinfo {author} {\bibfnamefont {D.~C.}\ \bibnamefont
  {Langreth}}\ and\ \bibinfo {author} {\bibfnamefont {J.~P.}\ \bibnamefont
  {Perdew}},\ }\href {https://doi.org/10.1103/PhysRevB.21.5469} {\bibfield
  {journal} {\bibinfo  {journal} {Physical Review B}\ }\textbf {\bibinfo
  {volume} {21}},\ \bibinfo {pages} {5469} (\bibinfo {year}
  {1980})}\BibitemShut {NoStop}%
\bibitem [{\citenamefont {Hohenberg}\ and\ \citenamefont
  {Kohn}(1964)}]{Hohenberg1964}%
  \BibitemOpen
  \bibfield  {author} {\bibinfo {author} {\bibfnamefont {P.}~\bibnamefont
  {Hohenberg}}\ and\ \bibinfo {author} {\bibfnamefont {W.}~\bibnamefont
  {Kohn}},\ }\href {https://doi.org/10.1103/PhysRev.136.B864} {\bibfield
  {journal} {\bibinfo  {journal} {Physical Review}\ }\textbf {\bibinfo {volume}
  {136}},\ \bibinfo {pages} {B864} (\bibinfo {year} {1964})}\BibitemShut
  {NoStop}%
\bibitem [{\citenamefont {Kohn}\ and\ \citenamefont {Sham}(1965)}]{Kohn1965}%
  \BibitemOpen
  \bibfield  {author} {\bibinfo {author} {\bibfnamefont {W.}~\bibnamefont
  {Kohn}}\ and\ \bibinfo {author} {\bibfnamefont {L.~J.}\ \bibnamefont
  {Sham}},\ }\href {https://doi.org/10.1103/PhysRev.140.A1133} {\bibfield
  {journal} {\bibinfo  {journal} {Physical Review}\ }\textbf {\bibinfo {volume}
  {140}},\ \bibinfo {pages} {A1133} (\bibinfo {year} {1965})}\BibitemShut
  {NoStop}%
\bibitem [{Note1()}]{Note1}%
  \BibitemOpen
  \bibinfo {note} {The $\lambda $-dependent xc energy (and xc kernel) is zero
  when $\lambda =0$ because the particles are non-interacting.}\BibitemShut
  {Stop}%
\bibitem [{\citenamefont {Martin}\ \emph {et~al.}(2016)\citenamefont {Martin},
  \citenamefont {Reining},\ and\ \citenamefont {Ceperley}}]{Martin2017}%
  \BibitemOpen
  \bibfield  {author} {\bibinfo {author} {\bibfnamefont {R.~M.}\ \bibnamefont
  {Martin}}, \bibinfo {author} {\bibfnamefont {L.}~\bibnamefont {Reining}},\
  and\ \bibinfo {author} {\bibfnamefont {D.~M.}\ \bibnamefont {Ceperley}},\
  }\href@noop {} {\emph {\bibinfo {title} {{Interacting Electrons Theory and
  Computational Approaches}}}}\ (\bibinfo  {publisher} {Cambridge University
  Press},\ \bibinfo {year} {2016})\BibitemShut {NoStop}%
\bibitem [{\citenamefont {Kubo}(1966)}]{Kubo1966}%
  \BibitemOpen
  \bibfield  {author} {\bibinfo {author} {\bibfnamefont {R.}~\bibnamefont
  {Kubo}},\ }\href {https://doi.org/10.1088/0034-4885/29/1/306} {\bibfield
  {journal} {\bibinfo  {journal} {Reports on Progress in Physics}\ }\textbf
  {\bibinfo {volume} {29}},\ \bibinfo {pages} {255} (\bibinfo {year}
  {1966})}\BibitemShut {NoStop}%
\bibitem [{\citenamefont {Dobson}\ \emph
  {et~al.}(1998{\natexlab{b}})\citenamefont {Dobson}, \citenamefont {Vignale},\
  and\ \citenamefont {Mukunda}}]{3Dobson1998}%
  \BibitemOpen
  \bibinfo {editor} {\bibfnamefont {J.}~\bibnamefont {Dobson}}, \bibinfo
  {editor} {\bibfnamefont {G.}~\bibnamefont {Vignale}},\ and\ \bibinfo {editor}
  {\bibfnamefont {D.}~\bibnamefont {Mukunda}},\ eds.,\ \href
  {https://link.springer.com/book/10.1007/978-1-4899-0316-7} {\emph {\bibinfo
  {title} {{Electronic Density Functional Theory: Recent Progress and New
  Directions}}}}\ (\bibinfo  {publisher} {Springer Science},\ \bibinfo {year}
  {1998})\ pp.\ \bibinfo {pages} {243--285}\BibitemShut {NoStop}%
\bibitem [{\citenamefont {Runge}\ and\ \citenamefont
  {Gross}(1984)}]{Runge1984}%
  \BibitemOpen
  \bibfield  {author} {\bibinfo {author} {\bibfnamefont {E.}~\bibnamefont
  {Runge}}\ and\ \bibinfo {author} {\bibfnamefont {E.~K.}\ \bibnamefont
  {Gross}},\ }\href {https://doi.org/10.1103/PhysRevLett.52.997} {\bibfield
  {journal} {\bibinfo  {journal} {Physical Review Letters}\ }\textbf {\bibinfo
  {volume} {52}},\ \bibinfo {pages} {997} (\bibinfo {year} {1984})}\BibitemShut
  {NoStop}%
\bibitem [{\citenamefont {van Leeuwen}(1999)}]{VanLeeuwen1999}%
  \BibitemOpen
  \bibfield  {author} {\bibinfo {author} {\bibfnamefont {R.}~\bibnamefont {van
  Leeuwen}},\ }\href {https://doi.org/10.1103/PhysRevLett.82.3863} {\bibfield
  {journal} {\bibinfo  {journal} {Physical Review Letters}\ }\textbf {\bibinfo
  {volume} {82}},\ \bibinfo {pages} {3863} (\bibinfo {year}
  {1999})}\BibitemShut {NoStop}%
\bibitem [{\citenamefont {Niquet}\ \emph {et~al.}(2003)\citenamefont {Niquet},
  \citenamefont {Fuchs},\ and\ \citenamefont {Gonze}}]{Niquet2003}%
  \BibitemOpen
  \bibfield  {author} {\bibinfo {author} {\bibfnamefont {Y.~M.}\ \bibnamefont
  {Niquet}}, \bibinfo {author} {\bibfnamefont {M.}~\bibnamefont {Fuchs}},\ and\
  \bibinfo {author} {\bibfnamefont {X.}~\bibnamefont {Gonze}},\ }\href
  {https://doi.org/10.1103/PhysRevA.68.032507} {\bibfield  {journal} {\bibinfo
  {journal} {Physical Review A - Atomic, Molecular, and Optical Physics}\
  }\textbf {\bibinfo {volume} {68}},\ \bibinfo {pages} {13} (\bibinfo {year}
  {2003})}\BibitemShut {NoStop}%
\bibitem [{\citenamefont {Hellgren}\ \emph {et~al.}(2012)\citenamefont
  {Hellgren}, \citenamefont {Rohr},\ and\ \citenamefont
  {Gross}}]{Hellgren2012}%
  \BibitemOpen
  \bibfield  {author} {\bibinfo {author} {\bibfnamefont {M.}~\bibnamefont
  {Hellgren}}, \bibinfo {author} {\bibfnamefont {D.~R.}\ \bibnamefont {Rohr}},\
  and\ \bibinfo {author} {\bibfnamefont {E.~K.}\ \bibnamefont {Gross}},\ }\href
  {https://doi.org/10.1063/1.3676174} {\bibfield  {journal} {\bibinfo
  {journal} {Journal of Chemical Physics}\ }\textbf {\bibinfo {volume} {136}},\
  \bibinfo {pages} {34106} (\bibinfo {year} {2012})},\ \Eprint
  {https://arxiv.org/abs/1110.6062} {arXiv:1110.6062} \BibitemShut {NoStop}%
\bibitem [{\citenamefont {Verma}\ and\ \citenamefont
  {Bartlett}(2012)}]{Verma2012}%
  \BibitemOpen
  \bibfield  {author} {\bibinfo {author} {\bibfnamefont {P.}~\bibnamefont
  {Verma}}\ and\ \bibinfo {author} {\bibfnamefont {R.~J.}\ \bibnamefont
  {Bartlett}},\ }\href {https://doi.org/10.1063/1.3678180} {\bibfield
  {journal} {\bibinfo  {journal} {Journal of Chemical Physics}\ }\textbf
  {\bibinfo {volume} {136}},\ \bibinfo {pages} {44105} (\bibinfo {year}
  {2012})}\BibitemShut {NoStop}%
\bibitem [{\citenamefont {Thierbach}\ and\ \citenamefont
  {G{\"{o}}rling}(2020)}]{Thierbach2020}%
  \BibitemOpen
  \bibfield  {author} {\bibinfo {author} {\bibfnamefont {A.}~\bibnamefont
  {Thierbach}}\ and\ \bibinfo {author} {\bibfnamefont {A.}~\bibnamefont
  {G{\"{o}}rling}},\ }\href {https://doi.org/10.1063/5.0021809} {\bibfield
  {journal} {\bibinfo  {journal} {Journal of Chemical Physics}\ }\textbf
  {\bibinfo {volume} {153}},\ \bibinfo {pages} {134113} (\bibinfo {year}
  {2020})}\BibitemShut {NoStop}%
\bibitem [{\citenamefont {G{\"{o}}rling}(2005)}]{Gorling2005}%
  \BibitemOpen
  \bibfield  {author} {\bibinfo {author} {\bibfnamefont {A.}~\bibnamefont
  {G{\"{o}}rling}},\ }\href {https://doi.org/10.1063/1.1904583} {\bibfield
  {journal} {\bibinfo  {journal} {Journal of Chemical Physics}\ }\textbf
  {\bibinfo {volume} {123}},\ \bibinfo {pages} {150901} (\bibinfo {year}
  {2005})}\BibitemShut {NoStop}%
\bibitem [{Note2()}]{Note2}%
  \BibitemOpen
  \bibinfo {note} {Electrons of like spin obey the Pauli exclusion principle,
  and exhibit features that would need a larger number of spin-half electrons
  to become apparent, e.g. two like-spin electrons experience the exchange
  effect, which is not the case for two spin-half electrons in an $S=0$
  state.}\BibitemShut {Stop}%
\bibitem [{\citenamefont {Ruggenthaler}\ \emph {et~al.}(2015)\citenamefont
  {Ruggenthaler}, \citenamefont {Penz},\ and\ \citenamefont {van
  Leeuwen}}]{Ruggenthaler2015}%
  \BibitemOpen
  \bibfield  {author} {\bibinfo {author} {\bibfnamefont {M.}~\bibnamefont
  {Ruggenthaler}}, \bibinfo {author} {\bibfnamefont {M.}~\bibnamefont {Penz}},\
  and\ \bibinfo {author} {\bibfnamefont {R.}~\bibnamefont {van Leeuwen}},\
  }\href {https://doi.org/10.1088/0953-8984/27/20/203202} {\bibfield  {journal}
  {\bibinfo  {journal} {Journal of Physics: Condensed Matter}\ }\textbf
  {\bibinfo {volume} {27}},\ \bibinfo {pages} {203202} (\bibinfo {year}
  {2015})}\BibitemShut {NoStop}%
\bibitem [{Note3()}]{Note3}%
  \BibitemOpen
  \bibinfo {note} {URL to be inserted}\BibitemShut {NoStop}%
\bibitem [{Note4()}]{Note4}%
  \BibitemOpen
  \bibinfo {note} {We note that the Gauss-Legendre scheme with $N$ grid points
  is able to exactly integrate a polynomial of degree less than or equal to
  $2N+1$. Therefore, an integral change of coordinates aims to transform the
  integrand into a low-degree polynomial.}\BibitemShut {Stop}%
\bibitem [{\citenamefont {{Dask Development Team}}(2016)}]{dask}%
  \BibitemOpen
  \bibfield  {author} {\bibinfo {author} {\bibnamefont {{Dask Development
  Team}}},\ }\href {https://dask.org} {\emph {\bibinfo {title} {Dask: Library
  for dynamic task scheduling}}} (\bibinfo {year} {2016})\BibitemShut {NoStop}%
\bibitem [{\citenamefont {Entwistle}\ \emph {et~al.}(2018)\citenamefont
  {Entwistle}, \citenamefont {Casula},\ and\ \citenamefont
  {Godby}}]{Entwistle2018}%
  \BibitemOpen
  \bibfield  {author} {\bibinfo {author} {\bibfnamefont {M.~T.}\ \bibnamefont
  {Entwistle}}, \bibinfo {author} {\bibfnamefont {M.}~\bibnamefont {Casula}},\
  and\ \bibinfo {author} {\bibfnamefont {R.~W.}\ \bibnamefont {Godby}},\ }\href
  {https://doi.org/10.1103/PhysRevB.97.235143} {\bibfield  {journal} {\bibinfo
  {journal} {Phys. Rev. B}\ }\textbf {\bibinfo {volume} {97}},\ \bibinfo
  {pages} {235143} (\bibinfo {year} {2018})}\BibitemShut {NoStop}%
\bibitem [{\citenamefont {Teale}\ \emph {et~al.}(2009)\citenamefont {Teale},
  \citenamefont {Coriani},\ and\ \citenamefont {Helgaker}}]{Teale2009}%
  \BibitemOpen
  \bibfield  {author} {\bibinfo {author} {\bibfnamefont {A.~M.}\ \bibnamefont
  {Teale}}, \bibinfo {author} {\bibfnamefont {S.}~\bibnamefont {Coriani}},\
  and\ \bibinfo {author} {\bibfnamefont {T.}~\bibnamefont {Helgaker}},\ }\href
  {https://doi.org/10.1063/1.3082285} {\bibfield  {journal} {\bibinfo
  {journal} {Journal of Chemical Physics}\ }\textbf {\bibinfo {volume} {130}},\
  \bibinfo {pages} {104111} (\bibinfo {year} {2009})}\BibitemShut {NoStop}%
\bibitem [{Note5()}]{Note5}%
  \BibitemOpen
  \bibinfo {note} {The probability that a particle resides at position $x$
  given that a particle has been measured at $x'$ is thus given by the
  conditional probability $n(x|x') = n(x) - n^\protect \text
  {hole}(x,x')$.}\BibitemShut {Stop}%
\bibitem [{\citenamefont {Dobson}\ \emph
  {et~al.}(1998{\natexlab{c}})\citenamefont {Dobson}, \citenamefont {Vignale},\
  and\ \citenamefont {Mukunda}}]{2Dobson1998}%
  \BibitemOpen
  \bibinfo {editor} {\bibfnamefont {J.}~\bibnamefont {Dobson}}, \bibinfo
  {editor} {\bibfnamefont {G.}~\bibnamefont {Vignale}},\ and\ \bibinfo {editor}
  {\bibfnamefont {D.}~\bibnamefont {Mukunda}},\ eds.,\ \href
  {https://link.springer.com/book/10.1007/978-1-4899-0316-7} {\emph {\bibinfo
  {title} {{Electronic Density Functional Theory: Recent Progress and New
  Directions}}}}\ (\bibinfo  {publisher} {Springer Science},\ \bibinfo {year}
  {1998})\ pp.\ \bibinfo {pages} {3--27}\BibitemShut {NoStop}%
\bibitem [{Note6()}]{Note6}%
  \BibitemOpen
  \bibinfo {note} {This means that the exchange hole will be
  exact.}\BibitemShut {Stop}%
\bibitem [{\citenamefont {Hodgson}\ \emph {et~al.}(2016)\citenamefont
  {Hodgson}, \citenamefont {Ramsden},\ and\ \citenamefont
  {Godby}}]{Hodgson2016}%
  \BibitemOpen
  \bibfield  {author} {\bibinfo {author} {\bibfnamefont {M.~J.~P.}\
  \bibnamefont {Hodgson}}, \bibinfo {author} {\bibfnamefont {J.~D.}\
  \bibnamefont {Ramsden}},\ and\ \bibinfo {author} {\bibfnamefont {R.~W.}\
  \bibnamefont {Godby}},\ }\href {https://doi.org/10.1103/PhysRevB.93.155146}
  {\bibfield  {journal} {\bibinfo  {journal} {Physical Review B}\ }\textbf
  {\bibinfo {volume} {93}},\ \bibinfo {pages} {155146} (\bibinfo {year}
  {2016})}\BibitemShut {NoStop}%
\bibitem [{\citenamefont {Hodgson}(2021)}]{Hodgson2021}%
  \BibitemOpen
  \bibfield  {author} {\bibinfo {author} {\bibfnamefont {M.~J.~P.}\
  \bibnamefont {Hodgson}},\ }\href {https://arxiv.org/abs/2104.04588v1}
  {\bibfield  {journal} {\bibinfo  {journal} {arXiv preprint arXiv:2104.04588}\
  } (\bibinfo {year} {2021})}\BibitemShut {NoStop}%
\bibitem [{\citenamefont {Almbladh}\ and\ \citenamefont {von
  Barth}(1985)}]{Almbladh1985}%
  \BibitemOpen
  \bibfield  {author} {\bibinfo {author} {\bibfnamefont {C.~O.}\ \bibnamefont
  {Almbladh}}\ and\ \bibinfo {author} {\bibfnamefont {U.}~\bibnamefont {von
  Barth}},\ }in\ \href@noop {} {\emph {\bibinfo {booktitle} {Density Functional
  Methods in Physics}}},\ \bibinfo {editor} {edited by\ \bibinfo {editor}
  {\bibfnamefont {R.~M.}\ \bibnamefont {Dreizler}}\ and\ \bibinfo {editor}
  {\bibfnamefont {J.}~\bibnamefont {da~Providencia}}}\ (\bibinfo  {publisher}
  {Plenum Press},\ \bibinfo {address} {New York and London},\ \bibinfo {year}
  {1985})\ p.\ \bibinfo {pages} {209}\BibitemShut {NoStop}%
\bibitem [{\citenamefont {Hellgren}\ and\ \citenamefont
  {Gross}(2012)}]{2Hellgren2012}%
  \BibitemOpen
  \bibfield  {author} {\bibinfo {author} {\bibfnamefont {M.}~\bibnamefont
  {Hellgren}}\ and\ \bibinfo {author} {\bibfnamefont {E.~K.~U.}\ \bibnamefont
  {Gross}},\ }\href {https://doi.org/10.1103/PhysRevA.85.022514} {\bibfield
  {journal} {\bibinfo  {journal} {Physical Review A}\ }\textbf {\bibinfo
  {volume} {85}},\ \bibinfo {pages} {022514} (\bibinfo {year}
  {2012})}\BibitemShut {NoStop}%
\bibitem [{\citenamefont {Hellgren}\ and\ \citenamefont
  {Gross}(2013)}]{Hellgren2013}%
  \BibitemOpen
  \bibfield  {author} {\bibinfo {author} {\bibfnamefont {M.}~\bibnamefont
  {Hellgren}}\ and\ \bibinfo {author} {\bibfnamefont {E.~K.~U.}\ \bibnamefont
  {Gross}},\ }\href {https://doi.org/10.1103/PhysRevA.88.052507} {\bibfield
  {journal} {\bibinfo  {journal} {Physical Review A}\ }\textbf {\bibinfo
  {volume} {88}},\ \bibinfo {pages} {052507} (\bibinfo {year}
  {2013})}\BibitemShut {NoStop}%
\bibitem [{\citenamefont {Hellgren}(2018)}]{Hellgren2018}%
  \BibitemOpen
  \bibfield  {author} {\bibinfo {author} {\bibfnamefont {M.}~\bibnamefont
  {Hellgren}},\ }\href {https://doi.org/10.1140/EPJB/E2018-90110-1} {\bibfield
  {journal} {\bibinfo  {journal} {The European Physical Journal B 2018 91:7}\
  }\textbf {\bibinfo {volume} {91}},\ \bibinfo {pages} {1} (\bibinfo {year}
  {2018})}\BibitemShut {NoStop}%
\bibitem [{\citenamefont {Maitra}\ \emph {et~al.}(2004)\citenamefont {Maitra},
  \citenamefont {Zhang}, \citenamefont {Cave},\ and\ \citenamefont
  {Burke}}]{Maitra2004}%
  \BibitemOpen
  \bibfield  {author} {\bibinfo {author} {\bibfnamefont {N.~T.}\ \bibnamefont
  {Maitra}}, \bibinfo {author} {\bibfnamefont {F.}~\bibnamefont {Zhang}},
  \bibinfo {author} {\bibfnamefont {R.~J.}\ \bibnamefont {Cave}},\ and\
  \bibinfo {author} {\bibfnamefont {K.}~\bibnamefont {Burke}},\ }\href
  {https://doi.org/10.1063/1.1651060} {\bibfield  {journal} {\bibinfo
  {journal} {The Journal of Chemical Physics}\ }\textbf {\bibinfo {volume}
  {120}},\ \bibinfo {pages} {5932} (\bibinfo {year} {2004})}\BibitemShut
  {NoStop}%
\bibitem [{\citenamefont {Maitra}(2021)}]{Maitra2021}%
  \BibitemOpen
  \bibfield  {author} {\bibinfo {author} {\bibfnamefont {N.~T.}\ \bibnamefont
  {Maitra}},\ }\href {https://arxiv.org/abs/2107.05600v1} {\bibfield  {journal}
  {\bibinfo  {journal} {(Preprint)}\ } (\bibinfo {year} {2021})},\ \Eprint
  {https://arxiv.org/abs/2107.05600} {arXiv:2107.05600} \BibitemShut {NoStop}%
\bibitem [{\citenamefont {Ruzsinszky}\ \emph {et~al.}(2020)\citenamefont
  {Ruzsinszky}, \citenamefont {Nepal}, \citenamefont {Pitarke},\ and\
  \citenamefont {Perdew}}]{Ruzsinszky2020}%
  \BibitemOpen
  \bibfield  {author} {\bibinfo {author} {\bibfnamefont {A.}~\bibnamefont
  {Ruzsinszky}}, \bibinfo {author} {\bibfnamefont {N.~K.}\ \bibnamefont
  {Nepal}}, \bibinfo {author} {\bibfnamefont {J.~M.}\ \bibnamefont {Pitarke}},\
  and\ \bibinfo {author} {\bibfnamefont {J.~P.}\ \bibnamefont {Perdew}},\
  }\href {https://doi.org/10.1103/PhysRevB.101.245135} {\bibfield  {journal}
  {\bibinfo  {journal} {Physical Review B}\ }\textbf {\bibinfo {volume}
  {101}},\ \bibinfo {pages} {245135} (\bibinfo {year} {2020})}\BibitemShut
  {NoStop}%
\bibitem [{Note7()}]{Note7}%
  \BibitemOpen
  \bibinfo {note} {\protect \url
  {https://doi.org/10.17863/CAM.75013}}\BibitemShut {NoStop}%
\end{thebibliography}%


\begin{thebibliography}{11}%
\makeatletter
\providecommand \@ifxundefined [1]{%
 \@ifx{#1\undefined}
}%
\providecommand \@ifnum [1]{%
 \ifnum #1\expandafter \@firstoftwo
 \else \expandafter \@secondoftwo
 \fi
}%
\providecommand \@ifx [1]{%
 \ifx #1\expandafter \@firstoftwo
 \else \expandafter \@secondoftwo
 \fi
}%
\providecommand \natexlab [1]{#1}%
\providecommand \enquote  [1]{``#1''}%
\providecommand \bibnamefont  [1]{#1}%
\providecommand \bibfnamefont [1]{#1}%
\providecommand \citenamefont [1]{#1}%
\providecommand \href@noop [0]{\@secondoftwo}%
\providecommand \href [0]{\begingroup \@sanitize@url \@href}%
\providecommand \@href[1]{\@@startlink{#1}\@@href}%
\providecommand \@@href[1]{\endgroup#1\@@endlink}%
\providecommand \@sanitize@url [0]{\catcode `\\12\catcode `\$12\catcode
  `\&12\catcode `\#12\catcode `\^12\catcode `\_12\catcode `\%12\relax}%
\providecommand \@@startlink[1]{}%
\providecommand \@@endlink[0]{}%
\providecommand \url  [0]{\begingroup\@sanitize@url \@url }%
\providecommand \@url [1]{\endgroup\@href {#1}{\urlprefix }}%
\providecommand \urlprefix  [0]{URL }%
\providecommand \Eprint [0]{\href }%
\providecommand \doibase [0]{https://doi.org/}%
\providecommand \selectlanguage [0]{\@gobble}%
\providecommand \bibinfo  [0]{\@secondoftwo}%
\providecommand \bibfield  [0]{\@secondoftwo}%
\providecommand \translation [1]{[#1]}%
\providecommand \BibitemOpen [0]{}%
\providecommand \bibitemStop [0]{}%
\providecommand \bibitemNoStop [0]{.\EOS\space}%
\providecommand \EOS [0]{\spacefactor3000\relax}%
\providecommand \BibitemShut  [1]{\csname bibitem#1\endcsname}%
\let\auto@bib@innerbib\@empty
\bibitem [{\citenamefont {Entwistle}\ \emph {et~al.}(2018)\citenamefont
  {Entwistle}, \citenamefont {Casula},\ and\ \citenamefont
  {Godby}}]{Entwistle2018}%
  \BibitemOpen
  \bibfield  {author} {\bibinfo {author} {\bibfnamefont {M.~T.}\ \bibnamefont
  {Entwistle}}, \bibinfo {author} {\bibfnamefont {M.}~\bibnamefont {Casula}},\
  and\ \bibinfo {author} {\bibfnamefont {R.~W.}\ \bibnamefont {Godby}},\ }\href
  {https://doi.org/10.1103/PhysRevB.97.235143} {\bibfield  {journal} {\bibinfo
  {journal} {Phys. Rev. B}\ }\textbf {\bibinfo {volume} {97}},\ \bibinfo
  {pages} {235143} (\bibinfo {year} {2018})}\BibitemShut {NoStop}%
\bibitem [{Note1()}]{Note1}%
  \BibitemOpen
  \bibinfo {note} {See Table II in Ref.~\protect \citep {Entwistle2018} for the
  optimal fit parameters}\BibitemShut {NoStop}%
\bibitem [{\citenamefont {Burke}(2007)}]{BurkeABC}%
  \BibitemOpen
  \bibinfo {editor} {\bibfnamefont {K.}~\bibnamefont {Burke}},\ ed.,\ \href
  {https://doi.org/https://dft.uci.edu/doc/g1.pdf} {\emph {\bibinfo {title}
  {{ABC of DFT}}}}\ (\bibinfo  {publisher} {Unpublished},\ \bibinfo {year}
  {2007})\ pp.\ \bibinfo {pages} {93--111}\BibitemShut {NoStop}%
\bibitem [{Note2()}]{Note2}%
  \BibitemOpen
  \bibinfo {note} {See Table III in Ref.~\protect \citep {Entwistle2018} for
  the optimal fit parameters}\BibitemShut {NoStop}%
\bibitem [{\citenamefont {Hodgson}\ \emph {et~al.}(2016)\citenamefont
  {Hodgson}, \citenamefont {Ramsden},\ and\ \citenamefont
  {Godby}}]{Hodgson2016}%
  \BibitemOpen
  \bibfield  {author} {\bibinfo {author} {\bibfnamefont {M.~J.~P.}\
  \bibnamefont {Hodgson}}, \bibinfo {author} {\bibfnamefont {J.~D.}\
  \bibnamefont {Ramsden}},\ and\ \bibinfo {author} {\bibfnamefont {R.~W.}\
  \bibnamefont {Godby}},\ }\href {https://doi.org/10.1103/PhysRevB.93.155146}
  {\bibfield  {journal} {\bibinfo  {journal} {Physical Review B}\ }\textbf
  {\bibinfo {volume} {93}},\ \bibinfo {pages} {155146} (\bibinfo {year}
  {2016})}\BibitemShut {NoStop}%
\bibitem [{\citenamefont {Almbladh}\ and\ \citenamefont {von
  Barth}(1985)}]{Almbladh1985}%
  \BibitemOpen
  \bibfield  {author} {\bibinfo {author} {\bibfnamefont {C.~O.}\ \bibnamefont
  {Almbladh}}\ and\ \bibinfo {author} {\bibfnamefont {U.}~\bibnamefont {von
  Barth}},\ }in\ \href@noop {} {\emph {\bibinfo {booktitle} {Density Functional
  Methods in Physics}}},\ \bibinfo {editor} {edited by\ \bibinfo {editor}
  {\bibfnamefont {R.~M.}\ \bibnamefont {Dreizler}}\ and\ \bibinfo {editor}
  {\bibfnamefont {J.}~\bibnamefont {da~Providencia}}}\ (\bibinfo  {publisher}
  {Plenum Press},\ \bibinfo {address} {New York and London},\ \bibinfo {year}
  {1985})\ p.\ \bibinfo {pages} {209}\BibitemShut {NoStop}%
\bibitem [{\citenamefont {Hellgren}\ and\ \citenamefont
  {Gross}(2012)}]{2Hellgren2012}%
  \BibitemOpen
  \bibfield  {author} {\bibinfo {author} {\bibfnamefont {M.}~\bibnamefont
  {Hellgren}}\ and\ \bibinfo {author} {\bibfnamefont {E.~K.~U.}\ \bibnamefont
  {Gross}},\ }\href {https://doi.org/10.1103/PhysRevA.85.022514} {\bibfield
  {journal} {\bibinfo  {journal} {Physical Review A}\ }\textbf {\bibinfo
  {volume} {85}},\ \bibinfo {pages} {022514} (\bibinfo {year}
  {2012})}\BibitemShut {NoStop}%
\bibitem [{\citenamefont {Hellgren}\ and\ \citenamefont
  {Gross}(2013)}]{Hellgren2013}%
  \BibitemOpen
  \bibfield  {author} {\bibinfo {author} {\bibfnamefont {M.}~\bibnamefont
  {Hellgren}}\ and\ \bibinfo {author} {\bibfnamefont {E.~K.~U.}\ \bibnamefont
  {Gross}},\ }\href {https://doi.org/10.1103/PhysRevA.88.052507} {\bibfield
  {journal} {\bibinfo  {journal} {Physical Review A}\ }\textbf {\bibinfo
  {volume} {88}},\ \bibinfo {pages} {052507} (\bibinfo {year}
  {2013})}\BibitemShut {NoStop}%
\bibitem [{\citenamefont {Hellgren}(2018)}]{Hellgren2018}%
  \BibitemOpen
  \bibfield  {author} {\bibinfo {author} {\bibfnamefont {M.}~\bibnamefont
  {Hellgren}},\ }\href {https://doi.org/10.1140/EPJB/E2018-90110-1} {\bibfield
  {journal} {\bibinfo  {journal} {The European Physical Journal B 2018 91:7}\
  }\textbf {\bibinfo {volume} {91}},\ \bibinfo {pages} {1} (\bibinfo {year}
  {2018})}\BibitemShut {NoStop}%
\bibitem [{\citenamefont {Maitra}(2021)}]{Maitra2021}%
  \BibitemOpen
  \bibfield  {author} {\bibinfo {author} {\bibfnamefont {N.~T.}\ \bibnamefont
  {Maitra}},\ }\href@noop {} {\bibinfo {title} {Double and charge-transfer
  excitations in time-dependent density functional theory}} (\bibinfo {year}
  {2021}),\ \Eprint {https://arxiv.org/abs/2107.05600} {arXiv:2107.05600
  [physics.chem-ph]} \BibitemShut {NoStop}%
\bibitem [{\citenamefont {Maitra}(2017)}]{Maitra2017}%
  \BibitemOpen
  \bibfield  {author} {\bibinfo {author} {\bibfnamefont {N.~T.}\ \bibnamefont
  {Maitra}},\ }\href {https://doi.org/10.1088/1361-648X/aa836e} {\bibfield
  {journal} {\bibinfo  {journal} {Journal of Physics Condensed Matter}\
  }\textbf {\bibinfo {volume} {29}},\ \bibinfo {pages} {423001} (\bibinfo
  {year} {2017})}\BibitemShut {NoStop}%
\end{thebibliography}%

\end{document}


\title{Accurate Total Energies from the Adiabatic-Connection Fluctuation-Dissipation Theorem (Supplemental Material)}
\author{N. D. Woods}
\email{nw361@cam.ac.uk}
\affiliation{Theory of Condensed Matter, Cavendish Laboratory, University of Cambridge, Cambridge, CB3 0HE, United Kingdom}
\author{M. T. Entwistle}
\affiliation{FU Berlin, Department of Mathematics and Computer Science, Arnimallee 12, 14195 Berlin, Germany}
\author{R. W. Godby}
\affiliation{Department of Physics, University of York, and European Theoretical Spectroscopy Facility, Heslington, York YO10 5DD, United Kingdom}

\date{\today}

\maketitle

\section{The Frequency Integration Scheme}
\label{sec:FreqIntScheme}

The $\omega$-dependent ACFDT correlation energy integrand can be written in the form
\begin{align}
E_\text{c}^\text{ACFD} = \int_0^\infty g(\omega) \ d\omega, \label{eq:FreqDepIntegrand}
\end{align}
where the full expression can be found in the second section of the main text. We now seek a change of coordinates $\omega \rightarrow \tilde{\omega}$ such that the transformed integral,
\begin{align}
\int_0^\infty g(\omega) \ d\omega = \int_{\tilde{\omega}(0)}^{\tilde{\omega}(\infty)} g(\omega(\tilde{\omega})) \frac{d\omega}{d \tilde{\omega}} \ d \tilde{\omega},
\end{align}
is more suited to the traditional Gauss-Legendre integration scheme. Inspection of the ACFDT correlation energy functional defined in the main text demonstrates that the $\omega$-dependence arises through $\chi^\lambda[n](i \omega)$ (and $\chi_0[n](i\omega)$). Therefore, 
it is sufficient to seek an accurate integration scheme for $\int_0^\infty \chi^\lambda(i\omega) \ d \omega$, which will ultimately yield an accurate integration scheme for the original integral Eq$.$ (\ref{eq:FreqDepIntegrand}).

Suppose we consider just the $\lambda=1$ response function, then we find
\begin{align}
\int_0^\infty \chi(x,x',i\omega) \ d\omega = -2\sum_{n} f_n(x) f_n(x') \int_0^\infty \frac{\Omega_n^2}{\omega^2 + \Omega_n^2} \ d\omega, \label{eq:ExactChiIntegration}
\end{align}
where $f_n(x) = \langle \Psi_0 | \hat{n}(x) | \Psi_n \rangle$ is the so-called excitation function and $\Omega_n$ is the $n-$th excitation energy. This integral has an analytic solution: $\pi / 2$ regardless of $\Omega_n$, but to utilize this fact requires us to be in possession of the exact excitation functions, and therefore the exact many-body wavefunctions. In practice, only an approximate form of the curve $g(\omega)$ is attained, the integral of which has no analytic solution, meaning we must construct some accurate and efficient approximate integration scheme.

Gauss-Legendre quadrature is able to exactly integrate a polynomial of degree less than or equal to $2N+1$ using just $N$ samples of the function to be integrated. However, the integrand $g(\omega)$ is far from polynomial. With this in mind, we propose a scheme that performs a change of coordinates such that one term in the Lehmann representation of $\chi(i\omega)$ is linear, i.e. integrated exactly with one grid point. In other words, we require
\begin{align}
\frac{a}{a^2 + \omega^2} \frac{d\omega}{d \tilde{\omega}} = \tilde{\omega},   
\end{align}
where $a$ is a parameter related to the excitation energies. Solution of this differential equation yields
\begin{align}
\omega = a \tan \left( \frac{a\tilde{\omega}^2}{2} \right),
\end{align}
meaning the integration limits are compressed $[0,\infty] \rightarrow [0,\sqrt{\pi / a}]$. Given an even sampling for $\tilde{\omega} \in [0,\sqrt{\pi / a}]$, this scheme heavily biases the low-$\omega$ region of $g(\omega)$, the importance of which is discussed in the final section of the main text. Applying this coordinate transform to the original integral Eq$.$ (\ref{eq:FreqDepIntegrand}) gives
\begin{align}
\int_0^\infty g(\omega) \ d\omega = \int_0^{\sqrt{\pi / a}} \frac{a^3 \tilde{\omega}}{(\cos (a \tilde{\omega}^2 / 2))^2} g(\omega(\tilde{\omega})) \ d \tilde{\omega}.
\end{align}
We find that the parameter $a$ should be chosen as the average of the excitation energies `close to' the first excitation energy (i.e. the gap), and, in case an estimate for this is not available, it is better to err on the side of larger rather than smaller $a$. As stated in the main text, this scheme outperforms certain other methods from literature for our systems, and it would be an interesting exercise to examine how this approach performs in practice. 

\section{LDA Parameterization}
In a previous paper \citep{Entwistle2018} we constructed a HEG-based LDA appropriate for our one-dimensional systems of like-spin electrons, within the framework of ground-state DFT. We now extend this HEG LDA by parameterizing it with respect to $\lambda$.

\subsection{Exchange functional}

In Ref.~\citep{Entwistle2018} we parameterized the exchange energy per electron using a seven-parameter fit \footnote{See Table II in Ref.~\citep{Entwistle2018} for the optimal fit parameters}:
\begin{equation} \label{eq:ex_lambda}
    \varepsilon_{\mathrm{x}}(n) = (A + B n + C n^{2} + D n^{3} + E n^{4} + F n^{5}) n^{G}.
\end{equation}
In ACFDT the $\lambda$ dependence of $\varepsilon_{\mathrm{x}}$ is trivial (linear) \citep{BurkeABC} and so we easily obtain a $\lambda$-dependent parameterization from Eq.~(\ref{eq:ex_lambda}):
\begin{equation}
    \varepsilon_{\mathrm{x}}^{\lambda}(n) = \lambda \ \varepsilon_{\mathrm{x}}(n),
\end{equation}
with the exchange part of the xc kernel following by taking the second derivative with respect to the density.

\subsection{Correlation functional}

In Ref.~\citep{Entwistle2018} we found the HEG LDA to be remarkably similar in many regards to a set of LDAs constructed from finite locally homogeneous systems (which we refer to as `slabs'). One such similarity is the nature of electron correlation across systems of different densities (see Fig. 4 in Ref.~\citep{Entwistle2018}). We now utilize this: by analyzing how electron correlation varies with $\lambda \in [0,1]$ in these finite systems, for which we can obtain results close to machine precision, we are able to parameterize $\varepsilon_{\mathrm{c}}^{\lambda}(n)$ in the HEG LDA.

To do so we define our $\lambda$-dependent many-body Hamiltonian as
\begin{align}
H(\lambda) = \hat{T} + \lambda \hat{v}_\text{ee} + \hat{v}_\text{ext} + \hat{v}_\text{dxm}(\lambda),
\end{align}
where $\hat{v}_\text{dxm}(\lambda)$ is the unique potential that ensures the ground-state density at all values of $\lambda \in [0,1]$ is equal to the ground-state electron density at $\lambda = 1$ (as discussed in the main text of the current paper).

For each value of $\lambda$, given by the Gauss-Legendre method, we generate a set of two-electron slab systems over a typical density range (in the same spirit as we did in Ref.~\citep{Entwistle2018}) and calculate the correlation energy per electron $\varepsilon_{\mathrm{c}}$ for each. We thus obtain a series of data points at each $\lambda, n$, and find the following parameterization to work well:
\begin{equation} \label{eq:ec_lambda}
\varepsilon_{\mathrm{c}}^{\lambda}(n) = \lambda^{h(n)} \ \varepsilon_{\mathrm{c}}^{\lambda=1}(n),
\end{equation}
where $h(n) = -\exp(-\{\theta n-\ln(2)\})+2$, with $\theta = 17.39 \pm 0.53$. 

The high-density limit (infinitely-weak correlation) of the parameterization is
\begin{equation}
\varepsilon_{\mathrm{c}}^{\lambda}(n \rightarrow \infty) = \lambda^{2} \ \varepsilon_{\mathrm{c}}^{\lambda=1}(n \rightarrow \infty),
\end{equation}
and its low-density limit (infinitely-strong correlation) is
\begin{equation}
\varepsilon_{\mathrm{c}}^{\lambda}(n \rightarrow 0) = \varepsilon_{\mathrm{c}}^{\lambda=1}(n \rightarrow 0),
\end{equation}
as expected.

In Ref.~\citep{Entwistle2018} we parameterized the correlation energy per electron for the HEG using a fit of the form  \footnote{See Table III in Ref.~\citep{Entwistle2018} for the optimal fit parameters}
\begin{equation}
\varepsilon_{\mathrm{c}}(r_{\mathrm{s}}) = -\frac{A_{\mathrm{RPA}} r_{\mathrm{s}} + E r_{\mathrm{s}}^{2}}{1 + B r_{\mathrm{s}} + C r_{\mathrm{s}}^{2} + D r_{\mathrm{s}}^{3}} \frac{\ln(1 + \alpha r_{\mathrm{s}} + \beta r_{\mathrm{s}}^{2})}{\alpha}, 
\end{equation}
where $r_{\mathrm{s}}$ is the Wigner-Seitz radius and is related to the density (in 1D) by $2 r_{\mathrm{s}} = 1/n$. We parameterize this with respect to lambda using Eq.~(\ref{eq:ec_lambda}), with the correlation part of the xc kernel following by taking the second derivative with respect to the density. 

\section{Prototype Systems}

Across all systems $N_x = 121$, $N_\omega = 30$, and $N_\lambda = 10$ grid points are used. The frequency integration parameter is set to $a = 1$. 

\subsection{Atom}

\begin{figure}[h!]
\begin{center}
\includegraphics[width=3.5in]{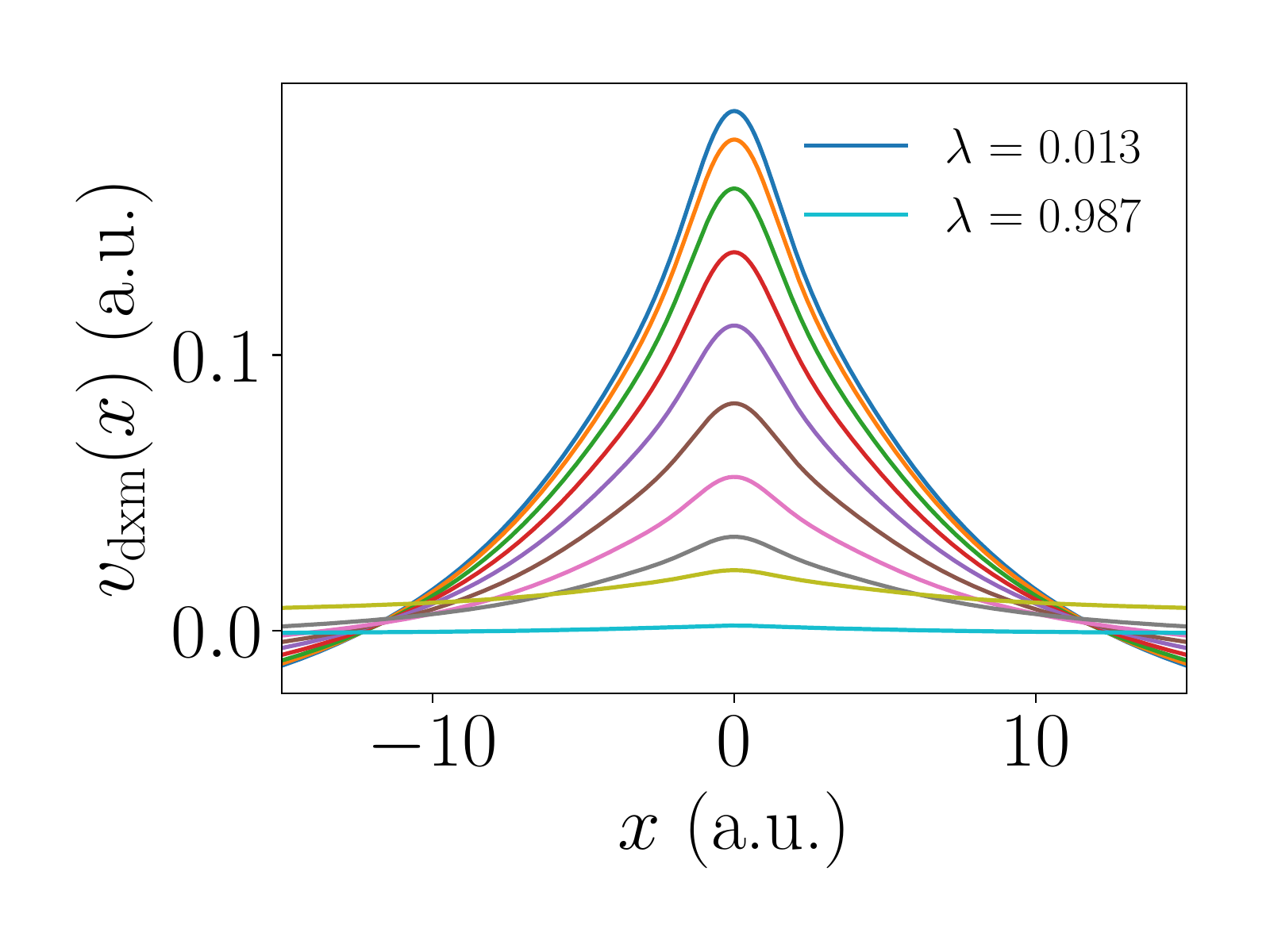}
\end{center}
\caption{The potential $v_\text{dxm}(x,\lambda)$ that ensures the $\lambda$-interacting system shares the same atomic density $n(x)$ as the $\lambda=1$ interacting system for $\lambda \in [0,1]$. Note that $v_\text{dxm}(x,\lambda = 1) = v_\text{xc}(x) + v_\text{H}(x)$, and $v_\text{dxm}(x,\lambda = 0) = 0$.}
\label{fig:AtomVdxm}
\end{figure}

The atom is defined with $v_\text{ext} = -1 / (|0.05 x| + 1)$ inside the domain $[-15,15]$ a.u., see main text. The potential $v_\text{dxm}(x,\lambda)$ at all points $\lambda$ along the adiabatic connection is given in Fig$.$ \ref{fig:AtomVdxm}. 

The data used to generate the main-text figure containing the absolute errors in the total energy across a range of approximate ACFDT calculations is given in Table \ref{Tbl:AtomEcRelError}. It can be observed that the exact total energy calculated via conventional means and the exact ACFDT total energy calculated using the algorithm presented in the main text agree to within $\mathcal{O}(10^{-10})$, thus verifying the numerical methods. 

\begin{table*}[t]\centering
\caption{Absolute error in the atomic total energy (a.u.) across a variety of approximate ACFDT total energies. It is clear that the exact $f_\text{xc}[n]$ in conjunction with the exact $v_\text{xc}[n]$ produces a total energy equal to within $\mathcal{O}(10^{-10})$ to the one computed in a conventional Kohn-Sham calculation (labeled `No ACFDT') when the exact $v_\text{xc}[n]$ is used. \\}
\label{Tbl:AtomEcRelError}
\noindent
\renewcommand{\arraystretch}{1.4}
\begin{tabular}{|l||*{5}{c|}}\hline
\backslashbox{$v_\text{xc}[n]$}{$f_\text{xc}[n]$}
&\makebox[6em]{RPA}&\makebox[6em]{ALDA}&\makebox[13em]{AE $f_\text{xc}[n](\omega=0)$}
&\makebox[7em]{Exact}&\makebox[7em]{No ACFDT}\\\hline\hline
Hartree & 0.049 & 0.037 & 0.0225 & 0.0226 & 0.34 \\\hline
Hartree-Fock & 0.024 & 0.017 & 0.0024 & 0.0021 & 0.0034 \\\hline
LDA & 0.052 & 0.043 & 0.00064 & 0.00061 & 0.0034 \\\hline
Exact KS & 0.047 & 0.039 & 0.00018 & 0.0 & 8.1 $\times 10^{-10}$ \\\hline
Self-consistent & 0.056 & 0.046 & 0.00018 & 2.22 $\times 10^{-16}$ & -- \\\hline
\end{tabular}
\renewcommand{\arraystretch}{1}
\end{table*}

\subsection{Infinite Potential Well}

\begin{figure}[h!]
\begin{center}
\includegraphics[width=3in]{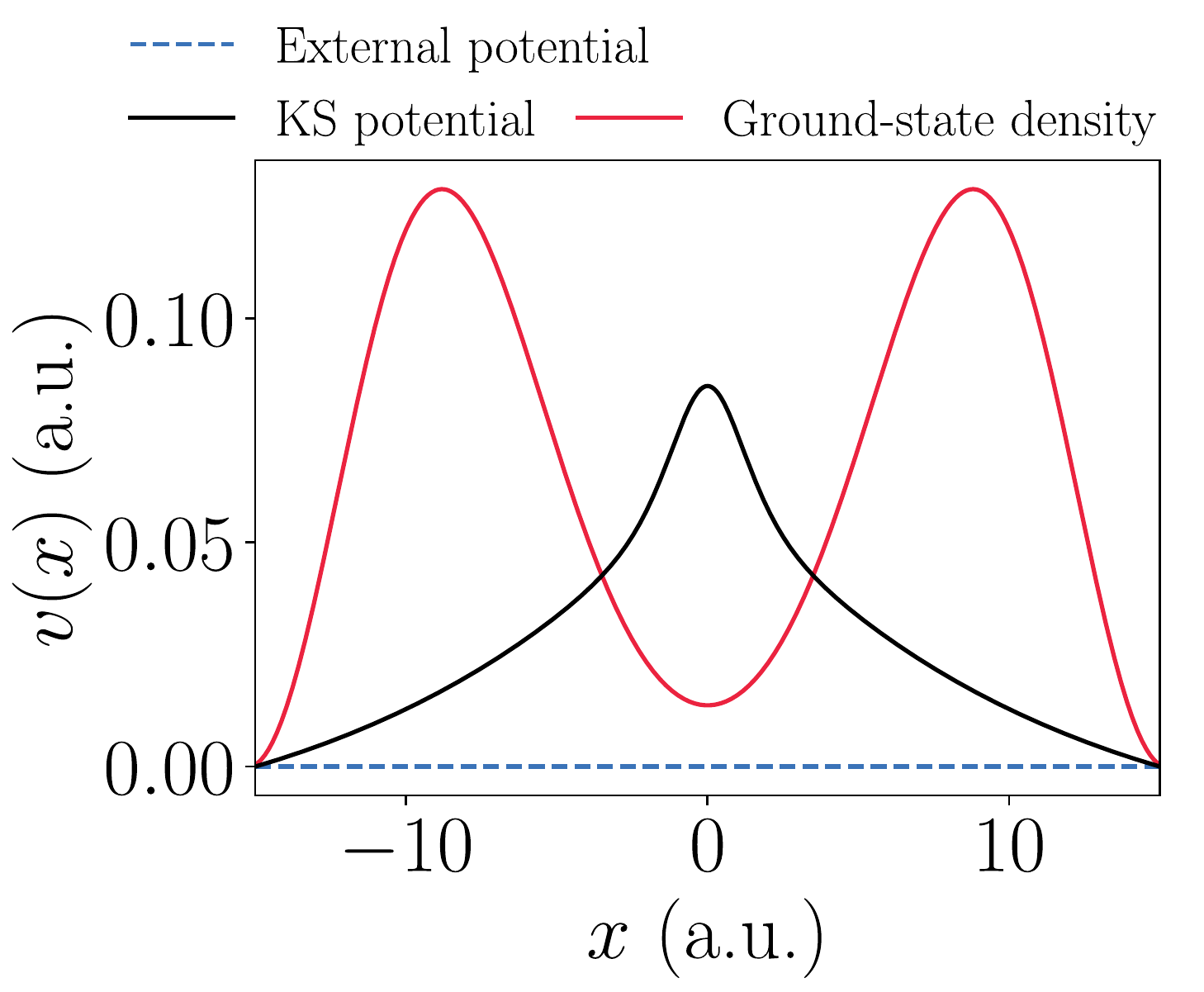}
\end{center}
\caption{The ground-state density, external potential, and reverse-engineered Kohn-Sham potential for the infinite potential well system. The external and Kohn-Sham potentials have been shifted for illustrative purposes.}
\label{fig:InfPotWell}
\end{figure}

The infinite potential well is defined with $v_\text{ext} = 0$ inside the domain $[-15,15]$ a.u., see Fig$.$ \ref{fig:InfPotWell}.

\begin{figure}[ht]
\begin{center}
\includegraphics[width=3.4in]{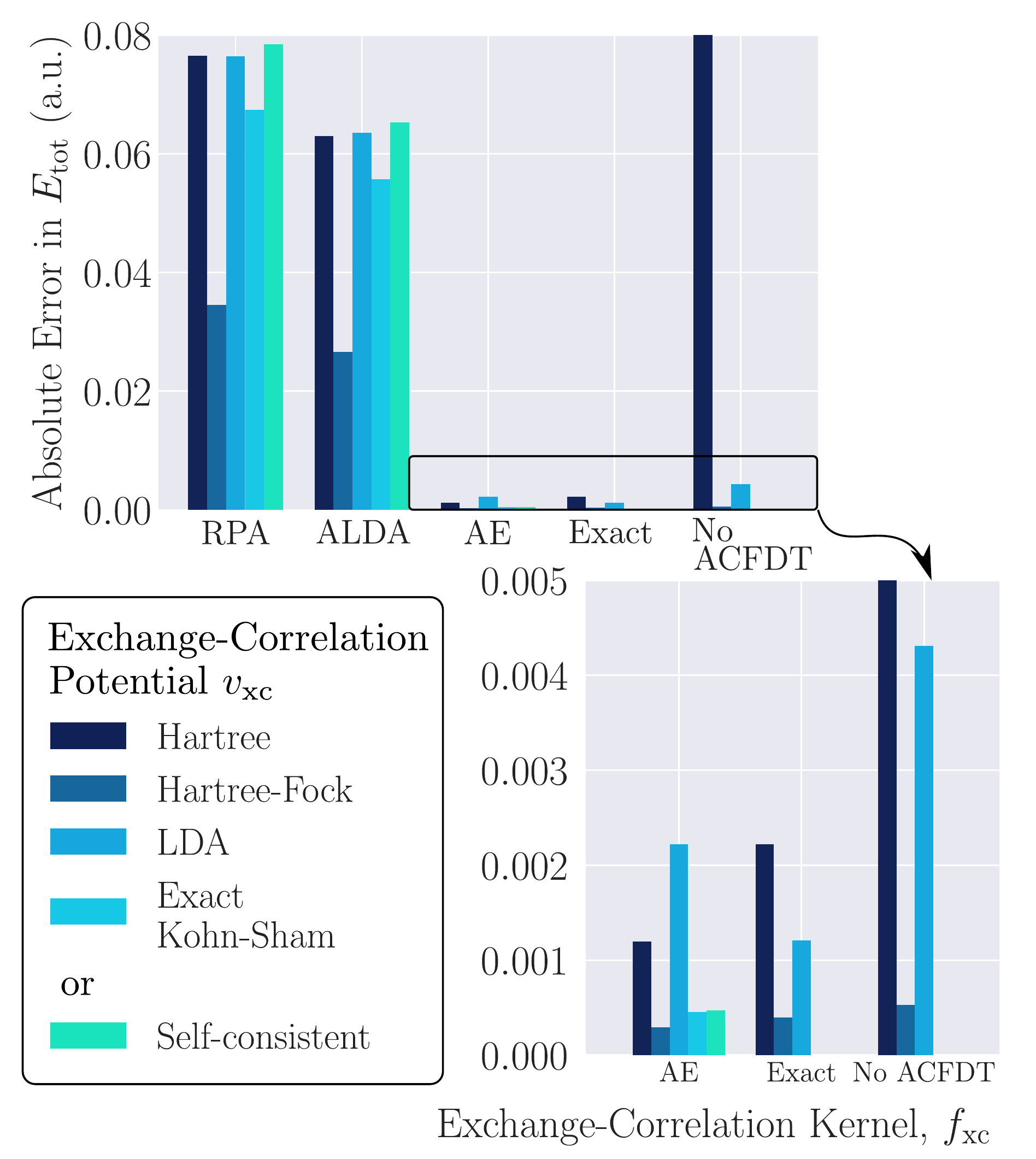}
\end{center}
\caption{Absolute error in the infinite potential well total energy across a variety of approximate ACFDT total energies, see caption of Figure 3 in the main text.}
\label{fig:InfPotWellErrors}
\end{figure}

This system has total energy $E_\text{tot} = 0.100$ a.u., correlation energy $E_\text{c} = -0.00086$ a.u., and exchange energy $E_\text{x} = -0.326$ a.u. The relative errors in the ACFDT total energy across the range of approximate xc potentials and xc kernels considered in this work are given in Fig$.$ \ref{fig:InfPotWellErrors}.
 
\subsection{Slab}
\label{sec:slab}

\begin{figure}[h!]
\begin{center}
\includegraphics[width=3.3in]{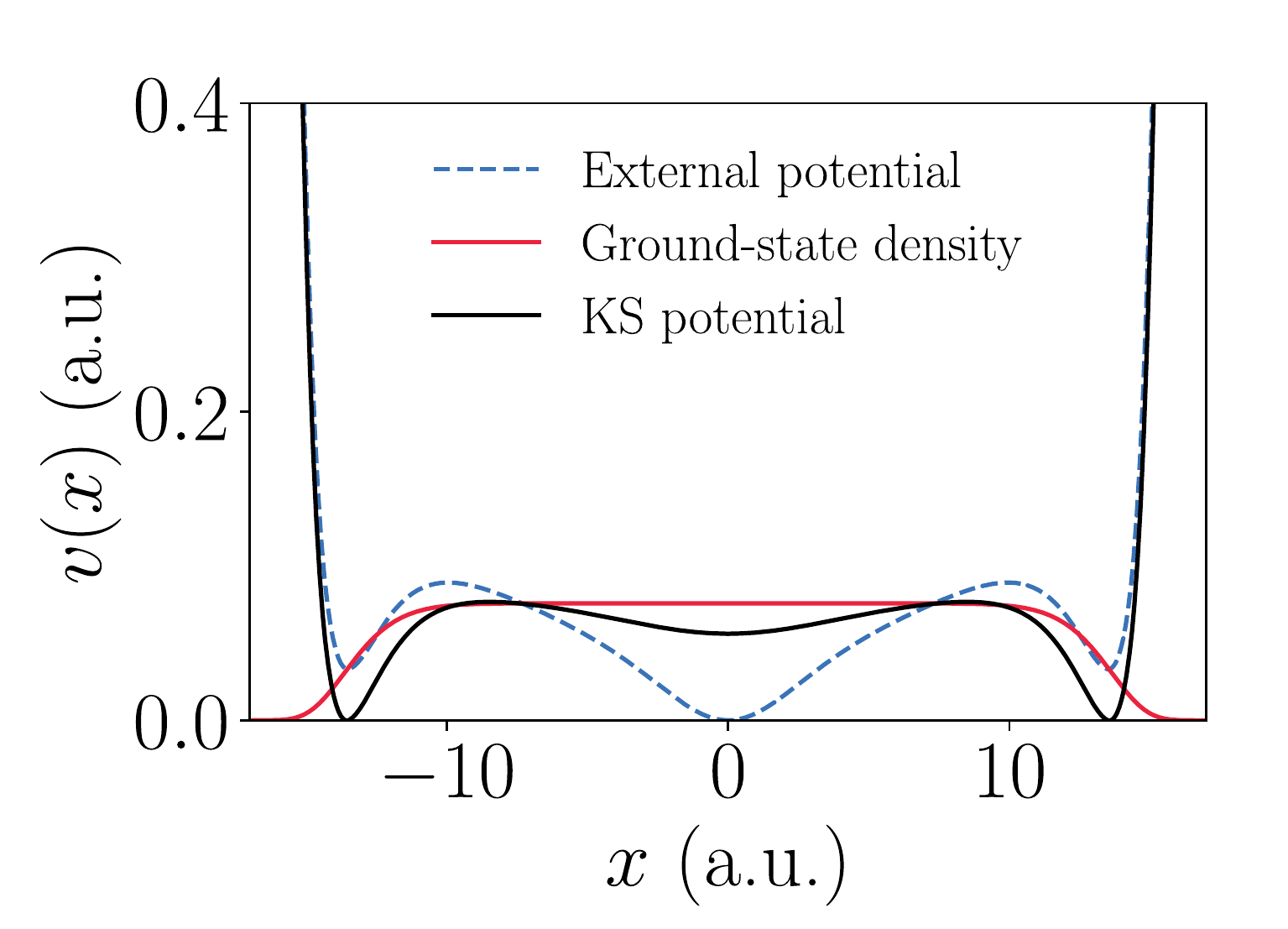}
\end{center}
\caption{The ground-state density, reverse-engineered external potential, and reverse-engineered Kohn-Sham potential for the slab system. The external and Kohn-Sham potentials have been shifted for illustrative purposes.}
\label{fig:Slab}
\end{figure}

\begin{figure}[ht]
\begin{center}
\includegraphics[width=3.4in]{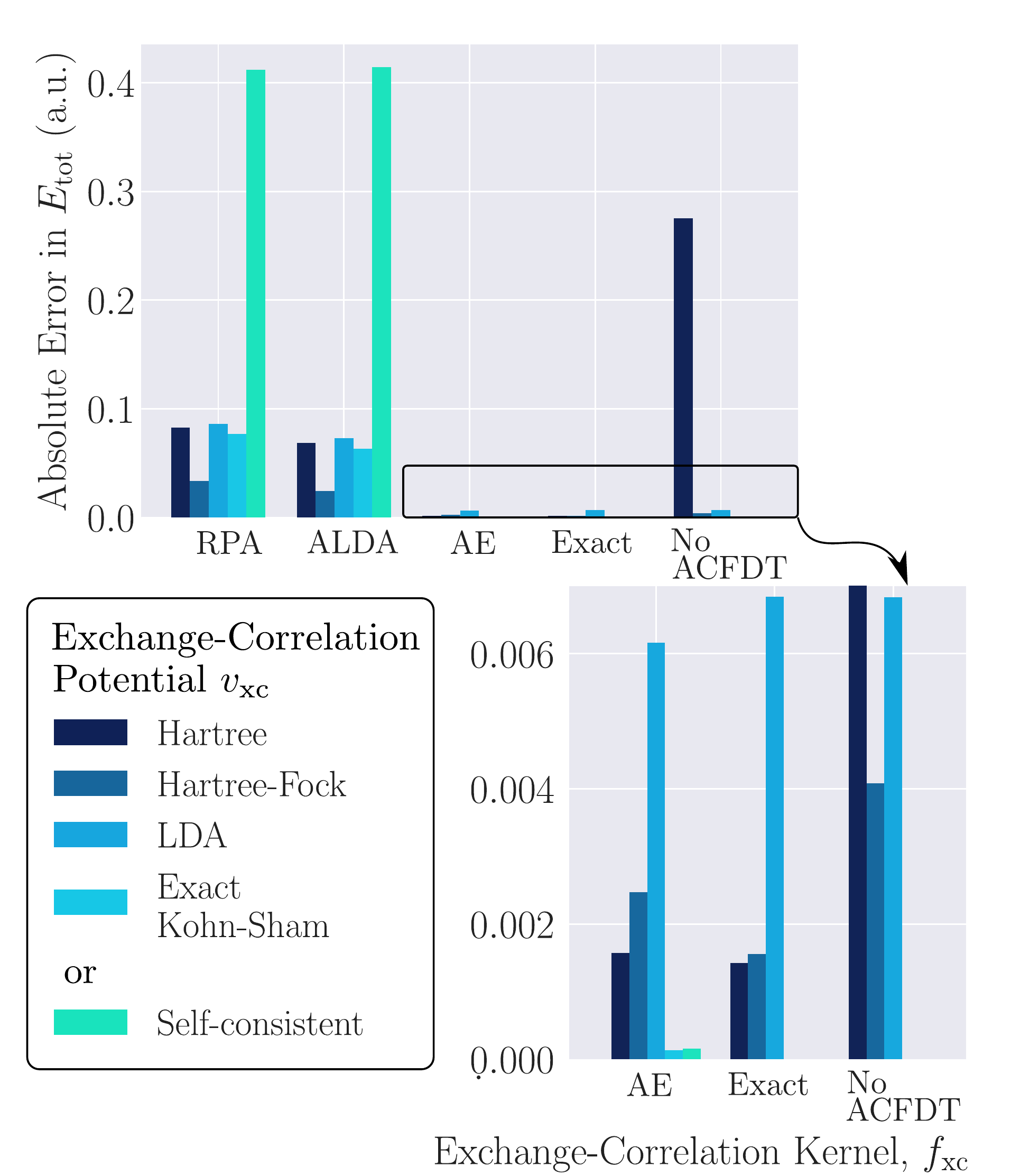}
\end{center}
\caption{Absolute error in the slab correlation energy across a variety of approximate ACFDT correlation energies, see caption of Figure 3 in the main text.}
\label{fig:SlabErrors}
\end{figure}

The slab system has an external potential that has been reverse-engineered in order to produce a slab-like density \citep{Entwistle2018} inside the domain $[-17,17]$ a.u., see Fig$.$ \ref{fig:Slab}. This system has total energy $E_\text{tot} = 0.221$ a.u., correlation energy $E_\text{c} = -0.00535$ a.u., and exchange energy $E_\text{x} = -0.277$ a.u. The relative errors in the ACFDT \textit{correlation} energy across the range of approximate xc potentials and xc kernels considered in this work is given in Fig$.$ \ref{fig:SlabErrors}.

\begin{figure}[ht]
\begin{center}
\includegraphics[width=3.3in]{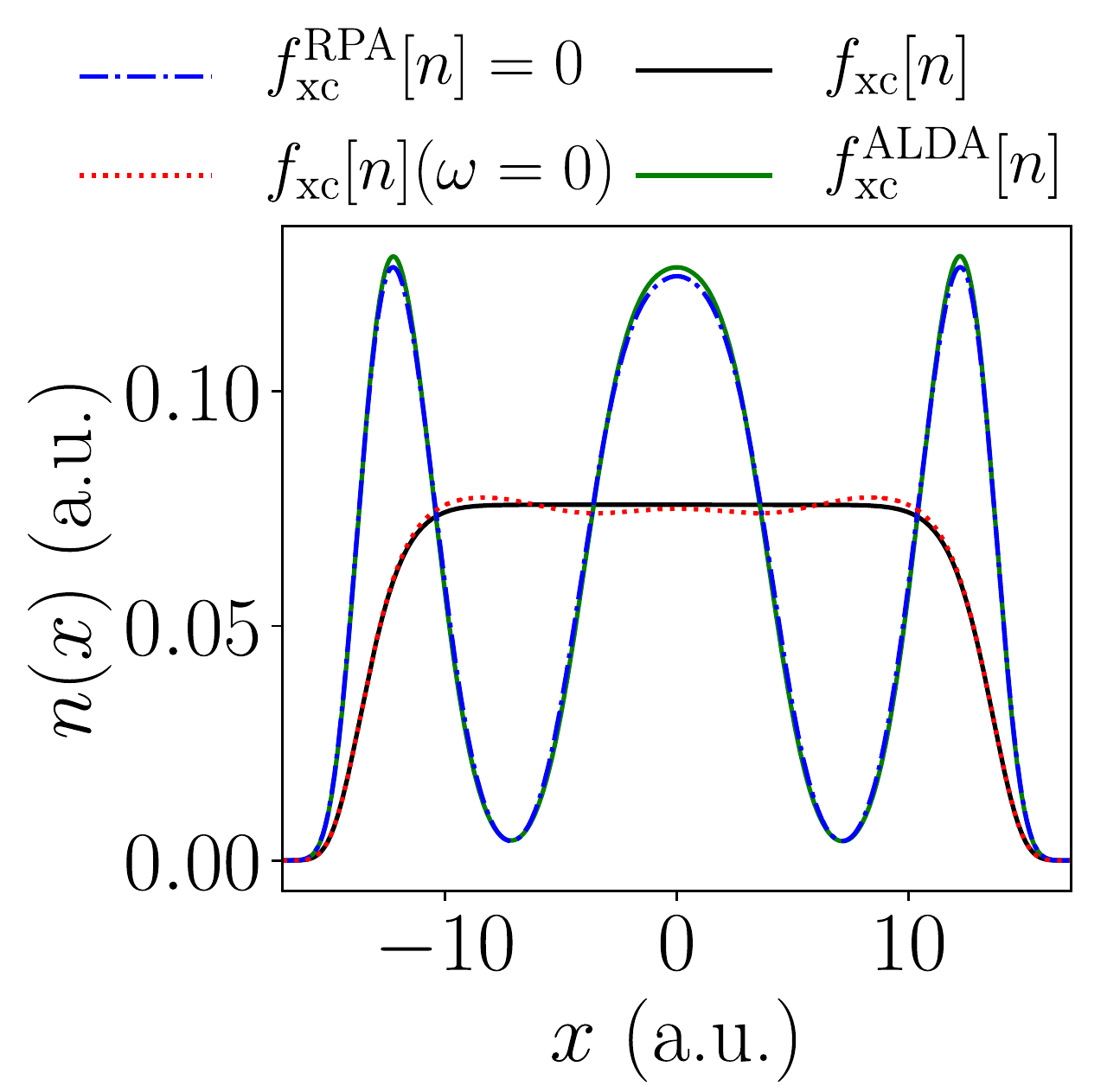}
\end{center}
\caption{The density $n$ that minimizes the slab ACFDT energy functional $E^\text{ACFD}[n]$ specified with some approximate $f_\text{xc}[n]$. The minimizing density when the \textit{exact} $f_\text{xc}[n]$ is used is the interacting ground-state density (red solid).}
\label{fig:SlabMinimisingDensities}
\end{figure}

The slab is a delicate system: small changes in the external potential yield large (qualitative) changes in the density. As such, the densities that minimize a given approximate ACFDT total energy functional and the interacting density differ the most in the case of the slab, see Fig$.$ \ref{fig:SlabMinimisingDensities}. This also manifests in the `self-consistent' errors from Fig$.$ \ref{fig:SlabErrors}. 

\subsection{Double Well}

The double well \citep{Hodgson2016,Almbladh1985} is defined with
\begin{align}
v_\text{ext} = -\frac{6}{5} e^{-\frac{1}{125} (x - 7)^4} - \frac{9}{10} e^{-\frac{1}{10} (x + 7)^2}
\end{align}
inside the domain $[-12,12]$ a.u., see main text. This system has total energy $E_\text{tot} = -1.71$ a.u., correlation energy $E_\text{c} = -1.33 \times 10^{-6}$ a.u., and exchange energy $E_\text{x} = -0.506$ a.u.

\begin{figure}[ht]
\begin{center}
\includegraphics[width=3.4in]{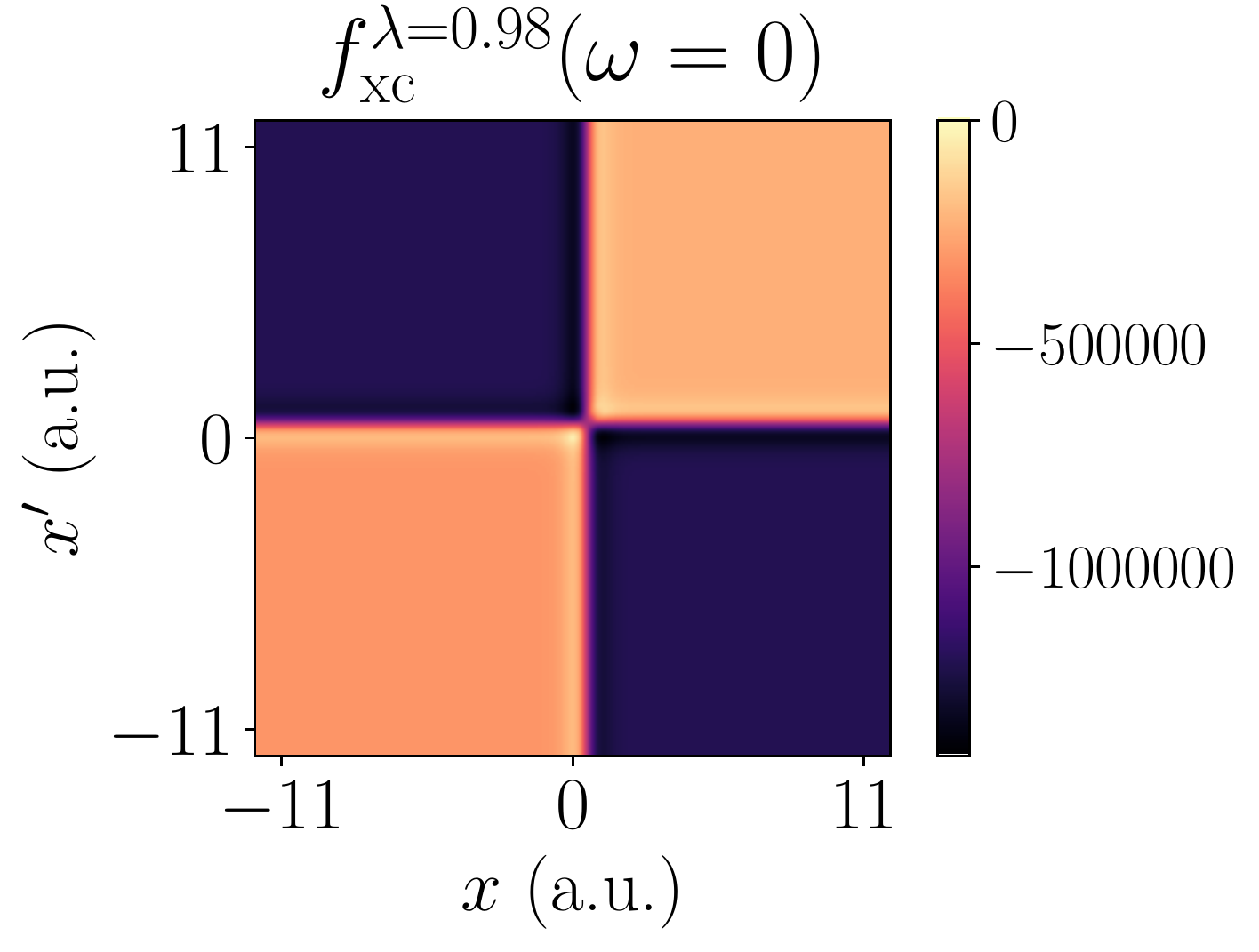}
\end{center}
\caption{The exact $f^\lambda_\text{xc}(x,x',\omega=0)$ for the double well system at $\lambda = 0.98$, i.e. at a discrete value of $\lambda$ that is sampled along the adiabatic connection. }
\label{fig:DoubleWellXCKernel}
\end{figure}

In similar fashion to the Kohn-Sham potential, the exact $f_\text{xc}[n]$ contains a discontinuous step feature, see Fig$.$ \ref{fig:DoubleWellXCKernel}, that has been studied in \citep{2Hellgren2012,Hellgren2013,Hellgren2018}. The magnitude of the step here is large, as can be seen in the labeled color bar in Fig$.$ \ref{fig:DoubleWellXCKernel}. One approach to understanding this behavior is that $f_\text{xc}$ must be able to capture charge-transfer excitations \citep{Maitra2021,Maitra2017} within the Casida equation. A simple expression derived from the Casida approximation is obtained by invoking the single-pole approximation, which shows that $f_\text{xc}$ must diverge in proportion to the increasing separation between the two subsystems.

\section{Analytic Frequency Integration}

\begin{figure}[ht]
\begin{center}
\includegraphics[width=3.4in]{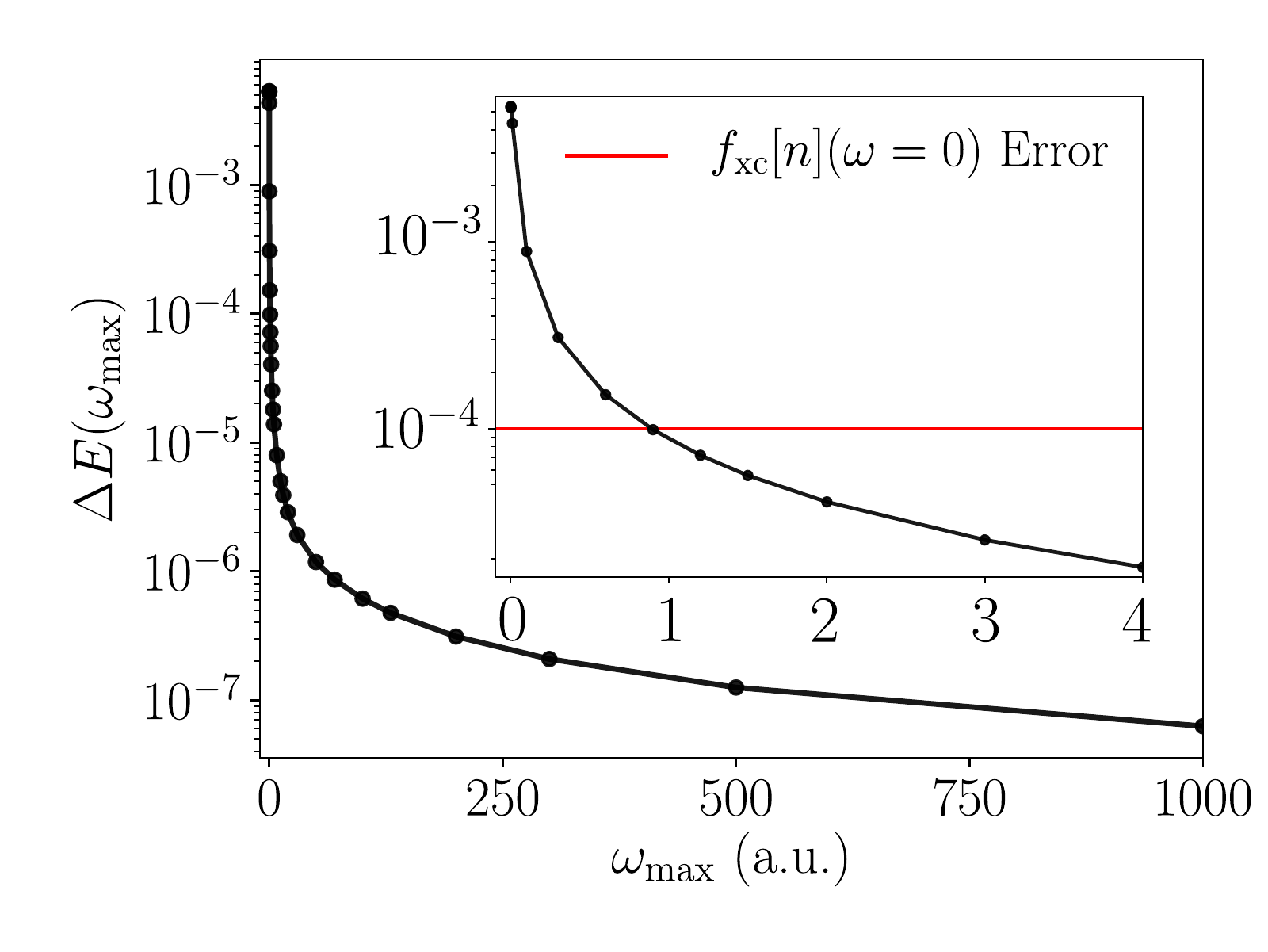}
\end{center}
\caption{Absolute error in the slab correlation energy (log scale) with increasing $\omega_\text{max}$ (upper bound in the ACFDT $\omega$-dependent integral). The curve tends toward zero in the $\omega_\text{max} \rightarrow \infty$ limit.  The inset enlarges the low-$\omega$ region, where it can be seen that the error involved in an AE one-shot ACFDT calculation (red) is contained within $0 < \omega < 1$.}
\label{fig:SlabFreqIntegrationError}
\end{figure}

Evaluating the $\omega$-dependent integral in $E_\text{c}^\text{ACFD}[n]$ is tantamount to evaluating the integral in Eq$.$ (\ref{eq:ExactChiIntegration}) from Section \ref{sec:FreqIntScheme}. Since we have access to the interacting wavefunctions and energies along the adiabatic connection, we also have access to the exact excitation functions $f^\lambda_n(x)$. The integral in Eq$.$ (\ref{eq:ExactChiIntegration}) can therefore be evaluated up to some $\omega_\text{max}$ as such
\begin{align}
\sum_{n} f_n(x) f_n(x') \int_0^{\omega_\text{max}}& \frac{\Omega_n^2}{\omega^2 + \Omega_n^2} \ d\omega =  \\
&\sum_n f_n(x) f_n(x') \arctan \left(\frac{\omega_\text{max}}{\Omega_n} \right). \nonumber
\end{align}
This allows us to isolate error coming solely from the finite upper limit $\omega_\text{max}$ in the $\omega$-dependent integral. The error in the correlation energy as a function of $\omega_\text{max}$ can be seen in Fig$.$ \ref{fig:SlabFreqIntegrationError}.

\bibliographystyle{apsrev4-2}
\bibliography{suppReferences}